Title Page

# Unveiling Normative Trajectories of Lifespan Brain Maturation Using Quantitative MRI

Authorship:

Xinjie Chen[1, 2, 3], Mario Ocampo-Pineda[1, 2, 3], Po-Jui Lu[1, 2, 3], Clara Ekerdt[4], Matthias Weigel[1, 2, 3, 5], Michelle G. Jansen[4], Alessandro Cagol[1, 2, 3, 6], Kwok-Shing Chan[7,8], Sabine Schädelin[1, 2, 3], Marcel Zwiers[4], Joukje M. Oosterman[4], David G. Norris[4], Johanna M. M. Bayer[4], Andre F. Marquand[4], Willeke M. Menks[4], Jens Kuhle[2, 3], Ludwig Kappos[2, 3], Lester Melie-Garcia[1, 2, 3], Cristina Granziera[1, 2, 3], José P. Marques[4]

Affiliations:

1. Translational Imaging in Neurology (ThINK) Basel, Department of Biomedical Engineering, Faculty of Medicine, University Hospital Basel and University of Basel, Basel, Switzerland
2. Department of Neurology, University Hospital Basel, Basel, Switzerland
3. Research Center for Clinical Neuroimmunology and Neuroscience Basel (RC2NB), University Hospital Basel and University of Basel, Basel, Switzerland
4. Donders Institute for Brain, Cognition and Behaviour, Radboud University, Nijmegen, the Netherlands
5. Division of Radiological Physics, Department of Radiology, University Hospital Basel, Basel, Switzerland
6. Department of Health Sciences, University of Genova, Genova, Italy
7. Athinoula A. Martinos Center for Biomedical Imaging, Charlestown, MA, United States
8. Department of Radiology, Harvard Medical School, Boston, MA, United States

Acknowledgments

The authors thank the participants of the MRI studies for their valuable contributions. The ABRIM study was possible thanks to the European FP7 program, FP7-PEOPLE-2013-ITN, Marie-Curie Action, and "Initial Training Networks" named "Advanced Brain Imaging with MRI" (no. 608123). The Variability in Language Learning study was funded by the Netherlands Organization for Scientific Research (NWO) Gravitation grant 'Language in Interaction' (grant number 024.001.006).

Conflict of Interests

Xinjie Chen has nothing to disclose. Mario Ocampo-Pineda has nothing to disclose. Po-Jui Lu has nothing to disclose. Clara Ekerdt has nothing to disclose. Matthias Weigel: has received research funding from Biogen for developing spinal cord MRI in the past. Michelle G. Jansen has nothing to disclose. Alessandro Cagol is supported by the Horizon 2020 Eurostar program (grant E!113682) and received speaker honoraria from Novartis. Kwok-Shing Chan has nothing to disclose. Sabine Schädelin has nothing to disclose. Marcel Zwiers has nothing to disclose. Joukje M. Oosterman has nothing to disclose. David G. Norris has nothing to disclose. Andre F. Marquand has nothing to disclose. Johanna Bayer has nothing to disclose. Willeke M. Menks has nothing to disclose. Jens Kuhle has nothing to disclose. Ludwig Kappos has nothing to disclose. Lester Melie-Garcia has nothing to disclose. Cristina Granziera as the employer of the University Hospital Basel (USB), has received the following fees which were used exclusively for research support: (i) advisory boards, and consultancy fees from Actelion, Novartis, Genzyme-Sanofi, GeNeuro, Hoffmann La Roche and Siemens; (ii) speaker fees from Biogen, Hoffmann La Roche, Teva, Novartis, Merck, Jannsen Pharmaceuticals and Genzyme-Sanofi; (iii) research grants: Biogen, Genzyme Sanofi, Hoffmann La Roche, GeNeuro. José P. Marques has nothing to disclose.

Ethics Approval

The ABRIM cohort adhered to the principles of the Helsinki Declaration. It was covered under the blanket ethics approval "Image Human Cognition" granted by the Commissie Mensgebonden Onderzoek Arnhem-Nijmegen (2014/288). Additionally, the study received approval from the Social Sciences Ethical Committee of Radboud University (ECSW 2017-3001-46), ensuring compliance with all local procedures and applicable national legislation. All participants provided written informed consent before their participation. The INsIDER study received approval from the Ethics Committee Northwest/Central Switzerland (EKNZ). All participants provided written consent before their enrollment, ensuring adherence to ethical guidelines. The Variability in Language Learning study was approved by the regional ethics committee Commissie Mensgebonden Onderzoek Arnhem-Nijmegen (2019/5975). All participants provided written informed consent. For participants younger than 18 years, the parents/guardians additionally provided written informed consent for their child's participation.

Data Availability Statement
The in-house scripts for image processing are available on GitHub (https://github.com/JosePMarques/MP2RAGE-related-scripts). In addition to the in-house scripts, other image-processing tasks were performed using software and codes including FreeSurfer (https://surfer.nmr.mgh.harvard.edu), FSL (https://fsl.fmrib.ox.ac.uk/fsl/fslwiki/), and SEPIA (https://github.com/kschan0214/sepia). To protect the privacy of the subjects, MRI and related data cannot be made publicly accessible. However, all codes used for statistical analysis from this manuscript can be obtained from the corresponding author, José Marques, and the first author, Xinjie Chen, upon reasonable request (jose.marques@donders.ru.nl; xinjie.chen@unibas.ch). All experiments and implementation details are thoroughly described in the Methods section and Supplementary Materials.

Corresponding author:

José P. Marques, Ph.D.

Kapittelweg 29

6525 EN, Nijmegen

The Netherlands

Tel number: +31 (0) 63 11 32 616

Email: jose.marques@donders.ru.nl

# Abstract

Background

Brain maturation and aging involve significant microstructural changes, resulting in functional and cognitive alterations. Quantitative MRI (qMRI) can measure this evolution, distinguishing the physiological effects of normal aging from pathological deviations.

Methods

We conducted a multicentre study using qMRI metrics (R1, R2*, and Quantitative Susceptibility Mapping) to model age trajectories across brain structures, including tractography-based white matter bundles (TWMB), superficial white matter (SWM), and cortical grey matter (CGM). MRI data from 537 healthy subjects, aged 8 to 79 years, were harmonized using two independent methods. We modeled age trajectories and performed regional analyses to capture maturation patterns and aging effects across the lifespan.

Findings

Our findings revealed a distinct brain maturation gradient, with early qMRI peak values in TWMB, followed by SWM, and culminating in CGM regions. This gradient was observed as a posterior-to-anterior maturation pattern in the cortex and an inferior-to-superior maturation pattern in white matter tracts. R1 demonstrated the most robust age trajectories, while R2* and susceptibility exhibited greater variability and different patterns. The normative modeling framework confirmed the reliability of our age-modelled trajectories across datasets.

Interpretation

Our study highlights the potential of multiparametric qMRI to capture complex, region-specific brain development patterns, addressing the need for comprehensive, age-spanning studies across multiple brain structures. Various harmonization strategies can merge qMRI cohorts, improving the robustness of qMRI-based age models and facilitating the understanding of normal patterns and disease-associated deviations.

# Introduction

Brain aging is characterized by fundamental microstructural alterations, including iron accumulation, myelin degradation, and morphological atrophy, leading to functional decline, increased risk for neurological disorders, and mortality (MacDonald & Pike, 2021; Rouault, 2013; Tomasi & Volkow, 2012, p. 4). Understanding the microstructural changes in brain maturation is crucial for elucidating the fundamental processes underlying neural aging and pathology (Ouyang et al., 2019). This knowledge provides insights into basic structural evolution, cognitive development, functional decline, and susceptibility to neurological disorders (Montine et al., 2019; Zilles et al., 2013), which hopefully can help clarify the biological mechanisms underpinning brain lifespan development (Sharma et al., 2013).

Brain maturation patterns, when described using cortical thickness and white matter (WM) volume changes, follow established distinct trajectories influenced by functional and anatomical hierarchies (Brenhouse & Andersen, 2011; Mills et al., 2014). Specifically, cortical grey matter (CGM) development follows a posterior-to-anterior maturation gradient, reaching early peak thickness in sensory and motor regions in the second and third decades (Tamnes et al., 2010, 2013). CGM maturation is characterized by an increase in synaptogenesis, dendritic complexity, and synaptic pruning, contributing to structural and functional development from infancy through adulthood (K. M. Harris & Weinberg, 2012; Paolicelli et al., 2011). WM maturation, as characterized by MRI techniques, follows a posterior-to-anterior, inferior-to-superior, and central-to-peripheral pattern (Lebel & Deoni, 2018; Yakovlev & Yakovlev, 1967), marked by myelination, axonal elongation, and glial proliferation (Nave & Werner, 2014), which enhances neural signal transmission (J. J. Harris & Attwell, 2012). Deep WM tracts, which facilitate long-range neural connectivity, have a rapid maturation during the first 2 years of development, (Dubois et al., 2014; Yu et al., 2020), which typically leads to an inversion of contrast between grey matter (GM) and WM in conventional MRI at birth. In contrast, superficial white matter (SWM), composed of U-fibers beneath the cortical mantle, matures more gradually into middle adulthood, enhancing connectivity between cortical areas and deeper WM, which supports the integration of higher-order functions (Friedrichs-Maeder et al., 2017; Wu et al., 2014, 2016).

Quantitative MRI (qMRI) extends conventional MRI by offering quantitative metrics for assessing microstructural changes associated with brain maturation patterns and the aging process (Granziera et al., 2021; Tofts, 2003) that can be reproduced and compared across subjects. While conventional MRI primarily provides qualitative images for anatomical visualization and morphological studies focused on regional volumes and cortical thickness (Krauss et al., 2018), qMRI metrics - such as $R_1$ (longitudinal relaxation rate, $R_1 = 1/T_1$), $R_2^*$ (apparent transverse relaxation rate, $R_2^* = 1/T_2^*$), and Quantitative Susceptibility Mapping (QSM) – can reliably and quantitatively measure a range of tissue properties (Granziera et al., 2021; Marques et al., 2017a). $R_1$ is largely influenced by water content, tissue density, and the size and rate of the exchange processes between intracellular and extracellular free water pools with the macromolecular pool (Van Gelderen et al., 2016; Y. Wang et al., 2020). Through this process, $R_1$ is sensitive to microstructural tissue changes associated with myelination in aging (Eminian et al., 2018; Grotheer et al., 2022; Kühne et al., 2021) and to a much smaller extent iron(Stüber et al., 2014). $R_2^*$ quantifies the rate of transverse magnetization decay, which is affected by local magnetic field inhomogeneities, primarily due to iron deposits (Barbosa et al., 2015). However, $R_2^*$ also exhibits residual sensitivity to myelin concentration and fibre bundle orientation, particularly in healthy WM (Gil et al., 2016; Oh et al., 2013). $R_2^*$ has been used in various studies to map iron distribution, particularly on deep grey matter (DGM) (Bagnato et al., 2018; Marques et al., 2017a). QSM estimates tissue magnetic susceptibility, reflecting the magnetization in an external magnetic field. This technique can be used to detect variations in the distribution of iron (which linearly increases local susceptibility) and myelin (which decreases local susceptibility), capturing the opposite signs of paramagnetic iron and diamagnetic myelin compared to $R_2^*$. (Hédouin et al., 2021; Marques et al., 2021). The distinct impact of myelin and iron on QSM and $R_2^*$ helps differentiate and understand the biophysical mechanisms seen on those qMRI metrics (Bagnato et al., 2018; QSM Consensus Organization Committee et al., 2024).

Despite the potential of qMRI, significant challenges persist in improving the accuracy of qMRI measurements and in linking them to brain microstructural changes (Gulani & Seiberlich, 2020). Previous research has often been limited to single-site datasets, which, although maintaining consistency in data sources, restricts the generalizability of findings and their applicability across different imaging protocols (Slater et al., 2019). Although qMRI provides quantitative metrics that should theoretically be comparable across MRI protocols, (Weiskopf et al., 2013) in practice, these measures are sensitive to the specifics of MRI acquisition and post-processing methods, particularly when assessing subtle variations in relaxation values (A G Teixeira et al., 2019, 2020; Karakuzu et al., 2022). To mitigate this issue, neuroCombat, a widely used harmonization method based on empirical Bayes statistics (EBS), can be effectively applied, though it has certain limitations, including sensitivity to assumptions of homogeneous variability across batches and the potential loss of site-specific nuances (Johnson et al., 2007). In contrast, another option based on the Hierarchical Bayesian Regression (HBR) framework provides a method that accounts for hierarchical data structures, site-specific variations, and heterogeneous variance across sites and covariates (Kia et al., n.d.). However, compared with EBS, HBR's increased computational complexity and requirement for large datasets (de Boer et al., 2024) present challenges. Nevertheless, HBR is expected to effectively address the challenges of multi-center site variance and protocol discrepancies in qMRI data analyses(Villalón-Reina et al., 2022).

Investigations using qMRI techniques into brain aging have primarily focused on single metrics within a restricted set of brain structures. For instance, Yeatman et al. used $R_1$ to model lifespan changes in brain tissue, demonstrating that $R_1$ development rates could predict degeneration rates, based on the assumption that $R_1$ in WM was closely associated with macromolecular tissue volume (Yeatman et al., 2014). Similarly, Zhang et al. and Wei et al. estimated age-related changes in iron and myelin content using QSM (Li et al., 2014; Zhang et al., 2018). Knight et al. employed $T_2$ mapping and diffusion tensor imaging (DTI) to access WM alterations in normal aging (Knight et al., 2016). While a substantial body of research has focused on qMRI changes in DGM and WM bundles, investigations into CGM and SWM changes remain limited. One study using multiparametric qMRI found that cortical $T_1$ values decreased and $T_2$ values increased with age, with $T_2$ mapping indicating significant global cortical iron deposition as a key process in normal aging (Seiler et al., 2020). Another study comparing $R_1$ and cortical thickness in predicting age-related changes in CGM, found that both metrics predicted age similarly, albeit with regional differences (Erramuzpe et al., 2021). Research on regional CGM remains underexplored, with most studies broadly categorizing the brain into major anatomical regions, such as frontal, parietal, temporal, and occipital lobes. However, a more detailed understanding of the development and degeneration of specific regional cortex areas, including primary and associative cortices, is essential for a comprehensive understanding of cortical maturation and aging. Additionally, many existing qMRI studies lack broad age spectrum coverage and are conducted in small cohorts (Giedd et al., 1996; Keuken et al., 2017; Weiskopf et al., 2013). Comprehensive studies spanning the entire age range are essential for a cross-sectional understanding of normal brain aging.

To address these knowledge gaps, this study aimed to (i) conduct a multiparametric mapping analysis to investigate their changes across ages in different brain structures, (ii) perform a comprehensive regional analysis to identify specific age-related patterns in different brain areas, and (iii) explore whether harmonization strategies can be effectively used to integrate large multicentric cohorts with non-overlapping age spectrums and mismatched qMRI protocols.  In this way, our research sought to provide a nuanced understanding of brain maturation and aging patterns across a broad demographic spectrum and a set of qMRI metrics.

# Methods

## Data and Ethics

*Study population*

This multi-center study included 541 healthy participants across three sites (Jansen et al., 2024; Menks et al., 2022; University Hospital, Basel, Switzerland, 2024). Healthy participants were defined as individuals with no known neurological disorders, major systemic illnesses, significant psychiatric conditions, or history of major surgeries affecting the brain or central nervous system. Due to data quality issues, 5 participants were excluded, resulting in a final sample size of 536 participants. The cohort's demographic characteristics are reported in Table 1.

*Ethics approval*

Site 1: The ABRIM cohort adhered to the principles of the Helsinki Declaration. It was covered under the blanket ethics approval "Image Human Cognition" granted by the Commissie Mensgebonden Onderzoek Arnhem-Nijmegen (2014/288). Additionally, the study received approval from the Social Sciences Ethical Committee of Radboud University (ECSW 2017-3001-46), ensuring compliance with all local procedures and applicable national legislation. All participants provided written informed consent before their participation.

Site 2: The INsIDER study received approval from the Ethics Committee Northwest/Central Switzerland (EKNZ). All participants provided written consent before their enrollment, ensuring adherence to ethical guidelines.

Site 3: The Variability in Language Learning study was approved by the regional ethics committee Commissie Mensgebonden Onderzoek Arnhem-Nijmegen (2019/5975). All participants provided written informed consent. For participants younger than 18 years, the parents/guardians additionally provided written informed consent for their child's participation.

## Image acquisition

The imaging protocols included Magnetization Prepared 2 Rapid Acquisition Gradient Echoes (MP2RAGE)(Marques et al., 2010) and multi-shell diffusion in all sites, and Multi-Echo Gradient Recalled Echo (ME-GRE) sequences (Y. Wang & Liu, 2015) in Site 1 and 2. Images were acquired with a 3T MRI system Magnetom Prisma (Siemens Healthcare, Erlangen, Germany) for Sites 1 and 2 and Magnetom Skyra (Siemens Healthcare, Erlangen, Germany) for Site 3. Acquisition parameters for each site are detailed in Table 2. Note that Site 3 acquired data from a younger population and as such used compressed sensing acceleration of the MP2RAGE acquisition to reduce the likelihood of movement artifacts in this relatively younger population (aged 8-25 years old; Table 1) (Mussard et al., 2020).

## Image preprocessing and computation of qMRI maps

Preprocessing steps were conducted separately for the MP2RAGE and ME-GRE sequences to generate $R_1$, $R_2^*$, and susceptibility maps. For the MP2RAGE data to generate the $T_1$ maps, preprocessing involved transmit field inhomogeneity correction (B1+ correction) to address variations in the radiofrequency field (Marques & Gruetter, 2013). From the $T_1$ maps, $R_1$ maps were calculated ($R_1=1/T_1$) and these values will be discussed throughout this work. Background noise removal of the MP2RAGE data was obtained using in-house scripts (O'Brien et al., 2014).

For the ME-GRE data, preprocessing included QSM phase unwrapping and background field removal, followed by $R_2^*$ and susceptibility maps reconstruction using the SEPIA toolbox (version 1.2.2.4)(Chan & Marques, 2021). The pipeline used has been extensively described in a publication using Site 1 data (Jansen et al., 2024) and relied on the following methods: ROMEO for field calculation (Dymerska et al., 2021); V-SHARP for background field removal (Li et al., 2011); LPCNN for dipole inversion (Lai et al., 2020); ARLO technique for $R_2^*$ maps computation (Pei et al., 2015).

$R_2^*$ and susceptibility maps were co-registered to the $R_1$ space. A rigid body transformation matrix between the first echo of the ME-GRE sequence and the second inversion time image from the MP2RAGE sequence was obtained using FSL (Jenkinson et al., 2012) and was subsequently applied to bring the $R_2^*$ and susceptibility maps to the $R_1$ map space.

Diffusion-weighted images (DWIs) were preprocessed using the default settings of QSIprep (0.18.0) (Cieslak et al., 2021) including denoising and motion correction.

## Cortical Parcellation and White Matter Tractography

*CGM and SWM parcellation*

For CGM and SWM segmentation, the FreeSurfer (version 6.0)(Fischl, 2012) recon-all pipeline was used on the MP2RAGE data, following the methodology outlined in (Fujimoto et al., 2014). Cortical parcellation was performed using the in-house refined PALS-B12 Brodmann Atlas originally provided by FreeSurfer, extracting 41 Brodmann areas (BAs) across the entire cortex (Van Essen, 2005). The BAs were systematically categorized into five principal regions: frontal, parietal, temporal, occipital, and mixed transitional areas (medial temporal lobe and proximity to DGM structures). Manual quality control steps including skull-stripping correction, pial surface error correction, and intensity normalization were implemented before and after segmentation to ensure the accuracy of the parcellation and the extraction of quantitative measurements.

*WM bundle identification*

To obtain bundle information, a pyAFQ (Kruper et al., 2021) pipeline of QSIprep was applied. This pipeline uses multi-shell and multi-tissue spherical deconvolution methods to estimate fiber orientations to extract a specific set of eighteen bundles, including Callosum Forceps Minor (FA), Callosum Forceps Major (FP), and the following bilateral tracts:

Arcuate (ARC), Posterior Arcuate Fasciculus (pARC), Thalamic Radiation (ATR), Cingulum Cingulate (CGC), Corticospinal (CST), Inferior Fronto-Occipital Fasciculus (IFO), Inferior Longitudinal Fasciculus (ILF), and Superior Longitudinal Fasciculus (SLF).

## Quantitative Mapping Across Brain Structures

*CGM and SWM analysis*

Quantitative surface maps were created by projecting qMRI maps onto the brain surface derived from FreeSurfer. qMRI values in CGW and SWM were extracted in the middle cortical layer between the white and pial surfaces, and below the white surface boundary using FreeSurfer with settings 0.5 and -0.5 respectively (Fischl, 2012).

Finally, the average qMRI measurements ($R_1$, $R_2^*$, and Susceptibility values) derived from the individual medians were calculated for each BA, considering only vertices that passed automatic quality assurance per region as described in a previous study ((Shams et al., 2019), with values for each BA averaged individually for two hemispheres.

*Tract-Based Analysis*

Fibre tracts were co-registered to $R_1$ space (in the same space with $R_2^*$ and QSM) using the transformation matrix computed from the flirt (Jenkinson et al., 2012) co-registration of DWI data (brain-masked and distortion-corrected images) with the brain-masked second inversion time images of the MP2RAGE. The SCILPY toolbox (*GitHub - Scilus/Scilpy: The Sherbrooke Connectivity Imaging Lab (SCIL) Python dMRI Processing Toolbox*, n.d.) was used to extract median qMRI values in each fiber tract. To ensure that the measurements were focused on the core WM tissue and devoid of GM partial volume effects (PVE), only the central 70% of each fiber tract was retained (excluding the distal 15% of each end).

## Age Trajectory Modelling

There are various models to study age trajectories. To evaluate which model is most appropriate for the qMRI data, we first applied the HBR framework (as implemented in the PCN toolkit (Marquand et al., 2016)) to test both a quadratic regression and a cubic B-spline (HBRbs) model. These were computed with a 5-fold cross-validation and their goodness of fit was evaluated using two key performance metrics: the Mean Standardized Log Loss (MSLL)(Williams & Rasmussen, 1995) and out-of-sample Explained Variance (EXPV). Details on the HBRbs model can be found in Supplementary Materials (S.1)(Betancourt & Girolami, 2015; de Boer et al., 2024). To compare the performance of the HBRbs and polynomial regression (PR) models, we used the EXPV Ratio (HBRbs/PR) to quantify the relative ability of the HBRbs and PR models to capture the variance related to age dependencies in qMRI metrics using only Site 1 data, which covered the widest age range and included the largest number of subjects. A higher out-of-sample EXPV Ratio indicates that the HBRbs model captures more variability compared to the PR model.

To facilitate the interpretation of the different trajectories per region of interest (ROI) across CGM, SWM, and TWMB, the aging effect on qMRI metrics was assessed employing a second-order PR model applied to the whole site 1 cohort:

$$qMRI_{ROI} = \beta_{0ROI} + \beta_{1ROI}*Age + \beta_{2ROI}*Age^2 + \beta_{3ROI}*Sex \quad [Eq. 1]$$

This model included age, age², and sex as covariates. $\beta_{1ROI}$ and $\beta_{2ROI}$ are the coefficients of the linear and quadratic terms, respectively. A likelihood-ratio test (LRT) was used to confirm the significance of the quadratic term.

The peak of the qMRI trajectory and its Standard Error (SE) were calculated from the quadratic regression equation and using error propagation ([Eq. 2-3]).

$$Age_{peak} = -\beta_{1ROI} / (2*\beta_{2ROI}) \quad [Eq. 2]$$

$$SE_{peak} = \sqrt{(\frac{SE_{\beta_{1ROI}}}{2\times\beta_{2ROI}})^2 + (\frac{\beta_{1ROI}\times SE_{\beta_{2ROI}}}{2\times\beta_{2ROI}^2})^2} \quad [Eq. 3]$$

When the quadratic coefficient was non-zero, the peak age was determined at the maximum or minimum of the quadratic curve. If the quadratic coefficient was not significant, the peak age was not defined.

Pearson correlation matrices (followed by Benjamini-Hochberg correction) were calculated using pairwise complete observations to explore intra- and inter-regional covariance and age-related influences within CGM and SWM. Correlation matrices were computed using $qMRI_{ROI_i}$ vs $qMRI_{ROI_j}$, where $qMRI$ stands for quantitative metric (separately performed for $R_1$, $R_2^*$, and QSM) and $ROI_i$ stands for the region of interest in CGM and SWM. A similar analysis was performed for $ZqMRI_{ROI_i}$ vs $ZqMRI_{RO_j}$, where $ZqMRI$ was related to the z score of a metric after accounting for age and gender using the [Eq. 1]. While the first analysis identified regions with similar aging time courses, the second analysis explored regions with shared covariance independently of aging.

## Harmonization of qMRI values across datasets

To evaluate the robustness of the single-cohort findings, we expanded the analysis by adding datasets from additional cohorts (Site 2 and Site 3).

Harmonization of qMRI values across datasets was performed using two methods. First, we used the EBS model implemented in neuroCombat (Johnson et al., 2007) to adjust for batch effects commonly found in multi-site studies. It modeled and corrected unwanted variability (e.g., differences in scanner type, protocol parameters, operator techniques, or environmental conditions), assuming that variability across batches was consistent. Age and sex were included as covariates to ensure that the harmonization process accounted for demographic factors.

Second, we employed the HBRbs model to estimate the mean and variance of each regional qMRI measure at a specific age and computed the corresponding z-score. Site 1, which covered the widest age range, was used as the reference dataset, while data from other sites (changing datasets) were projected into the reference space of site 1 by back-transforming z-scores, effectively correcting for site effects and allowing the pooling of data across sites. In this step, stratified five-fold cross-validation was used to ensure consistency in both site data and sex ratio across folds after data pooling from different sites. In each fold, 80% of the data was used for model training (20% for validation). The root mean square error was employed to identify the best model among all folds. Harmonized data in the changing datasets were then obtained by inversely transforming the corresponding z-scores to the reference dataset model. The effectiveness of different harmonization methods was evaluated based on the mean, standard deviation, and coefficient of variation (CV) at the overall level.

For interpretability purposes, PR models were used to describe qMRI trajectories across datasets. To assess the performance of the two harmonization approaches in modeling age, we conducted pairwise Spearman correlation analysis on derived peak ages. These peak ages were defined as the chronological ages corresponding to the maximum or minimum values of each qMRI trajectory [Eq. 2]. This analysis allowed us to evaluate the consistency of overall age trajectories. For $R_1$, raw data were pooled from the three cohorts with minimal differences in the MRI sequence parameter. Peak ages were excluded based on the following criteria: regions exhibiting extreme values due to image noise (BA 25 in Site 3 data), significant deviation from the median peak age in the group level (with a standardized deviation exceedingly twice the standard error) or lying outside the studied age range (with the trajectory continuously increased or decreased). In contrast, for $R_2^*$ and QSM, data were derived exclusively from Site 1, which offered isotropic resolution and a broader age range. This analysis compared the peak ages derived from raw, HBRbs-harmonized, and EBS-harmonized datasets to evaluate the impact of harmonization in the presence of data with non-overlapping age ranges ($R_1$) or protocols with systematic biases ($R_2^*$ and QSM).

# Results

Single Cohort Analysis of Distribution Patterns and Normative Trajectories of R1, R2*, and QSM

*$R_1$ Relaxation Patterns Across Cortical Areas*

The analysis revealed distinct $R_1$ distribution patterns (Fig. 1). The primary sensory cortices, including the primary visual cortex (V1, BA 17), the primary somatosensory cortex (S1, BA 1, 2, 3), and the primary auditory cortex (A1, BA 41, 42), exhibited higher $R_1$ relaxation rates. Notably, BA 4 (M1) in the primary motor cortex had the highest $R_1$ mean value of 0.76 (±0.019). In contrast, the associative cortices, encompassing the prefrontal cortex (BA 9, 10, 11, 46), temporal association areas (BA 20, 21, 22, 37), parietal association cortex (BA 5, 7), occipital association cortex (BA 18, 19), temporal pole (BA 38), and anterior cingulate cortex (BA 24, 33), displayed lower $R_1$ relaxation rates ranging from 0.66 to 0.70. The lowest $R_1$ mean value of 0.64 (±0.01) was observed in BA 25, part of the associative cortex (anterior cingulate cortex). For detailed distributions of average $R_1$ values across different brain cortical areas, refer to Fig. 1 (a, d, and g).

*$R_2$* and QSM Distribution Patterns in Cortical and Superficial Regions*

The analysis of $R_2$* and susceptibility values (Fig. 1b and 1c respectively) across cortical regions showed distinct patterns. In the primary motor cortex (BA 4) and primary visual cortex (BA 17), both $R_2$* and susceptibility values were high. In contrast, lower $R_2$* values of 16.56 (± 0.76) were noted in the frontal cortex (BA 9). Increased susceptibility values were observed in the prefrontal cortex (BA 33), primary auditory cortex (BA 42), and posterior cingulate cortex (BA 23). Notably, the anterior cingulate cortex (BA 24) exhibited high susceptibility values of 0.0011 (± 0.002). These results highlighted the distinct distribution patterns of $R_2$* and susceptibility values across CGM (See Fig. 1 b-c).

A distinct pattern was observed in SWM (Fig. 1d-f) compared to CGM (Fig. 1a-c). Increased $R_1$, $R_2$*, and susceptibility values were evident in the primary visual cortex (BA 17). In the lateral part of the frontal lobe, SWM showed increased $R_1$, $R_2$*, and susceptibility values. Conversely, in the interior frontal lobe, there was an increase in $R_1$ and susceptibility values while $R_2$* was decreased.

*Analysis of $R_1$, $R_2$*, and QSM in White Matter Fiber Bundles*

The $R_1$, $R_2$*, and susceptibility values analysis in TWMB revealed distinct patterns across brain regions. The highest $R_1$ value was observed in FA, while the lowest was in the CSTL. For $R_2$*, the highest value was in FP, with the lowest in the CSTL. In QSM, the CGCR had the highest value, whereas the FP had the lowest.

FA and FP showed high $R_1$ and $R_2$* values as commissural tracts. Among association tracts, particularly IFOL exhibited elevated $R_2$* (21.24 ±1.12) and susceptibility (-0.02±0.00) values. CGCL and CGCR showed the highest susceptibility values but lower $R_2$*. Conversely, CST demonstrated lower values in all metrics.

*Age Model Comparisons*
Normative models derived from the quadratic PR and HBRbs approaches largely overlapped across the lifespan for various metrics and regions (Fig. 2). Overall, HBRbs models offer greater flexibility and provide robust fits across most regions, particularly for $R_1$ in TWMB, where the quadratic behavior is less pronounced. However, the small sample size increased sensitivity to outliers, occasionally leading to unrealistic behavior, as seen in the HBRbs fitting example for SWM (Fig. 2). In contrast, the predefined quadratic regression model, while being more robust due to fewer parameters, sometimes forced a quadratic behavior, and thus overestimated rate changes at the extremes, especially in CGM (Fig. 2).

A more objective comparison of model performance was made using out-of-sample explained variance (EXPV). For $R_1$ metrics, the differences between the models were small, with both showing stable fits and an EXPV ratio (HBRbs/PR) of 0.99. However, for $R_2^*$ and QSM models, significantly higher or lower EXPV ratios (HBRbs/PR) were observed, particularly for QSM, indicating less stability in age model fitting for QSM and $R_2^*$ compared to $R_1$, as reflected by an average ratio of 0.77 for $R_2^*$ in TWMB and 0.46 for QSM in SWM. Detailed model fits for regions with significant quadratic age dependency are provided in S. 2.

*Age Dependencies in $R_1$, $R_2^*$, and QSM*
PR models revealed significant quadratic age dependencies in $R_1$ across all structures. BA 44 in the inferior frontal gyrus demonstrated the best fit for aging effects in CGM ($R^2 = 0.47$, $p < 0.001$), as well as the strongest age dependency in SWM ($R^2 = 0.37$, $p < 0.001$). The IFOL, connecting the frontal and occipital lobes, showed the strongest age relationship with $R_1$ among TWMBs ($R^2 = 0.44$, $p < 0.001$).

For $R_2^*$, quadratic age dependencies were significant in most cortical regions except for BA 29 and 26 in CGM, where strong linear dependencies were nevertheless detected. BA 6, part of the premotor cortex, exhibited the strongest quadratic aging effect ($R^2 = 0.40$, $p < 0.001$). The highest quadratic dependency in SWM was noted in the primary motor cortex (BA 4, $R^2 = 0.27$, $p < 0.001$), while the ATRL exhibited the strongest effect among TWMBs ($R^2 = 0.22$, $p < 0.001$).

QSM analysis revealed distinct age dependency patterns across different regions. In CGM, most regions exhibited quadratic age dependencies with curves resembling those observed in $R_1$ and $R_2^*$. Due to the negative quadratic coefficient in the PR model, susceptibility decreases after reaching a maximum value. However, certain areas (Frontal area: BA 10, 11, 33, 47; Parietal area: BA 3, 29; Temporal area: BA 20, 21, 38; Mixed transitional: BA 25, 26, 27, 28, 35, 36) showed only linear relationships with age in the LRT test ($p > 0.05$). In SWM, the quadratic curves tended to be flatter or transition to linear, with the most pronounced quadratic dependency observed in BA 10 ($R^2 = 0.15$, $p < 0.001$).

Supplementary Materials (S. 3) provide details of the age quadratic regression models in $R_1$, $R_2^*$, and QSM.

*Brain Maturation across Structures*

The analysis of peak ages using quadratic models for $R_1$ and $R_2^*$ metrics revealed a progression of qMRI measurement peaks from outer CGM to SWM and then to deeper WM in TWMB. Specifically, $R_1$ maturation peaks occurred earliest in TWMB, averaging at 37.55 ± 2.33 years, followed by SWM at 42.69 ± 1.69 years, and latest in CGM, peaking at 56.93 ± 1.33 years. $R_2^*$ maturation followed a similar pattern, with peaks at 44.48 years (± 12.17 years) in TWMB, 52.52 ± 2.15 years in SWM, and 59.62 ± 3.41 years in CGM. $R_1$ peak ages were generally more concentrated and stable across different brain structures than $R_2^*$ (see Fig. 3) but showed greater variability in the medial temporal regions and areas near the corpus callosum in mixed transitional areas. In contrast, $R_2^*$ metrics displayed more significant bootstrapping standard errors in specific regions, such as the peak ages of CGM in BA 20 and BA 25, and SWM in BA 26, indicating less stability in these regional measurements. In QSM analysis, however, some regional measurements failed in quadratic fitting with age, either showing significance only on the linear fitting or completely lacking age dependencies, as detailed in Supplementary Materials (S. 4a-b).

Correlation analysis revealed a significant relationship between the ages at peak values of $R_1$ in CGM and SWM (r = 0.37, $p < 0.05$) across individual brain regions, while $R_2^*$ showed a trend towards significance ($p$ = 0.058), indicating a nuanced linkage in maturation timing between these structures.

*Age Covariation in Quantitative Connectivity Analysis*

Inter- and intra-connectivity analyses were used to assess whether the regional qMRI measurements were independent or formed networks of joint maturation, myelination, and iron deposition in CGM and SWM. The intra-connectivities ($R_1$, $R_2^*$ and susceptibility) within SWM and CGM remained generally stable, regardless of age effects (Fig. 4a-b, S. 4c). However, significantly reduced $R_1$ inter-connectivity between SWM and CGM was observed, as indicated by the decreased explained variance in the diagonal squares (Fig. 4a). Notably, the auditory cortex's SWM (BA 41 and BA 42) showed a substantial reduction in general connectivities under aging effects. Furthermore, the frontal, parietal, and occipital lobes displayed substantial influence from age. In contrast, the temporal lobe and the medial temporal structures in the mixed category showed relatively minimal age-related impact.

Compared with $R_1$ inter-connectivities, $R_2^*$ showed stronger correlations between the same regions in both SWM and CGM (Fig. 4b), as indicated by the higher correlation coefficients in the diagonal matrices of both the upper-left and lower-right sections. Within the intra-connectivity matrix for CGM and SWM, similar to $R_1$, the lowest correlations were observed in BA 26–29 and 33–36, which correspond to splenial and entorhinal areas, rather than neocortical regions.

Given the reduced age dependence observed for QSM in most CGM and SWM regions (S. 4c), the expected outcome showed stable correlation matrices after accounting for age. Unlike the $R_1$, $R_2^*$ correlation matrices, where statistically significant correlations were consistently positive, most susceptibility correlations were positive for intra-connectivities within CGM and SWM, but negative for inter-connectivities.

## Age-Related Dynamics of Quantitative Metrics Across Multi-Centre Cohorts

*Comparative Analysis of Harmonization Methods in Quantitative Metrics*

The comparison of raw and harmonized data for $R_1$, $R_2^*$, and susceptibility metrics across CGM, SWM, and TWMB showed that both EBS and HBRbs models effectively reduce the CV, thereby achieving model harmonization (For CV details, see S. 5). Specifically, both methods consistently lowered the CV in $R_1$ and $R_2^*$ metrics, with EBS slightly outperforming HBRbs in CGM and SWM for $R_1$. However, in QSM, particularly in CGM where the data was noisier, HBRbs demonstrated a more significant reduction in CV.

PR models were fitted to each combined dataset to compare the two harmonization strategies. In $R_1$, across the three cohorts in CGM, EBS adjustments tended to centralize the data from Site 3 around the average levels, removing cohort-specific features and attenuating the substantial early increase in myelination typically observed. In contrast, HBRbs preserved these site-specific characteristics, effectively capturing the sharp rise in $R_1$ during the younger age range (see example region in Fig. 5a). A similar effect was observed in WM, where EBS indicated that by age 8, $R_1$ was already near its maximum, significantly altering previous observations that identified the peak ARCR age as 39 years (Fig. 3).

*Comparative Validation of Age Modeling Across Harmonized MRI Metrics*

The effectiveness of different harmonization methods in modeling age trajectories was evaluated through pairwise correlation analysis of peak ages derived from significant quadratic PR models in $R_1$, $R_2^*$, and susceptibility across various brain structures (Fig. 6). The comparison showed that $R_1$ peak ages from raw data and harmonized data using HBRbs and EBS methods exhibited a high degree of consistency across CGM, SWM, and TWMB, as indicated by strong correlations and minimal dispersion, and the fact that the slope across classes remained more consistent (Fig. 6a).

The harmonization results for $R_2^*$ and susceptibility metrics revealed a similar pattern of peak age correlations across CGM, SWM, and TWMB (Fig. 6b-c). Peak ages derived from raw data showed strong alignment and significant correlations with those obtained from both HBRbs and EBS harmonization methods in CGM and SWM, indicating that both methods effectively stabilized and preserved age-related changes in these structures. However, in TWMB, where the trajectory peaks occur closer to the younger end of Site 1's lifespan, the correlation between raw data and HBRbs harmonized data remained robust and significant, while the alignment with EBS harmonized data was weaker, with correlations failing to reach significance (as detailed in S. 7). Consequently, further analysis was performed on data harmonized using the HBRbs model.

*Multicentre Evaluation of $R_1$ Age Trajectories in Quantitative MRI Metrics*

The HBRbs-based harmonization and quadratic PR model fitting of $R_1$, $R_2^*$, and QSM are shown in figures (Fig. 7a, 8a-b), with each column displaying a representative lobe and each row showing different brain structures. In CGM, the average $R_1$ peak age was 58.55 ± 3.69 years, reflecting a later phase of maturation typically associated with mid-to-late adulthood. For SWM, maturation occurred earlier, with an average peak age of 37.58 ± 7.67 years (S. 6). TWMB demonstrated the earliest maturation, with an average peak age of 33.55 ± 4.97 years, indicating a rapid $R_1$ increase and early maturation during early adulthood in deeper WM tissue. Within CGM and SWM, we observed that temporal lobe regions exhibit the latest maturation, with peaks at 61.29 ± 2.94 years and 38.66 ± 9.65 years, respectively, followed by the ventral parts of the frontal lobe. In contrast, the occipital lobe regions showed the earliest maturation in both CGM and SWM, peaking at 54.63 ± 0.41 years and 33.96 ± 2.57 years, respectively. A similar pattern was observed in the motor cortex. In TWMB, the pARCL and pARCR exhibited the latest peaks, at 39.48 ± 0.97 years and 39.98 ± 0.75 years, respectively.

These results show that the comprehensive age range present in Site 1 data means adding additional datasets does not significantly affect the outcomes, illustrated by a gradient in $R_1$ peaks, beginning in the brain's WM bundles (Fig. 7d), progressing to SWM (Fig. 7c), and culminating in CGM (Fig. 7b). Within CGM (Fig. 7b), the peak age gradient began earlier in the visual and somatosensory cortices, spreading to the frontal and temporal cortex.

*Cross-Centre Study of $R_2$* and Susceptibility Trajectories in Quantitative MRI Metrics*
For $R_2^*$, peak ages in TWMB occur significantly earlier, ranging from 23.36 to 49.87 years, with an average of 43.52 (± 6.25) years. In SWM, peak ages ranged from 43.38 to 59.62 years, averaging 52.25 ± 3.22 years. In CGM, excluding BA 26 and 29 due to an LRT p-value > 0.05 favoring linear over quadratic models, peaks occurred later, between 54.61 and 68.02 years, with an average of 59.30 ±3.19 years (S. 6). This indicated a notable shift towards younger ages for maturation as it progresses from CGM to SWM, and finally to deeper WM.

For QSM, certain regions in SWM, such as BA 35 in the medial temporal areas, exhibited higher variability and noise. Compared to $R_1$ and $R_2^*$, QSM often exhibited weaker age modeling (with non-significant linear or quadratic age dependencies), particularly in the temporal lobe, inner temporal lobe in splenial and entorhinal areas (S. 6). These areas frequently displayed linear trends, and in some cases, even failed to achieve significant linear fits due to the increased noise (corroborating the findings in Fig. 4). Despite these variations, a U-shaped age trajectory was predominantly observed in most TWMB regions, peaking at 47.65 ± 7.62 years (Fig. 8b). In SWM and CGM, susceptibility peak ages occurred later, averageing 48.81 ± 9.29 years and 50.97 ± 4.54 years, respectively (Fig. 8b).

# Discussion

This multicentre study in a large healthy cohort population showed a maturation gradient across different brain structures, with peak values of qMRI age trajectories initially manifesting in deeper WM tracts during early adulthood, progressing to cortical-adjacent WM in middle adulthood, and culminating in the GM of cortical regions in the late adulthood. This gradient was consistently observed across quantitative MRI metrics, demonstrating similar age-related patterns in different brain tissues. Additionally, we validated the stability of age modeling for qMRI metrics across datasets with varying protocol parameters. Hereby, we demonstrated the importance of data harmonization, improving the robustness and reliability of the observed age trajectories with the HBRbs normative model.

The comparison of $R_1$, $R_2^*$, and susceptibility metrics revealed distinct performances in modeling age-related changes across different brain structures. $R_1$, derived from data across three centers and encompassing a broad age range, showed the most robust and stable age trajectories, with a consistent maturation gradient from TWMB to SWM and CGM. The observed stability of $R_1$ values was attributed to the robustness of the MP2RAGE sequence (Marques et al., 2010) against motion and respiration artifacts, especially when compared to the ME-GRE sequence. Additionally, the acquisition of $R_1$ at matched resolution (1 mm isotropic) across sites contributed to the stability of the results across studies, as it was previously shown using other acquisition strategies to map $R_1$ (Leutritz et al., 2020; Weiskopf et al., 2013). In contrast, $R_2^*$ and susceptibility metrics exhibited more significant variability across sites, which is probably due to differences in echo time, shimming performance, and PVE caused by varying spatial resolutions (Leutritz et al., 2020; R. Wang et al., 2017). Specifically, the large slice thickness (3mm) used at Site 2 caused issues when studying thin structures such as the neocortex with its average thickness of approximately 2.5 mm.

$R_1$ relaxation rates vary between primary and associative cortical regions similar to what has been previously reported for "Myelin-sensitive" approaches (Glasser et al., 2014). The distribution of $R_2^*$ and susceptibility values across brain regions reflected their associations with iron and myelin content. Specifically, the elevated $R_1$, $R_2^*$, and susceptibility values in the primary motor and visual cortices suggested substantial concentrations of both myelin and iron. We observed increased $R_2^*$ values within the SWM in the primary and supplemental visual cortex and lateral frontal and temporal areas. In $R_1$, there was a significant relative reduction in the primary visual cortex compared to other cortical areas. Both these observations suggested that at 3.0 T MRI, our measurements were sensitive to iron-rich oligodendrocytes in SWM (Kirilina et al., 2020). $R_2^*$ and susceptibility maps in CGM showed a very strong correlation as it has also been observed in both high-field cortical gray matter (Marques et al., 2017b) and in deep GM (Treit et al., 2021, p. 498). Yet in SWM, this behavior changed substantially, while $R_2^*$ values were increased in SWM of primary motor and supplemental motor, those regions showed below-average values in QSM. This observation suggested that, in those regions, $R_2^*$ and susceptibility metrics might be dominated by the diamagnetic contribution of myelin (R. Wang et al., 2017; Y. Wang & Liu, 2015). Specifically, the diamagnetic dominance of myelin in WM contrast was clear on the susceptibility trajectories of TWMB, where a U-shape trajectory was observed (Fig. 2, 5, and 8).

$R_2^*$-derived maturation peak ages were more uniformly distributed across various brain structures (Fig. 3), with some regions in the parietal and medial-temporal lobes having comparable peak ages in CGM and SWM. The fitting age models showed that generally the peak ages of $R_2^*$ happened later than those of $R_1$ in the cortex. It is well-established that the brain lacks effective mechanisms for removing iron deposits (Möller et al., 2019), which are expected to accumulate progressively throughout adulthood (Rouault, 2013). Our observation of a later peak ages in $R_2^*$ could be explained by the increased contribution of iron as a function of age which shifts the peak ages associated with myelination. The average gap in the peak ages between $R_1$ and $R_2^*$ in SWM increased to 14.7 years, which suggested either an accelerated rate of iron deposition or a reduced variation in myelination in these structures with age. $R_2^*$ supported the observed maturation (from WM to GM) gradient but exhibited increased sensitivity to sex differences in specific BA regions (S. 6), which has not been reported in previous studies on $R_2^*$ age modeling (Ghadery et al., 2015; Slater et al., 2019). QSM's performance was less consistent in modeling the age dependencies, with some regions displaying flat or nonsignificant trends, likely due to its higher susceptibility to artifacts close to the brain boundaries (Burgetova et al., 2021; Lao et al., 2023).

There is substantial agreement between some of our findings and previous data available in the literature. Considerable evidence supports the pattern of $R_1$ in WM development progressing from central to peripheral regions with earlier maturation in central compared to peripheral areas (Lebel & Deoni, 2018). In addition, it has been shown that $R_1$ values for various WM tracts peak between 30 and 50 years (Yeatman et al., 2014), which aligns with our observation of $R_1$ maturation in deeper WM tracts during early adulthood. Furthermore, $R_1$ peaks around 40 years in TWMB have been reported, along with the greater variability in $R_2^*$ (Slater et al., 2019), which is consistent with our data. By contrast, the SWM, which lies just beneath the cortical mantle and consists mainly of slow-myelinating short-association fibers known as U-fibers, matures into the fourth decade of life (Wu et al., 2014). Our findings confirm that SWM matures in mid-adulthood, a pattern distinct from the earlier maturation of TWMB and the later maturation of CGM. The later maturation peak in CGM may be due to iron deposition, water content, and lipid composition (Filo et al., 2019; Seiler et al., 2020). Previous studies showed that global cortical $T_1$ values, which are inversely related to $R_1$, decreased with age and negatively correlated from the 3rd to the 8th decade of life (Erramuzpe et al., 2021). These studies documented $R_1$ peak maturation occurring between 40 and 60 years, aligning with our results of $R_1$ maturation progressing from SWM in mid-adulthood to CGM in late adulthood. These consistent observations across multiple studies support a sequential maturation trajectory from deeper WM to cortical regions.

Our results are also aligned with existing research on lobe-specific cortical maturation patterns within the brain, as well as findings on WM tracts. In CGM, we observed that the temporal lobe, crucial for complex auditory processing, language, and memory, matures later, in line with the extended developmental timeline into early adulthood previously reported (Gogtay et al., 2004; Sowell et al., 2003). The frontal and parietal lobes exhibited intermediate maturation, reflecting their roles in higher-order cognitive functions and sensory integration, with the parietal structures maturing earlier than the frontal ones, corroborating previous findings (Shaw et al., 2008). It is interesting to note that CGM, SWM, as well as TWMB regions associated with the motor cortex (Fig. 5 b-d), had the earliest peak ages derived from $R_1$, reflecting rapid motor processing development (and earlier decline). This finding aligned with the earlier maturation observed in the motor-related CST bundle, which occurred at approximately 26 years of age in the harmonized data fitting model (S. 6), and in the visual-related FP, which reached its peak maturation earlier than the beginning of our study age range of 8 years. This supports the inferior-to-superior developmental theory, where early maturation facilitates motor functions and visual areas. Within $R_1$, the next earlier peak was observed in the FA (28.4 years), which connected pre-frontal cortices involved in higher-level motor control, and the IFO, implicated in visual processing, reading, and semantic processing. In contrast, temporal lobe regions supporting higher-order cognitive processes matured later, with other associative fibers maturing in the late third decade. This gradient demonstrates that basic sensory and motor regions mature earlier, supporting both inferior-to-superior and posterior-to-anterior brain development models, and provides a sequential understanding of specific regional brain maturation (Friedrichs-Maeder et al., 2017; Yeatman et al., 2014; Yu et al., 2020).

Compared to previous research, this qMRI study provides in-depth, reliable new findings based on a more comprehensive and novel research framework beyond traditional morphology. One important aspect is the observation that these qMRI-derived trajectories peak significantly later in cortical grey matter (~60 years old), than what has been estimated using cortical thickness(Brenhouse & Andersen, 2011; Mills et al., 2014; Tamnes et al., 2010), showing that thickness and qMRI unveil different aspects of tissue maturation. We modeled detailed region-based aging patterns in both GM and WM, and uniquely included SWM to observe that the lessons learned remain unchanged once considering harmonized multi-site data (see Fig. 6). Based on the structural spatial localizations of specific regions and their corresponding aging patterns, we pioneeringly elucidated comprehensive brain maturation gradients under qMRI. Finally, we used the quantitative $R_1$, $R_2^*$, and susceptibility mapping trajectories to qualitatively explain the alterations in iron and myelin components in normal aging.

In the quantitative connectivity analysis, the intra-connectivity patterns of SWM and CGM remained largely stable across $R_1$, $R_2^*$, and susceptibility metrics even after removing the age effect (correlating the quantitative metrics or the age-adjusted z-scores). This suggested that many developmental patterns observed with age are still evident after adjusting for chronological age, indicating the potential utility of the quantitative metrics subject-specific brain aging studies (Wu et al., 2014) and that much of the variance in those metrics across CGM and SWM could be driven by external factors associated with, for example, a lifestyle rather than regional/functional specific variations in the population. $R_2^*$ exhibited significant inter-regional correlations between SWM and CGM in the same BAs which were not detected in $R_1$, showing that iron deposition follows a similar pattern in adjacent SWM and CGM regions. The differing connectivity patterns observed in $R_2^*$ and susceptibility metrics highlight their distinct tissue sensitivities. $R_2^*$ (as $R_1$) predominantly showed positive correlations, while QSM exhibited primarily negative correlations between SWM and CGM regions. This can be attributed to QSM being primarily influenced by the diamagnetic effects of myelin in SWM (Phillips et al., 2016), whereas in CGM it is predominantly affected by the paramagnetic effects of iron concentration (Barbosa et al., 2015). Notably, these two quantities were generally correlated.

One important question addressed in this manuscript is how to combine qMRI metrics from different studies to make the most of new cohort studies of limited size and/or demographic variation. Our study evaluated both the HBR framework and EBS models for harmonization. In this study, both EBS and HBRbs were effective at mitigating site-specific biases as evaluated by qMRI age peak estimation. Specifically, the biases stemmed from the inclusion of a new dataset with an age range distinct from the two pre-existing, closely aligned age-range datasets and having different acquisition protocols. Leveraging a probabilistic framework, HBRbs harmonization proved particularly important in not distorting peak maturation ages close to our various cohorts' boundary ages (25-35). Nevertheless, it has to be acknowledged that HBRbs also exhibited some limitations in this study, such as fluctuations in the aging trajectory (Fig. 2), which were attributed to the relatively small sample data size of our study compared to the larger datasets where this framework has been previously applied (Kia et al., 2022; Villalón-Reina et al., 2022). Unfortunately, the systematic acquisition of larger qMRI datasets (as opposed to morphological and resting state functional MRI data typically acquired in large-scale open datasets including HCP, ABCD, or UK Biobank) is just starting. This marks a promising start for future research. One possible alternative to overcome data shortage is to perform multiple scans of a single subject at each site, then compare and analyze the results to estimate the noise of each qMRI metric per region, allowing site-specific data augmentation. Future work should focus on better characterizing the size requirements of new datasets to fully explore the advantages of this normative framework.

The scope of this study is limited by certain constraints in the dataset, particularly for the younger subjects aged 8-18, where data originate from one site only. While the consistency in MP2RAGE protocol parameters and the stability of structural images have allowed for the direct merging of $R_1$ data without inducing significant model bias, as demonstrated in our harmonization comparisons, the absence of infant and older age groups naturally restricts the ability to characterize age trajectories outside the study's age range fully. Acquiring younger age cohorts is particularly important, as evidence showed significant changes in myelination and iron accumulation during this period (Grotheer et al., 2022; Kühne et al., 2021; Mandine et al., 2023). Yet we have created the framework to which the younger cohort data can be added, but in the presence of such data quadratic models are less likely to describe trajectory and care should be taken interpreting peak ages.

Finally, another potential limitation is related to the accuracy of the applied BA parcellation. It is known that this does not completely overlap with functional myeloarchitectural distributions (Glasser et al., 2014). A functional-based atlas may be more suitable when the objective is to examine the relationship between specific cognitive abilities (e.g., visual acuity, reaction times, language fluency) and qMRI metrics. However, in this study, we opted for an intermediate level of granularity across the various lobes to better capture maturation gradients. This approach avoids the increased risk of noise sensitivity that could arise from using an atlas with too many cortical regions. Finally, here qMRI metrics in TWMBs were analyzed separately for each hemisphere, supported by evidence of distinct aging trajectories in the left and right hemispheres (Slater et al., 2019) with the right bundles systematically peaking earlier in R1. In contrast, we have an averaging of measurements across hemispheres for cortical areas, although following the empirical approach of most current studies and based on evidence indicating minimal hemispheric differences in morphometrics and quantitative measurements (Giedd et al., 1996; Raz et al., 1995), may overlook subtle lateralization effects.

This is a purely cross-sectional study, and the modeling trajectories represent those of the mean population. Therefore, this novel qMRI normative model framework might be exploited to follow individual maturation curves and their changes through learning or lifestyle interventions. The proposed models and framework set an important ground for future studies in both health and pathology, opening the possibility of monitoring intervention-driven changes in brain maturation patterns.

# Figures

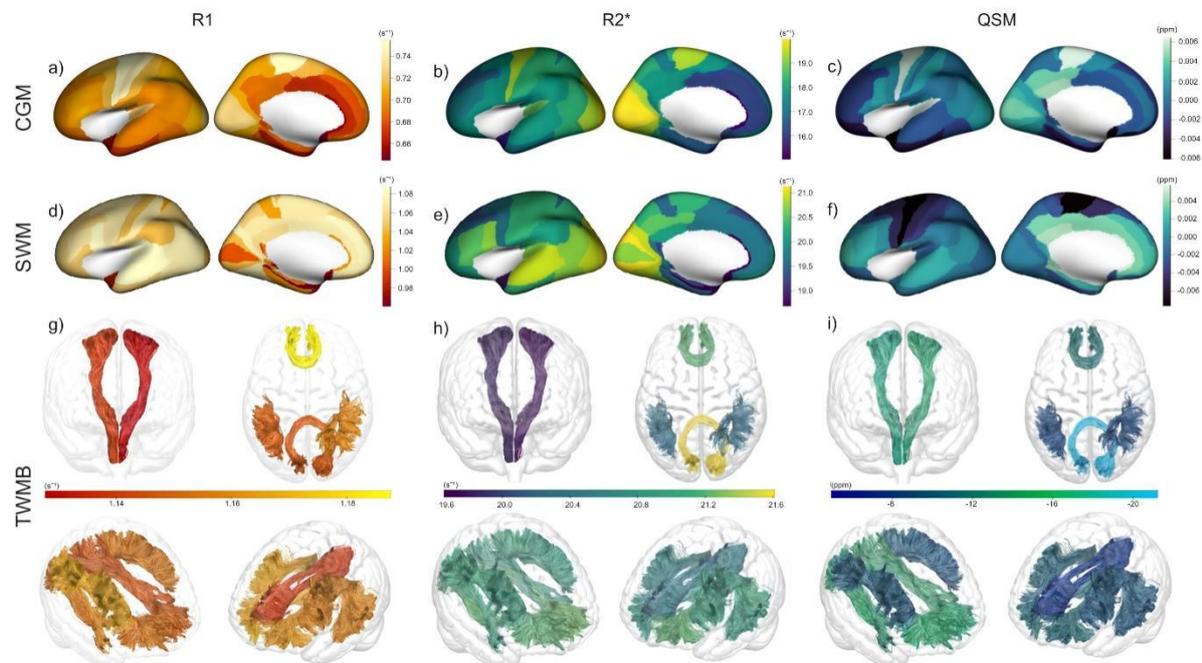

**Fig. 1 Mapping Quantitative Structures for Regional Averages**

Panels a-c display surface maps of cortical grey matter (CGM) for R1, R2*, and Susceptibility, respectively. Panels e-f show surface maps of superficial white matter (SWM) while Panels g-i depict tractography white matter bundles (TWMB) for the same metric. These metrics were derived from Site 1. Outliers beyond three times the median were excluded before calculating regional averages based on brain parcellation to account for potential vessel artifacts or segmentation errors. The color bars in each panel represent the distribution range of average metric values across different brain regions, with warmer colors indicating higher values and cooler colors indicating lower values. CGM: Cortical Grey Matter, SWM: Superficial White Matter, TWMB: Tractography White Matter Bundles. QSM: quantitative susceptibility mapping.

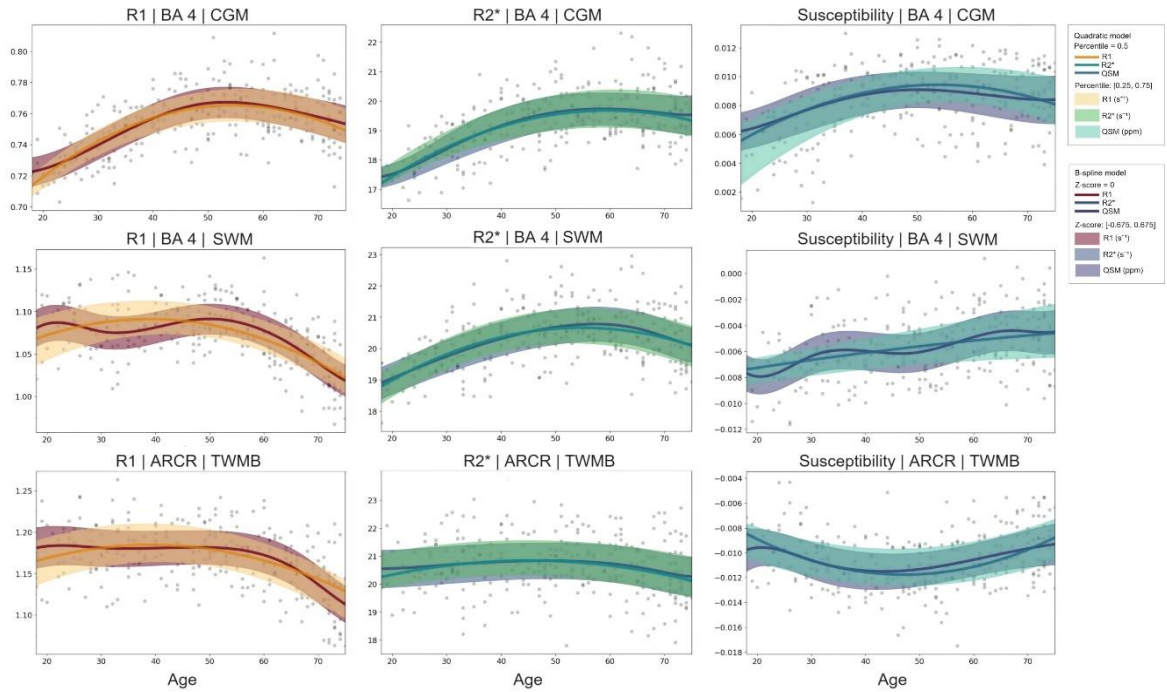

**Fig. 2 Modeling aging Effect Across Brain Regions**

Age trajectories for R1, R2*, and QSM in cortical grey matter (CGM), superficial white matter (SWM), and tractography white matter bundles (TWMB) for selected example regions (BA 4 / ARCR). The figure shows trends using Hierarchical Bayesian-based B-spline (HBRbs) model and Polynomial Regression (PR) models to capture age dependencies. Data points represent individual measurements and were derived from Site 1. For the PR model, the shaded areas represent the 25th to 75th percentile range, with the middle line showing the 50th percentile curve. For the HBRbs model, the shaded areas correspond to the equivalent range for z-scores (from -0.675 to 0.675), with the middle line representing the z-score of 0. CGM: Cortical Grey Matter, SWM: Superficial White Matter, TWMB: Tractography White Matter Bundles. BA: Brodmann Area. ARCR: Right Arcuate. HBRbs: Hierarchical Bayesian-based B-spline. PR: Polynomial Regression. QSM: quantitative susceptibility mapping.

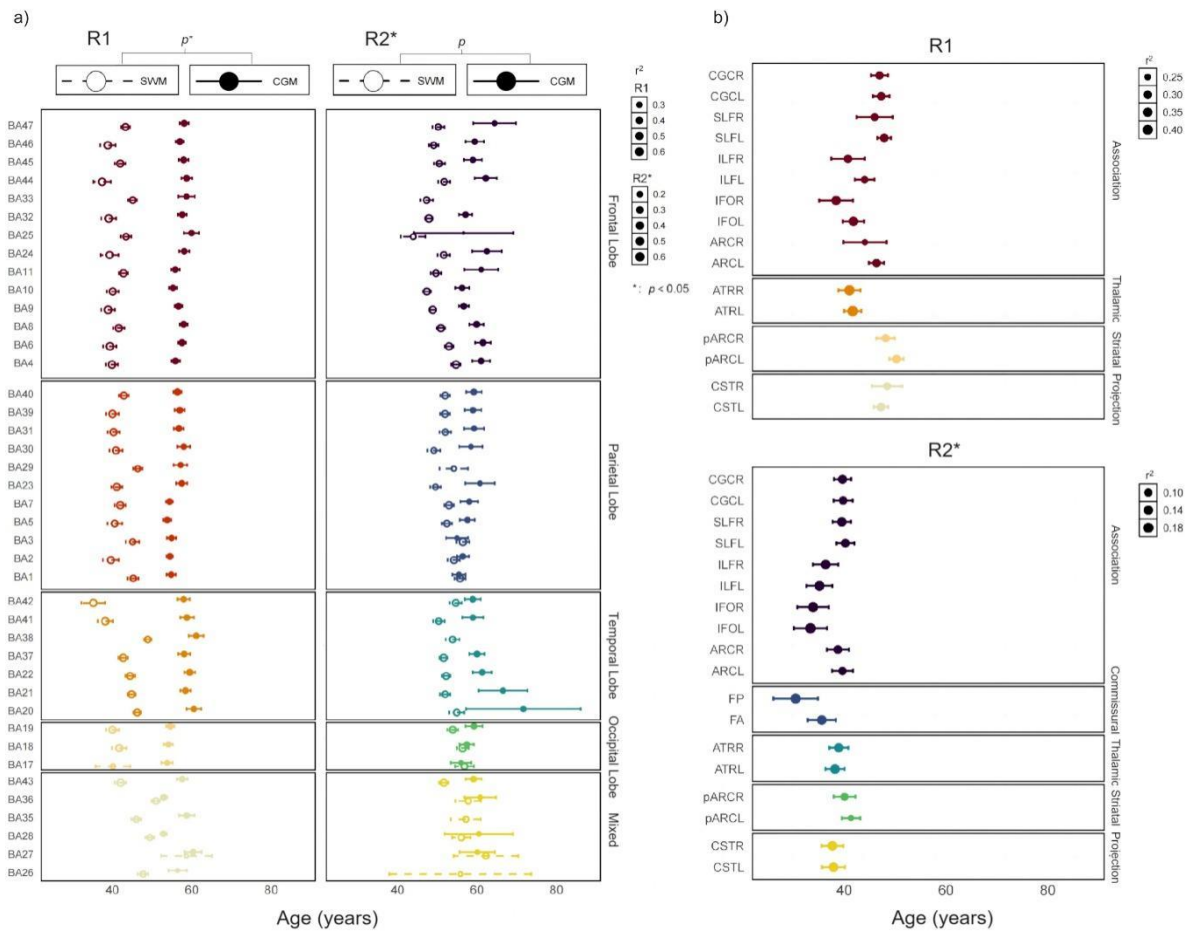

**Fig. 3 Brain Maturation Among Multiple qMRI Parameters**

Panel a shows the age peaks determined by significant LTRs from quadratic regression models ($p < 0.05$) for R1 and R2* metrics in CGM and SWM across BAs. Each data point represents the mean value of a specific BA, with an error bar showing standard error derived from 1000 bootstrap iterations. Spearman correlation analysis of age peaks between CGM and SWM was conducted in both R1 and R2*, with *p*-values marked with an asterisk indicating significant results. Panel b showed age peaks extracted among TWMBs in R1 and R2*. FA and FP were removed due to excessively large bootstrapping standard errors and peak ages exceeding the specified age range. The data are organized based on the anatomical functional characteristics of different brain lobes or white matter tracts. CGM: Cortical Grey Matter, SWM: Superficial White Matter, BA: Brodmann Area, FA: Callosum Forceps Minor, FP: Callosum Forceps Major, pARC: Posterior Arcuate Fasciculus, ILF: Inferior Longitudinal Fasciculus, SLF: Superior Longitudinal Fasciculus, IFO: Inferior Fronto-Occipital Fasciculus, ARC: Arcuate, ATR: Thalamic Radiation, CGC: Cingulum Cingulate, CST: Corticospinal, L: Left, R: Right.

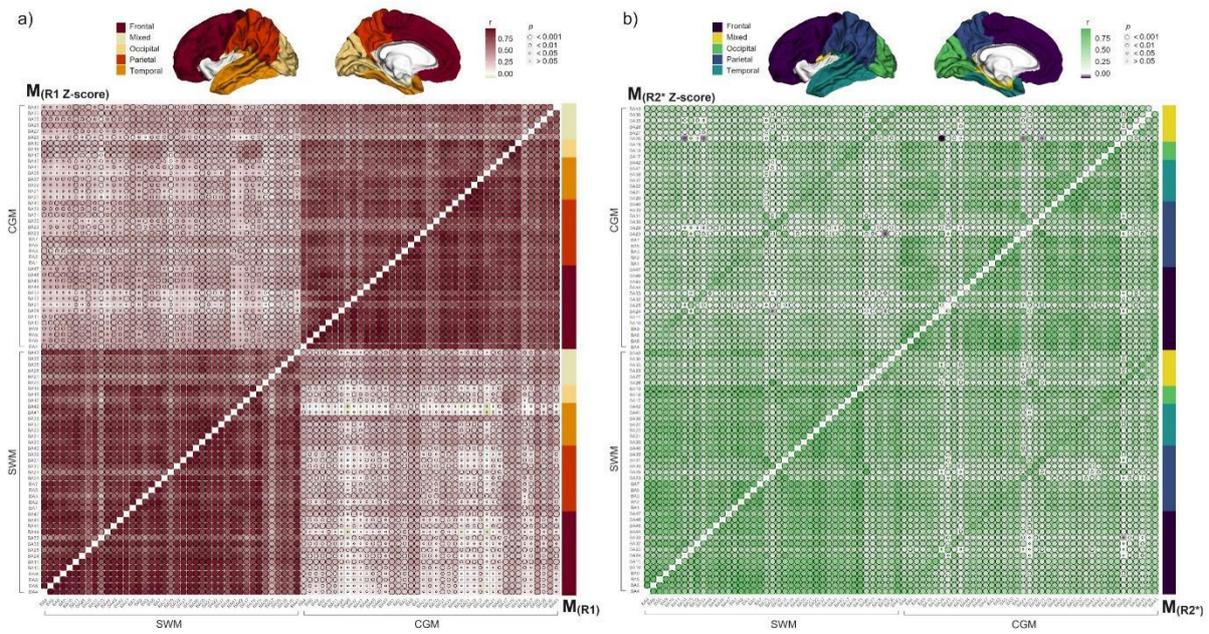

**Fig. 4 Connectivity Analysis of Brain Cortical Grey Matter and Superficial White Matter**

The figure presents Pearson correlation matrices for (a) R1, and (b) R2* showing both intra- and inter-regional connectivity patterns within and between CGM and SWM. In each matrix, the upper triangles depict z-scores derived from the residuals of significant quadratic models adjusted for aging effects, illustrating connections that are independent of age. The lower triangles display the raw average values for each region, highlighting the overall connectivity. The circle size increases as the P-value decreases, and the color deepens with a higher correlation. Only regions with significant quadratic regression fits were included in the matrix calculations. Each matrix organizes data into five lobes based on the anatomical locations of the selected BAs: frontal, parietal, temporal, occipital, and mixed regions. CGM: Cortical Grey Matter, SWM: Superficial White Matter, BA: Brodmann Area.

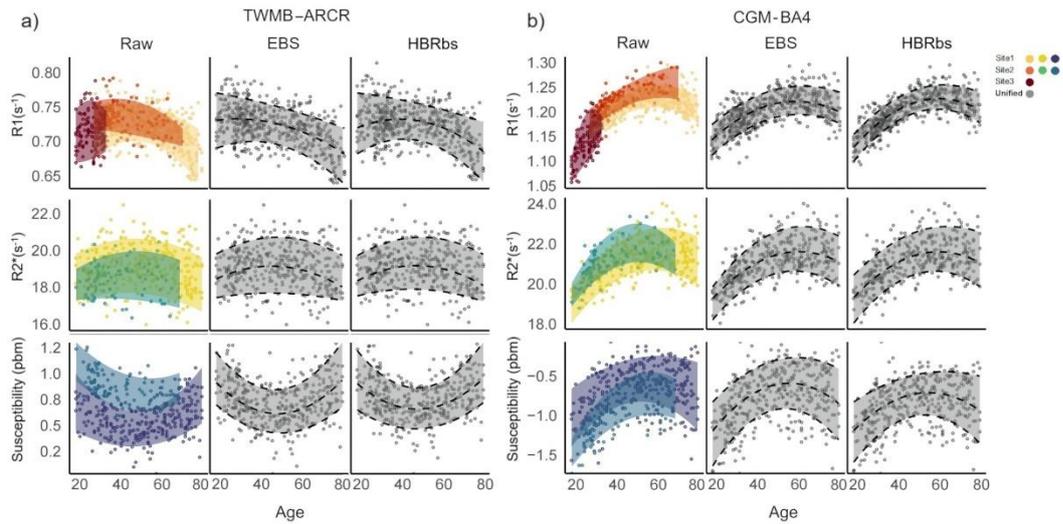

**Fig. 5 Comparison of Raw and Harmonized Data for qMRI Metrics in Selected Regions**

The figure shows age trajectories for R1, R2*, and QSM metrics in the example white matter region (ARCR, Panel a) and the example grey matter region (BA 4, Panel b). For each region, raw data (left) from different sites are color-coded, with the shaded area representing the 10th to 90th percentile range based on separate-fit polynomial regression models. Harmonized data (middle and right column) are shown using the EBS model as method 1 and the HBRbs model as method 2 respectively. These illustrate unified trajectories where shaded areas denote the 10th, 50th (solid line), and 90th percentiles. This comparison highlights how harmonization aligns with raw data trends, preserving age-related patterns while reducing site-specific variations. ARCR: Right Arcuate, CGM: Cortical Grey Matter, SWM: Superficial White Matter, TWMB: Tractography White Matter Bundles, BA: Brodmann Area, EBS: Empirical Bayes Statistics, HBRbs: Hierarchical Bayesian-based B-spline, QSM: quantitative susceptibility mapping.

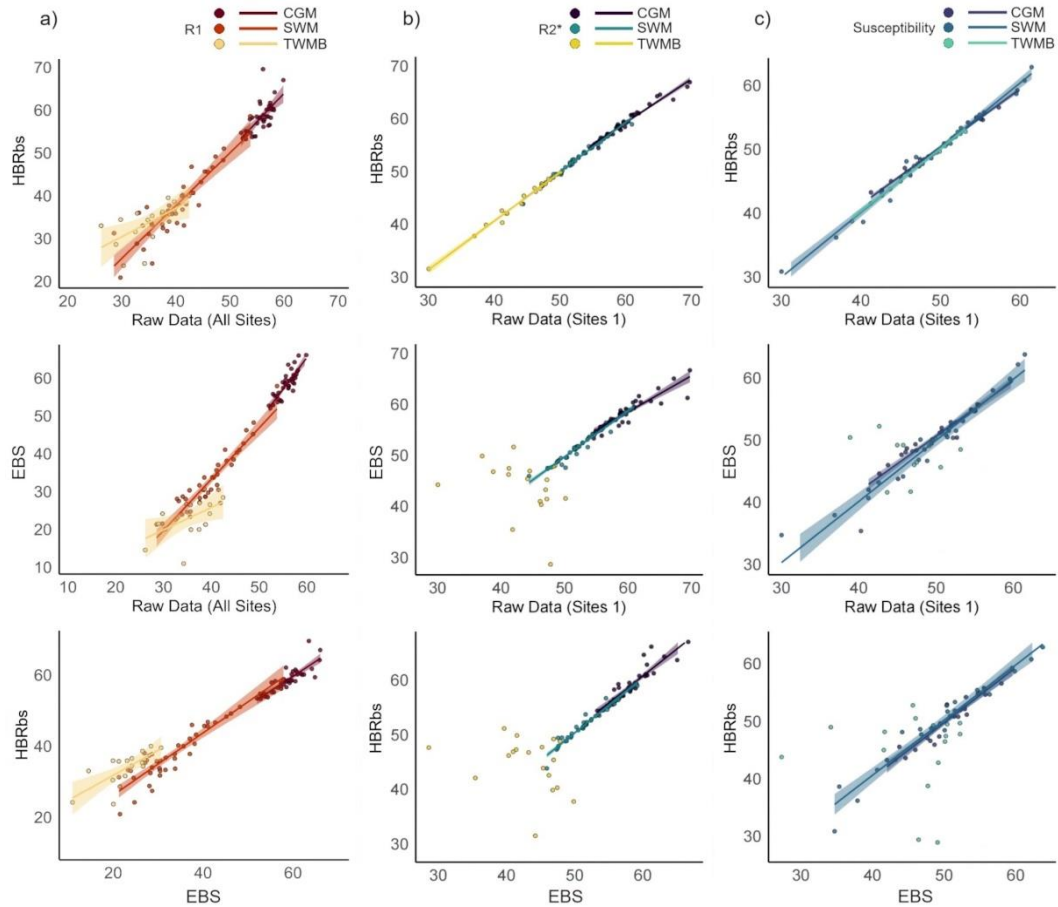

**Fig. 6 Correlation of Age Peaks Between Raw and Harmonized Data**

Pairwise correlation analysis of age peaks derived from R1 (panel a), R2* (panel b), and susceptibility values (panel c) between raw data, HBRbs harmonized data, and EBS harmonized data. Age peaks are derived using quadratic regression models across different brain structures: CGM, SWM, and TWMB. Data for R1 is from three datasets (Site1, Site2, Site3), while data for R2* and susceptibility is from two datasets (Site1, Site2) with Site1 as the reference for the raw data. Fitted lines are displayed when correlations are significant. Data points are color-coded to distinguish different structures. CGM: Cortical Grey Matter, SWM: Superficial White Matter, TWMB: Tractography White Matter Bundles, HBRbs: Hierarchical Bayesian-based B-spline; EBS: Empirical Bayes Statistics.

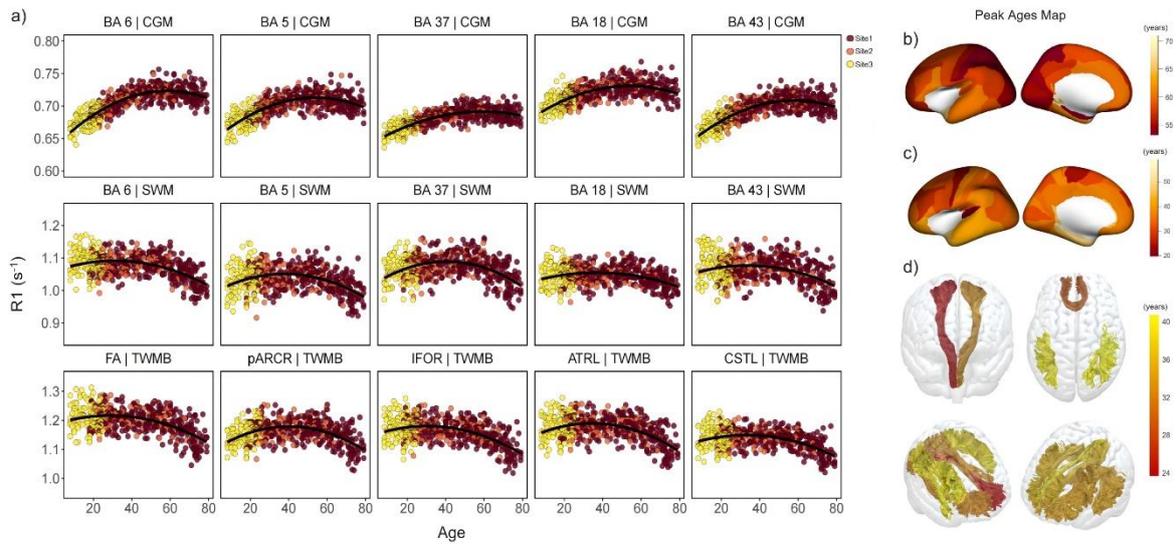

**Fig. 7 Age Trajectories of R1 Across Different Brain Structures**

Panel a shows age trajectories of R1 in representative regions from CGM, SWM, and TWMB, using harmonized data from three sites based on a Hierarchical Bayesian-based B-spline (HBRbs) model. The figure illustrates R1 across different brain structures showing how R1 changes with age. Representative regions were selected for each lobe or bundle category. Data points are color-coded according to different sites. Each row and column were aligned to facilitate direct comparison across different regions and structures. Panels b, c, and d represent the distribution of age peaks in different regions under CGM, SWM, and TWMB, respectively, with color bars showing the ranges. The Callosum Forceps Major was removed in panel d due to continuously decreased age peaks located outside the studied age range. HBRbs: Hierarchical Bayesian-based B-spline, CGM: Cortical Grey Matter, SWM: Superficial White Matter, TWMB: Tractography White Matter Bundles. FA: Callosum Forceps Minor, pARCR: Right Posterior Arcuate Fasciculus, IFOR: Right Inferior Fronto-Occipital Fasciculus, ATRL: Left Thalamic Radiation, CSTL: Left Corticospinal Tract.

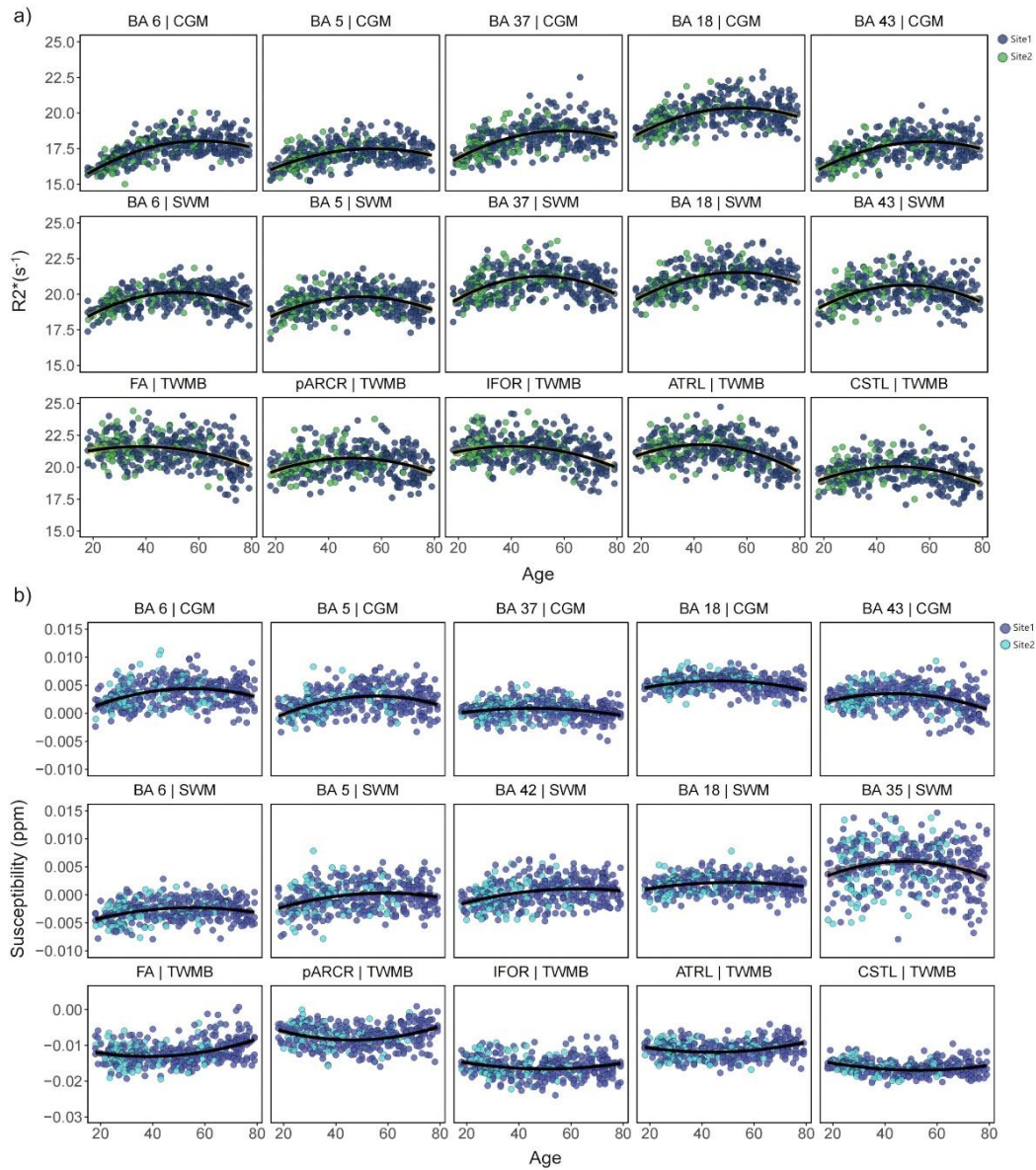

**Fig. 8 Age Trajectories of R2* and QSM Across Different Brain Structures**

Age trajectories for R2* and QSM in representative regions from CGM, SWM, and TWMB, using harmonized data from two sites based on a Hierarchical Bayesian-based B-spline (HBRbs) model. Panel A illustrates the R2* trajectories, while Panel B shows QSM trajectories across different brain structures demonstrating how these metrics change with age. Data points are color-coded according to different sites. Each row and column are aligned to facilitate direct comparison across different regions and structures. HBRbs: Hierarchical Bayesian-based B-spline, CGM: Cortical Grey Matter, SWM: Superficial White Matter, TWMB: Tractography White Matter Bundles. FA: Callosum Forceps Minor, pARCR: Right Posterior Arcuate Fasciculus, IFOR: Right Inferior Fronto-Occipital Fasciculus, ATRL: Left Thalamic Radiation, CSTL: Left Corticospinal Tract. QSM: quantitative susceptibility mapping.

# Table

**Table 1 Participant Demographics and qMRI Data Availability by Site**

| Description | Site 1 | Site 2 | Site 3 | Total |
|---|---|---|---|---|
| Participant, n | 299 | 101 | 136 | 536 |
| R1 | 299 | 101 | 136 | 536 |
| R2* | 293 | 101 | - | 394 |
| QSM | 293 | 101 | - | 394 |
| Female/Male, % | 53.5/46.5 | 55.4/44.6 | 65.4/34.6 | 56.9/43.1 |
| Mean (SD) age, years | 51.3 (17.2) | 37.7 (12.9) | 17.4 (5.1) | 40.1 (20.1) |
| Age range, years | 18 - 79 | 18 - 68 | 8 - 25 | 8 - 79 |
| Age Group Distribution, n | | | | |
| < 20 years | 8 | 2 | 92 | 102 |
| 20 - 39 years | 84 | 63 | 44 | 191 |
| 40 - 59 years | 99 | 30 | 0 | 129 |
| ≥ 60 years | 108 | 6 | 0 | 114 |

SD: standard deviation, QSM: quantitative susceptibility mapping.

**Table 2 MRI Acquisition Parameters**

| Sequence | Parameter | Site 1 | Site 2 | Site 3 |
|---|---|---|---|---|
| MP2RAGE-T1 | Resolution (mm³) | 1 × 1 × 1 | 1 × 1 × 1 | 1 × 1 × 1 |
| | S/P | 176 | 176 | 224 |
| | TR/TI1, TI2 (ms) | 6000/700/2400 | 5000/700/2500 | 5000/700/2500 |
| | FA (°) | 6, 6 | 4, 5 | 4, 5 |
| | ST (min) | 7:32 | 8:20 | 4:00 |
| ME-GRE-T2, QSM | Resolution (mm³) | 0.8 × 0.8 × 0.8 | 0.75 × 0.75 × 3 | - |
| | TR/TE1/ΔTE (ms) | 44/6.14/4 | 49/6.69/4.06 | - |
| | ST (min) | 9.25 | 4.42 | - |
| DWI | Resolution (mm³) | 1.8 × 1.8 × 1.8 | 1.8 × 1.8 × 1.8 | 1.8 × 1.8 × 1.8 |
| | B values | 0/1250/2500 | 0/700/1000/2000/3000 | 0/925/1850 |
| | Orientation | 11/86/85 | 12/6/20/45/66 | 6/40/40 |
| | ST (min) | 9:39 | 4:36 | 4:53 |

MP2RAGE: Magnetization Prepared 2 Rapid Acquisition Gradient Echoes, ME-GRE: Multi-Echo Gradient Recalled Echo, QSM: Quantitative Susceptibility Mapping, DWI: diffusion-weighted imaging, S/P: Slice/partitions, TR: Repetition Time, TI1: Inversion Time 1, TI2: Inversion Time 2, TE1: Echo Time 1, ΔTE: Delta Echo Time, FA: Flip Angle(s), ST: Scan Time.

# Contributors

XC was responsible for conceptualization, methodology, software, data curation, visualization, writing the original draft, and review & editing. MOP and PJL contributed to methodology, formal analysis, conceptualization, and review & editing. CE contributed to data provision and review & editing. MZ, MW, MGJ, and JMO handled methodology, data curation, and review & editing. KSC, AC, WMM, and SS were involved in methodology, conceptualization, and review & editing. JB and AFM provided statistical support, and review & editing. DGN, JK, and LK focused on conceptualization and review & editing. LMG provided supervision, conceptualization, methodology, and review & editing. CG was involved in supervision, project administration, funding acquisition, conceptualization, methodology, and review & editing. JPM contributed to supervision, project administration, funding acquisition, conceptualization, methodology, and review & editing.

# Supplementary Materials

**S. 1 Statistical Framework of the HBRbs Model**

The HBRbs model imposes a hierarchical structure to explain variation in the data. At the top level, the $n^{th}$ observation $y_n$ is assumed to vary around a mean $\mu_n$ with variance $\sigma_n$:

$$y_n = N(y_n|\mu_n, \sigma_n)$$

with $n \in (1 .. n$ samples.$)$ On a second level, $\mu_n$ and $\sigma_n$ are modeled as the output of linear regression based on the input matrix:

$$\mu_n = w_\mu^T \phi + \tau_\mu$$

$$\sigma_n = w_\sigma^T \phi + \tau_\sigma$$

Here, $\phi$ describes the full, dummy coded regression matrix including additional bspline columns based on the input matrix $X$; $w$ describes a vector containing the corresponding (slope) regression parameters, and $\tau$ describes the intercept offset of that regression of the input matrix onto $\mu$ and $\sigma$, respectively.

Batch effects (sex, site) are modeled by allowing the intercept ($\tau$) and slope ($w$) parameters of $\mu$ and $\sigma$ to vary with a group standard deviation $\sigma$, multiplied by an offset $v_b$ around a mean $\mu$, for all batches b. This so-called non-center sampling approach helps to avoid certain errors during sampling (82). Further, priors are placed over each set of parameters $\theta$ containing batch effects, allowing them to learn from each other (random effect):

$$\theta_b = \mu_\theta + \sigma_\theta \, v_b$$

$$v_b = N(0, 1)$$

$$\mu_\theta = N(\mu_\theta | 0,1)$$

$$\sigma_\theta = N^+(\sigma_\theta | 1)$$

where $\theta_b \in \{w_\mu, w_\sigma, \tau_\mu, \tau_\sigma\}$ for batch b $\in \{1... n\_batches\}$.

Lastly, a set of standard Gaussian priors was placed over all remaining parameters. For more details on the default settings and priors on all parameters see in a published study (47).

Model estimation and inference were performed employing the No U-Turn sampler encoded in pyMC (version 5.4.1). The model was estimated in 4 chains, drawing 1500 samples each, from which the first 500 were discarded as a warm-up.

**S. 2 B-spline and Polynomial Regression Model Fitting for Site 1 Data**

| Region | Bspline MSLL | Bspline EXPV | Polynomial MSLL | Polynomial EXPV | EXPV Ratio (Bspline/Polynomial) |
|---|---|---|---|---|---|
| **R1-CGM** | | | | | |
| BA1 | -0.18 | 0.27 | -0.18 | 0.26 | 1.04 |
| BA10 | -0.25 | 0.40 | -0.24 | 0.40 | 1.00 |
| BA11 | -0.24 | 0.40 | -0.26 | 0.41 | 0.97 |
| BA17 | -0.14 | 0.29 | -0.14 | 0.29 | 0.97 |
| BA18 | -0.21 | 0.37 | -0.20 | 0.36 | 1.02 |
| BA19 | -0.21 | 0.37 | -0.22 | 0.39 | 0.95 |
| BA2 | -0.30 | 0.48 | -0.34 | 0.53 | 0.91 |
| BA20 | -0.27 | 0.43 | -0.27 | 0.43 | 0.99 |
| BA21 | -0.27 | 0.43 | -0.25 | 0.41 | 1.04 |
| BA22 | -0.36 | 0.52 | -0.37 | 0.52 | 0.99 |
| BA23 | -0.31 | 0.47 | -0.33 | 0.49 | 0.96 |
| BA24 | -0.28 | 0.45 | -0.31 | 0.49 | 0.92 |
| BA25 | -0.17 | 0.33 | -0.20 | 0.35 | 0.95 |
| BA26 | -0.07 | 0.15 | -0.07 | 0.16 | 0.93 |
| BA27 | -0.20 | 0.35 | -0.18 | 0.35 | 0.99 |
| BA28 | -0.21 | 0.43 | -0.20 | 0.39 | 1.12 |
| BA29 | -0.23 | 0.39 | -0.14 | 0.23 | 1.68 |
| BA3 | -0.18 | 0.29 | -0.20 | 0.29 | 0.98 |
| BA30 | -0.17 | 0.37 | -0.22 | 0.39 | 0.94 |
| BA31 | -0.30 | 0.46 | -0.32 | 0.48 | 0.96 |
| BA32 | -0.30 | 0.45 | -0.31 | 0.47 | 0.98 |
| BA33 | -0.17 | 0.30 | -0.19 | 0.32 | 0.93 |
| BA35 | -0.17 | 0.35 | -0.20 | 0.35 | 1.01 |
| BA36 | -0.34 | 0.56 | -0.32 | 0.52 | 1.07 |
| BA37 | -0.28 | 0.46 | -0.27 | 0.44 | 1.06 |
| BA38 | -0.19 | 0.32 | -0.19 | 0.31 | 1.02 |
| BA39 | -0.29 | 0.49 | -0.32 | 0.50 | 0.98 |
| BA4 | -0.29 | 0.45 | -0.31 | 0.45 | 0.99 |
| BA40 | -0.38 | 0.56 | -0.39 | 0.56 | 1.01 |
| BA41 | -0.24 | 0.40 | -0.24 | 0.40 | 1.00 |
| BA42 | -0.25 | 0.44 | -0.25 | 0.44 | 1.00 |
| BA43 | -0.29 | 0.47 | -0.29 | 0.49 | 0.95 |
| BA44 | -0.35 | 0.53 | -0.37 | 0.54 | 0.99 |

| Region | Bspline MSLL | Bspline EXPV | Polynomial MSLL | Polynomial EXPV | EXPV Ratio (Bspline/Polynomial) |
|---|---|---|---|---|---|
| BA45 | -0.33 | 0.48 | -0.31 | 0.46 | 1.04 |
| BA46 | -0.30 | 0.45 | -0.36 | 0.53 | 0.86 |
| BA47 | -0.34 | 0.51 | -0.36 | 0.52 | 0.98 |
| BA5 | -0.25 | 0.43 | -0.27 | 0.44 | 0.97 |
| BA6 | -0.40 | 0.58 | -0.43 | 0.60 | 0.96 |
| BA7 | -0.26 | 0.42 | -0.28 | 0.41 | 1.01 |
| BA8 | -0.40 | 0.56 | -0.40 | 0.57 | 0.98 |
| BA9 | -0.35 | 0.50 | -0.33 | 0.48 | 1.04 |
| **R1-SWM** | | | | | |
| BA1 | -0.15 | 0.28 | -0.14 | 0.25 | 1.12 |
| BA10 | -0.24 | 0.40 | -0.26 | 0.43 | 0.94 |
| BA11 | -0.21 | 0.36 | -0.24 | 0.39 | 0.92 |
| BA17 | -0.08 | 0.14 | -0.06 | 0.11 | 1.26 |
| BA18 | -0.12 | 0.21 | -0.19 | 0.34 | 0.63 |
| BA19 | -0.20 | 0.34 | -0.28 | 0.45 | 0.75 |
| BA2 | -0.17 | 0.30 | -0.20 | 0.36 | 0.85 |
| BA20 | -0.16 | 0.22 | -0.18 | 0.27 | 0.81 |
| BA21 | -0.19 | 0.34 | -0.18 | 0.29 | 1.14 |
| BA22 | -0.17 | 0.30 | -0.21 | 0.38 | 0.79 |
| BA23 | -0.24 | 0.35 | -0.22 | 0.34 | 1.03 |
| BA24 | -0.26 | 0.42 | -0.24 | 0.40 | 1.04 |
| BA25 | -0.16 | 0.33 | -0.19 | 0.34 | 0.96 |
| BA26 | -0.07 | 0.12 | -0.10 | 0.17 | 0.67 |
| BA27 | -0.02 | 0.05 | -0.03 | 0.05 | 1.02 |
| BA28 | -0.09 | 0.10 | -0.08 | 0.17 | 0.58 |
| BA29 | -0.09 | 0.19 | -0.16 | 0.28 | 0.69 |
| BA3 | -0.04 | 0.12 | -0.09 | 0.23 | 0.51 |
| BA30 | -0.23 | 0.40 | -0.27 | 0.46 | 0.87 |
| BA31 | -0.30 | 0.45 | -0.30 | 0.46 | 0.98 |
| BA32 | -0.34 | 0.49 | -0.28 | 0.45 | 1.08 |
| BA33 | -0.09 | 0.16 | -0.11 | 0.19 | 0.86 |
| BA35 | -0.14 | 0.26 | -0.13 | 0.24 | 1.07 |
| BA36 | -0.13 | 0.26 | -0.10 | 0.16 | 1.64 |
| BA37 | -0.19 | 0.40 | -0.26 | 0.44 | 0.92 |
| BA38 | -0.10 | 0.15 | -0.14 | 0.22 | 0.70 |
| BA39 | -0.23 | 0.41 | -0.25 | 0.45 | 0.92 |
| BA4 | -0.22 | 0.33 | -0.27 | 0.44 | 0.76 |
| BA40 | -0.19 | 0.36 | -0.23 | 0.41 | 0.87 |
| BA41 | -0.19 | 0.23 | -0.20 | 0.26 | 0.89 |
| BA42 | -0.21 | 0.28 | -0.21 | 0.33 | 0.86 |
| BA43 | -0.14 | 0.20 | -0.15 | 0.23 | 0.89 |
| BA44 | -0.29 | 0.44 | -0.30 | 0.49 | 0.91 |
| BA45 | -0.25 | 0.42 | -0.23 | 0.39 | 1.06 |
| BA46 | -0.23 | 0.38 | -0.24 | 0.45 | 0.86 |
| BA47 | -0.22 | 0.39 | -0.22 | 0.39 | 0.98 |
| BA5 | -0.20 | 0.33 | -0.21 | 0.38 | 0.87 |
| BA6 | -0.25 | 0.38 | -0.28 | 0.48 | 0.78 |
| BA7 | -0.21 | 0.35 | -0.24 | 0.42 | 0.83 |
| BA8 | -0.21 | 0.34 | -0.24 | 0.43 | 0.78 |
| BA9 | -0.25 | 0.40 | -0.27 | 0.47 | 0.86 |
| **R1-TWMB** | | | | | |
| ARCL | -0.26 | 0.39 | -0.21 | 0.37 | 1.06 |
| ARCR | -0.27 | 0.43 | -0.24 | 0.41 | 1.06 |
| ATRL | -0.31 | 0.48 | -0.27 | 0.44 | 1.09 |
| ATRR | -0.29 | 0.45 | -0.24 | 0.41 | 1.11 |
| CGCL | -0.16 | 0.27 | -0.16 | 0.26 | 1.04 |
| CGCR | -0.27 | 0.43 | -0.26 | 0.41 | 1.04 |
| CSTL | -0.36 | 0.50 | -0.20 | 0.33 | 1.48 |
| CSTR | -0.34 | 0.50 | -0.32 | 0.48 | 1.06 |
| FA | -0.36 | 0.50 | -0.18 | 0.28 | 1.82 |
| FP | -0.28 | 0.44 | -0.26 | 0.38 | 1.16 |
| IFOL | -0.41 | 0.55 | -0.33 | 0.49 | 1.12 |
| IFOR | -0.36 | 0.52 | -0.32 | 0.49 | 1.06 |
| ILFL | -0.36 | 0.52 | -0.35 | 0.49 | 1.04 |
| ILFR | -0.33 | 0.49 | -0.31 | 0.45 | 1.08 |
| pARCL | -0.16 | 0.27 | -0.16 | 0.26 | 1.04 |
| pARCR | -0.15 | 0.25 | -0.15 | 0.24 | 1.06 |
| SLFL | -0.27 | 0.41 | -0.18 | 0.28 | 1.46 |
| SLFR | -0.32 | 0.47 | -0.29 | 0.45 | 1.03 |
| **R2\*-CGM** | | | | | |
| BA1 | -0.13 | 0.24 | -0.13 | 0.22 | 1.08 |
| BA10 | -0.08 | 0.14 | -0.06 | 0.10 | 1.39 |
| BA11 | -0.05 | 0.11 | -0.04 | 0.09 | 1.22 |
| BA17 | -0.03 | 0.04 | -0.04 | 0.07 | 0.62 |
| BA18 | -0.16 | 0.28 | -0.16 | 0.26 | 1.10 |
| BA19 | -0.17 | 0.28 | -0.05 | 0.07 | 4.30 |
| BA2 | -0.14 | 0.25 | -0.14 | 0.25 | 0.97 |
| BA20 | -0.03 | 0.02 | -0.17 | 0.31 | 0.05 |
| BA21 | 0.00 | -0.02 | 0.00 | -0.04 | 0.57 |
| BA22 | -0.18 | 0.36 | -0.21 | 0.37 | 0.97 |
| BA23 | -0.06 | 0.12 | -0.07 | 0.16 | 0.72 |
| BA24 | -0.01 | 0.15 | -0.08 | 0.15 | 1.00 |
| BA25 | -0.01 | 0.08 | 0.01 | 0.03 | 2.66 |
| BA27 | 0.00 | -0.04 | 0.02 | -0.10 | 0.41 |
| BA28 | -0.08 | 0.19 | -0.07 | 0.18 | 1.11 |
| BA3 | -0.08 | 0.15 | -0.08 | 0.13 | 1.20 |
| BA30 | -0.03 | 0.06 | -0.04 | 0.08 | 0.74 |
| BA31 | -0.07 | 0.12 | -0.07 | 0.16 | 0.77 |
| BA32 | -0.12 | 0.21 | -0.12 | 0.23 | 0.93 |
| BA33 | -0.02 | 0.05 | -0.02 | 0.03 | 1.61 |
| BA35 | 0.05 | -0.14 | 0.03 | -0.19 | 0.74 |
| BA36 | -0.15 | 0.29 | -0.13 | 0.25 | 1.17 |
| BA37 | -0.10 | 0.17 | -0.09 | 0.15 | 1.12 |
| BA38 | -0.04 | 0.06 | -0.05 | 0.07 | 0.81 |
| BA39 | -0.22 | 0.40 | -0.22 | 0.38 | 1.05 |
| BA4 | -0.30 | 0.46 | -0.31 | 0.46 | 1.01 |
| BA40 | -0.24 | 0.41 | -0.25 | 0.42 | 0.96 |
| BA41 | -0.14 | 0.28 | -0.16 | 0.30 | 0.93 |
| BA42 | -0.06 | 0.13 | -0.08 | 0.14 | 0.93 |
| BA43 | -0.16 | 0.31 | -0.17 | 0.30 | 1.01 |
| BA44 | -0.32 | 0.51 | -0.31 | 0.50 | 1.03 |
| BA45 | -0.12 | 0.28 | -0.12 | 0.28 | 0.98 |
| BA46 | -0.20 | 0.35 | -0.21 | 0.36 | 0.98 |
| BA47 | -0.07 | 0.20 | -0.09 | 0.21 | 1.00 |
| BA5 | -0.12 | 0.23 | -0.10 | 0.19 | 1.22 |
| BA6 | -0.38 | 0.52 | -0.35 | 0.54 | 0.97 |
| BA7 | -0.20 | 0.34 | -0.20 | 0.35 | 0.98 |
| BA8 | -0.33 | 0.50 | -0.34 | 0.52 | 0.97 |
| BA9 | -0.26 | 0.46 | -0.26 | 0.43 | 1.06 |
| **R2*-SWM** | | | | | |
| BA1 | -0.17 | 0.32 | -0.17 | 0.32 | 1.00 |
| BA10 | -0.07 | 0.17 | -0.08 | 0.15 | 1.13 |
| BA11 | -0.08 | 0.15 | -0.07 | 0.13 | 1.13 |
| BA17 | 0.01 | -0.06 | -0.01 | -0.03 | 2.13 |
| BA18 | -0.14 | 0.25 | -0.15 | 0.27 | 0.92 |
| BA19 | -0.10 | 0.20 | -0.11 | 0.20 | 0.99 |
| BA2 | -0.09 | 0.18 | -0.10 | 0.18 | 0.98 |
| BA20 | -0.07 | 0.12 | -0.10 | 0.18 | 0.66 |
| BA21 | -0.11 | 0.20 | -0.07 | 0.14 | 1.49 |
| BA22 | -0.08 | 0.17 | -0.11 | 0.22 | 0.75 |
| BA23 | -0.03 | 0.10 | -0.04 | 0.11 | 0.93 |
| BA24 | -0.09 | 0.17 | -0.06 | 0.12 | 1.41 |
| BA25 | -0.06 | 0.15 | -0.03 | 0.07 | 2.01 |
| BA26 | 0.00 | 0.05 | 0.03 | -0.04 | -1.21 |
| BA27 | -0.02 | -0.02 | 0.01 | -0.07 | 0.23 |
| BA28 | -0.12 | 0.30 | -0.09 | 0.25 | 1.18 |
| BA29 | -0.01 | 0.06 | 0.00 | 0.02 | 2.95 |
| BA3 | -0.19 | 0.33 | -0.19 | 0.32 | 1.02 |
| BA30 | -0.03 | 0.10 | -0.02 | 0.08 | 1.27 |
| BA31 | -0.05 | 0.12 | -0.05 | 0.11 | 1.07 |
| BA32 | -0.07 | 0.12 | -0.08 | 0.14 | 0.86 |
| BA33 | -0.03 | 0.05 | -0.04 | 0.10 | 0.56 |
| BA35 | -0.01 | -0.02 | 0.01 | -0.08 | 0.32 |
| BA36 | -0.07 | 0.18 | -0.07 | 0.18 | 0.98 |
| BA37 | -0.12 | 0.24 | -0.12 | 0.23 | 1.04 |
| BA38 | -0.05 | 0.07 | -0.05 | 0.07 | 1.03 |
| BA39 | -0.08 | 0.17 | -0.09 | 0.19 | 0.92 |
| BA4 | -0.20 | 0.35 | -0.20 | 0.36 | 0.97 |
| BA40 | -0.12 | 0.23 | -0.12 | 0.24 | 0.96 |
| BA41 | -0.06 | 0.15 | -0.05 | 0.13 | 1.18 |
| BA42 | 0.02 | -0.19 | 0.00 | -0.11 | 1.80 |
| BA43 | -0.06 | 0.11 | -0.06 | 0.09 | 1.20 |
| BA44 | -0.09 | 0.17 | -0.10 | 0.15 | 1.10 |
| BA45 | -0.08 | 0.16 | -0.08 | 0.16 | 1.00 |
| BA46 | -0.08 | 0.12 | -0.08 | 0.12 | 1.02 |
| BA47 | -0.02 | 0.10 | -0.03 | 0.06 | 1.80 |
| BA5 | -0.04 | 0.08 | -0.04 | 0.08 | 1.01 |
| BA6 | -0.18 | 0.32 | -0.18 | 0.31 | 1.04 |
| BA7 | -0.09 | 0.19 | -0.09 | 0.18 | 1.05 |
| BA8 | -0.11 | 0.23 | -0.11 | 0.21 | 1.10 |
| BA9 | -0.15 | 0.30 | -0.14 | 0.26 | 1.12 |
| **R2\*-TWMB** | | | | | |
| ARCL | -0.10 | 0.18 | -0.09 | 0.14 | 1.32 |
| ARCR | -0.02 | 0.05 | -0.02 | 0.05 | 1.18 |
| ATRL | -0.09 | 0.19 | -0.08 | 0.16 | 1.19 |
| ATRR | -0.13 | 0.33 | -0.13 | 0.32 | 1.04 |
| CGCL | -0.02 | 0.06 | -0.02 | 0.04 | 1.51 |
| CGCR | -0.04 | 0.10 | -0.05 | 0.13 | 0.80 |
| CSTL | -0.02 | 0.03 | -0.03 | 0.06 | 0.52 |
| CSTR | 0.00 | -0.02 | -0.01 | 0.01 | -1.79 |
| FA | -0.19 | 0.22 | -0.17 | 0.19 | 1.16 |
| FP | -0.08 | 0.18 | -0.08 | 0.18 | 0.97 |
| IFOL | -0.04 | 0.05 | -0.04 | 0.06 | 0.90 |
| IFOR | -0.05 | 0.12 | -0.03 | 0.06 | 1.79 |
| ILFL | -0.01 | 0.01 | -0.01 | -0.01 | -1.66 |
| ILFR | -0.02 | 0.04 | -0.03 | 0.06 | 0.63 |
| pARCL | -0.05 | 0.10 | -0.05 | 0.10 | 1.06 |
| pARCR | 0.00 | 0.01 | 0.00 | 0.02 | 0.57 |
| SLFL | -0.09 | 0.20 | -0.09 | 0.16 | 1.21 |
| SLFR | -0.02 | 0.04 | -0.01 | 0.03 | 1.45 |
| **QSM-CGM** | | | | | |
| BA1 | 0.00 | -0.06 | -0.04 | 0.08 | -0.73 |
| BA17 | -0.01 | 0.04 | 0.00 | 0.01 | 3.46 |
| BA18 | -0.12 | 0.15 | -0.13 | 0.15 | 0.98 |
| BA19 | 0.02 | -0.07 | -0.02 | -0.04 | 1.52 |
| BA2 | -0.02 | -0.03 | 0.00 | -0.10 | 0.29 |
| BA22 | -0.01 | 0.03 | -0.03 | 0.04 | 0.89 |
| BA23 | -0.11 | 0.20 | -0.11 | 0.18 | 1.10 |
| BA24 | -0.01 | 0.02 | 0.00 | 0.01 | 1.48 |
| BA30 | -0.01 | 0.03 | -0.01 | 0.05 | 0.67 |
| BA31 | 0.12 | -0.25 | 0.09 | -0.12 | 2.10 |
| BA32 | -0.02 | 0.07 | -0.04 | 0.03 | 2.69 |
| BA37 | 0.01 | 0.06 | 0.01 | 0.04 | 1.59 |
| BA39 | -0.05 | 0.19 | -0.05 | 0.15 | 1.29 |
| BA4 | 0.02 | -0.20 | 0.03 | -0.24 | 0.84 |

| Region | | | | | |
|---|---|---|---|---|---|
| BA40 | -0.10 | 0.25 | -0.09 | 0.23 | 1.06 |
| BA41 | -0.07 | 0.16 | 0.02 | -0.06 | -2.82 |
| BA42 | 0.02 | 0.01 | 0.03 | -0.04 | -0.23 |
| BA43 | -0.06 | 0.12 | -0.10 | 0.22 | 0.57 |
| BA44 | 0.05 | -0.11 | 0.05 | -0.14 | 0.79 |
| BA45 | -0.03 | 0.08 | -0.02 | 0.05 | 1.40 |
| BA46 | 0.04 | -0.09 | 0.01 | -0.04 | 2.46 |
| BA5 | 0.08 | -0.29 | 0.05 | -0.18 | 1.61 |
| BA6 | -0.11 | 0.24 | -0.13 | 0.27 | 0.89 |
| BA7 | 0.01 | -0.04 | -0.02 | 0.01 | -2.85 |
| BA8 | -0.06 | 0.21 | -0.09 | 0.23 | 0.89 |
| BA9 | -0.04 | 0.14 | -0.04 | 0.11 | 1.30 |
| **QSM-SWM** | | | | | |
| BA10 | -0.01 | 0.02 | -0.05 | 0.07 | 0.23 |
| BA17 | -0.02 | 0.08 | -0.03 | 0.05 | 1.41 |
| BA18 | -0.01 | 0.05 | -0.02 | 0.09 | 0.51 |
| BA2 | -0.02 | 0.13 | -0.03 | 0.14 | 0.93 |
| BA23 | -0.03 | 0.10 | -0.04 | 0.10 | 1.03 |
| BA24 | -0.02 | 0.05 | -0.05 | 0.09 | 0.56 |
| BA25 | -0.06 | 0.15 | -0.08 | 0.17 | 0.85 |
| BA27 | -0.02 | 0.01 | -0.01 | 0.02 | 0.56 |
| BA28 | 0.14 | 0.01 | 0.03 | 0.05 | 0.25 |
| BA31 | -0.02 | 0.07 | -0.02 | 0.08 | 0.93 |
| BA32 | 0.01 | -0.02 | -0.02 | 0.02 | -1.55 |
| BA33 | -0.08 | 0.18 | -0.07 | 0.13 | 1.34 |
| BA35 | -0.02 | -0.05 | -0.02 | 0.02 | -2.34 |
| BA36 | 0.01 | -0.06 | 0.01 | -0.03 | 1.97 |
| BA42 | -0.04 | 0.06 | -0.03 | 0.07 | 0.95 |
| BA45 | 0.01 | -0.07 | 0.02 | -0.03 | 2.26 |
| BA46 | -0.05 | 0.10 | 0.03 | -0.05 | -2.26 |
| BA5 | -0.05 | 0.11 | -0.06 | 0.13 | 0.85 |
| BA6 | -0.03 | 0.09 | -0.03 | 0.09 | 0.93 |
| BA8 | 0.00 | -0.02 | -0.02 | 0.04 | -0.51 |
| BA9 | -0.01 | 0.04 | -0.02 | 0.07 | 0.66 |
| **QSM-TWMB** | | | | | |
| ARCL | -0.08 | 0.16 | -0.05 | 0.12 | 1.27 |
| ARCR | -0.10 | 0.22 | -0.10 | 0.23 | 0.98 |
| ATRL | -0.15 | 0.28 | -0.03 | 0.06 | 4.39 |
| ATRR | -0.04 | 0.08 | -0.05 | 0.09 | 0.92 |
| CSTL | -0.02 | 0.00 | -0.04 | 0.15 | 0.01 |
| CSTR | -0.04 | -0.03 | -0.09 | 0.14 | -0.23 |
| FA | -0.09 | 0.17 | -0.09 | 0.19 | 0.92 |
| IFOR | 0.00 | 0.02 | -0.02 | 0.03 | 0.47 |
| ILFR | -0.03 | 0.12 | -0.03 | 0.11 | 1.05 |
| pARCL | -0.07 | 0.16 | -0.07 | 0.14 | 1.15 |
| pARCR | -0.04 | 0.13 | -0.06 | 0.14 | 0.95 |
| SLFL | -0.02 | 0.02 | -0.01 | 0.01 | 1.65 |
| SLFR | -0.12 | 0.22 | -0.11 | 0.21 | 1.09 |

MSLL: Mean Standardized Log-Loss, EXPV: Explained Variance, CGM: Cortical Grey Matter, SWM: Superficial White Matter, TWMB: Tractography White Matter Bundle, BA: Brodmann Area, FA: Callosum Forceps Minor, FP: Callosum Forceps Major, pARC: Posterior Arcuate Fasciculus, ILF: Inferior Longitudinal Fasciculus, SLF: Superior Longitudinal Fasciculus, IFO: Inferior Fronto-Occipital Fasciculus, ARC: Arcuate, ATR: Thalamic Radiation, CGC: Cingulum Cingulate, CST: Corticospinal, L: Left, R: Right.

**S. 3 Polynomial Regression Analysis for Site 1 Data**

| Region | β(age) | p (age) | β(age$^2$) | p (age$^2$) | R$^2$ |
|---|---|---|---|---|---|
| **R1-CGM** | | | | | |
| BA10 | 2.73E-03 | 3.99E-23 | -2.48E-05 | <0.0001 | 0.33 |
| BA11 | 2.60E-03 | 3.18E-24 | -2.33E-05 | <0.0001 | 0.36 |
| BA24 | 1.99E-03 | 6.94E-21 | -1.72E-05 | <0.0001 | 0.36 |
| BA25 | 1.60E-03 | 3.31E-14 | -1.34E-05 | <0.0001 | 0.30 |
| BA32 | 2.42E-03 | 1.56E-24 | -2.10E-05 | <0.0001 | 0.40 |
| BA33 | 1.65E-03 | 2.50E-13 | -1.42E-05 | <0.0001 | 0.25 |
| BA44 | 2.73E-03 | 8.23E-29 | -2.33E-05 | <0.0001 | 0.47 |
| BA45 | 2.95E-03 | 7.40E-29 | -2.55E-05 | <0.0001 | 0.46 |
| BA46 | 2.81E-03 | 7.73E-28 | -2.47E-05 | <0.0001 | 0.42 |
| BA47 | 2.97E-03 | 1.80E-28 | -2.56E-05 | <0.0001 | 0.46 |
| BA1 | 3.59E-03 | 2.45E-23 | -3.29E-05 | <0.0001 | 0.32 |
| BA2 | 3.03E-03 | 2.43E-29 | -2.79E-05 | <0.0001 | 0.39 |
| BA3 | 3.10E-03 | 7.69E-21 | -2.83E-05 | <0.0001 | 0.29 |
| BA5 | 2.74E-03 | 8.43E-23 | -2.55E-05 | <0.0001 | 0.30 |
| BA7 | 2.84E-03 | 4.62E-28 | -2.62E-05 | <0.0001 | 0.37 |
| BA23 | 2.25E-03 | 1.84E-22 | -1.96E-05 | <0.0001 | 0.37 |
| BA29 | 2.33E-03 | 2.23E-15 | -2.05E-05 | <0.0001 | 0.26 |
| BA30 | 2.20E-03 | 6.82E-17 | -1.90E-05 | <0.0001 | 0.30 |
| BA31 | 2.27E-03 | 9.48E-24 | -2.01E-05 | <0.0001 | 0.36 |
| BA39 | 2.45E-03 | 8.12E-26 | -2.16E-05 | <0.0001 | 0.40 |
| BA40 | 2.68E-03 | 2.72E-31 | -2.38E-05 | <0.0001 | 0.45 |
| BA20 | 1.70E-03 | 3.84E-17 | -1.41E-05 | <0.0001 | 0.35 |
| BA21 | 2.12E-03 | 2.19E-22 | -1.82E-05 | <0.0001 | 0.39 |
| BA22 | 2.50E-03 | 4.08E-27 | -2.11E-05 | <0.0001 | 0.47 |
| BA37 | 2.01E-03 | 4.92E-20 | -1.74E-05 | <0.0001 | 0.34 |
| BA38 | 1.66E-03 | 1.70E-17 | -1.36E-05 | <0.0001 | 0.37 |
| BA41 | 2.28E-03 | 4.56E-18 | -1.94E-05 | <0.0001 | 0.33 |
| BA42 | 2.66E-03 | 3.30E-17 | -2.30E-05 | <0.0001 | 0.30 |
| BA17 | 2.24E-03 | 1.66E-13 | -2.09E-05 | <0.0001 | 0.18 |
| BA18 | 2.36E-03 | 4.53E-18 | -2.18E-05 | <0.0001 | 0.24 |
| BA19 | 2.31E-03 | 5.69E-23 | -2.12E-05 | <0.0001 | 0.31 |
| BA26 | 2.47E-03 | 2.58E-08 | -2.20E-05 | <0.0001 | 0.12 |
| BA27 | 2.04E-03 | 4.30E-14 | -1.70E-05 | <0.0001 | 0.30 |
| BA28 | 3.37E-03 | 6.14E-21 | -3.19E-05 | <0.0001 | 0.28 |
| BA35 | 2.04E-03 | 7.50E-15 | -1.74E-05 | <0.0001 | 0.28 |
| BA36 | 2.46E-03 | 7.27E-26 | -2.33E-05 | <0.0001 | 0.34 |
| BA43 | 2.72E-03 | 5.12E-23 | -2.37E-05 | <0.0001 | 0.38 |
| **R1-SWM** | | | | | |
| BA10 | 4.53E-03 | 4.79E-10 | -5.64E-05 | <0.0001 | 0.35 |
| BA11 | 5.57E-03 | 2.26E-13 | -6.51E-05 | <0.0001 | 0.33 |
| BA24 | 3.70E-03 | 4.87E-07 | -4.69E-05 | <0.0001 | 0.29 |
| BA25 | 6.26E-03 | 4.88E-10 | -7.21E-05 | <0.0001 | 0.23 |
| BA32 | 4.18E-03 | 1.24E-08 | -5.34E-05 | <0.0001 | 0.35 |
| BA33 | 5.78E-03 | 2.45E-10 | -6.41E-05 | <0.0001 | 0.20 |
| BA44 | 3.60E-03 | 9.87E-08 | -4.78E-05 | <0.0001 | 0.37 |
| BA45 | 4.61E-03 | 3.14E-10 | -5.49E-05 | <0.0001 | 0.28 |
| BA46 | 3.69E-03 | 2.27E-07 | -4.74E-05 | <0.0001 | 0.31 |
| BA47 | 5.33E-03 | 1.45E-12 | -6.15E-05 | <0.0001 | 0.29 |
| BA1 | 4.19E-03 | 2.36E-10 | -4.62E-05 | <0.0001 | 0.18 |
| BA2 | 3.74E-03 | 1.27E-07 | -4.72E-05 | <0.0001 | 0.29 |
| BA3 | 3.10E-03 | 3.27E-07 | -3.43E-05 | <0.0001 | 0.13 |
| BA5 | 4.11E-03 | 5.26E-08 | -5.05E-05 | <0.0001 | 0.27 |
| BA7 | 4.58E-03 | 1.90E-10 | -5.44E-05 | <0.0001 | 0.29 |
| BA23 | 5.23E-03 | 1.59E-11 | -6.34E-05 | <0.0001 | 0.34 |
| BA29 | 6.47E-03 | 7.96E-11 | -6.97E-05 | <0.0001 | 0.17 |
| BA30 | 4.93E-03 | 1.28E-09 | -6.02E-05 | <0.0001 | 0.31 |
| BA31 | 4.69E-03 | 2.79E-10 | -5.81E-05 | <0.0001 | 0.35 |
| BA39 | 4.26E-03 | 2.84E-09 | -5.32E-05 | <0.0001 | 0.33 |
| BA40 | 5.01E-03 | 1.08E-11 | -5.86E-05 | <0.0001 | 0.30 |
| BA20 | 6.75E-03 | 3.29E-16 | -7.30E-05 | <0.0001 | 0.26 |
| BA21 | 6.30E-03 | 1.09E-14 | -7.03E-05 | <0.0001 | 0.29 |
| BA22 | 5.42E-03 | 2.41E-12 | -6.10E-05 | <0.0001 | 0.24 |
| BA37 | 5.57E-03 | 1.42E-12 | -6.52E-05 | <0.0001 | 0.31 |
| BA38 | 6.90E-03 | 4.30E-16 | -7.06E-05 | <0.0001 | 0.21 |
| BA41 | 3.84E-03 | 8.96E-08 | -5.00E-05 | <0.0001 | 0.34 |
| BA42 | 2.94E-03 | 2.11E-05 | -4.13E-05 | <0.0001 | 0.35 |
| BA17 | 2.12E-03 | 1.03E-03 | -2.58E-05 | <0.0001 | 0.11 |
| BA18 | 3.15E-03 | 1.84E-07 | -3.77E-05 | <0.0001 | 0.22 |
| BA19 | 4.07E-03 | 1.17E-09 | -5.07E-05 | <0.0001 | 0.34 |
| BA26 | 6.77E-03 | 6.03E-11 | -7.09E-05 | <0.0001 | 0.15 |
| BA27 | 3.92E-03 | 9.16E-05 | -3.41E-05 | <0.001 | 0.08 |
| BA28 | 5.14E-03 | 3.43E-12 | -5.20E-05 | <0.0001 | 0.15 |
| BA35 | 4.12E-03 | 2.91E-09 | -4.48E-05 | <0.0001 | 0.15 |
| BA36 | 5.55E-03 | 1.04E-14 | -5.45E-05 | <0.0001 | 0.18 |
| BA43 | 5.06E-03 | 2.54E-10 | -6.01E-05 | <0.0001 | 0.28 |
| **R1-TWMB** | | | | | |
| ARCL | 3.96E-03 | 1.41E-07 | -4.96E-05 | <0.0001 | 0.29 |
| ARCR | 3.85E-03 | 2.82E-07 | -4.95E-05 | <0.0001 | 0.31 |
| ATRL | 4.71E-03 | 1.60E-08 | -6.15E-05 | <0.0001 | 0.38 |
| ATRR | 4.75E-03 | 1.79E-08 | -6.07E-05 | <0.0001 | 0.34 |
| CGCL | 4.48E-03 | 3.34E-08 | -5.60E-05 | <0.0001 | 0.30 |
| CGCR | 4.54E-03 | 9.03E-09 | -5.69E-05 | <0.0001 | 0.33 |
| CSTL | 3.62E-03 | 5.45E-08 | -4.76E-05 | <0.0001 | 0.37 |
| CSTR | 3.72E-03 | 2.90E-08 | -4.90E-05 | <0.0001 | 0.38 |
| FA | 3.78E-03 | 8.47E-06 | -5.27E-05 | <0.0001 | 0.36 |
| FP | 2.68E-03 | 6.85E-04 | -4.29E-05 | <0.0001 | 0.40 |
| pARCL | 4.30E-03 | 7.34E-08 | -5.18E-05 | <0.0001 | 0.25 |
| pARCR | 4.30E-03 | 8.46E-08 | -5.34E-05 | <0.0001 | 0.28 |
| IFOL | 3.48E-03 | 9.51E-06 | -5.15E-05 | <0.0001 | 0.44 |
| IFOR | 3.38E-03 | 2.06E-05 | -4.95E-05 | <0.0001 | 0.40 |
| ILFL | 3.52E-03 | 5.48E-06 | -4.98E-05 | <0.0001 | 0.39 |
| ILFR | 3.76E-03 | 1.04E-06 | -5.14E-05 | <0.0001 | 0.37 |
| SLFL | 4.40E-03 | 5.19E-09 | -5.43E-05 | <0.0001 | 0.31 |
| SLFR | 4.52E-03 | 2.00E-09 | -5.70E-05 | <0.0001 | 0.36 |
| **R2*-CGM** | | | | | |
| BA4 | 1.67E-01 | 4.17E-17 | -1.37E-03 | <0.0001 | 0.36 |
| BA6 | 1.44E-01 | 8.83E-19 | -1.18E-03 | <0.0001 | 0.40 |
| BA8 | 1.36E-01 | 2.39E-18 | -1.14E-03 | <0.0001 | 0.37 |
| BA9 | 1.34E-01 | 9.43E-19 | -1.18E-03 | <0.0001 | 0.31 |
| BA10 | 1.20E-01 | 5.06E-12 | -1.07E-03 | <0.0001 | 0.20 |
| BA11 | 1.14E-01 | 9.82E-07 | -9.44E-04 | <0.0001 | 0.16 |
| BA24 | 1.05E-01 | 4.91E-07 | -8.50E-04 | <0.0001 | 0.17 |
| BA25 | 1.01E-01 | 5.76E-03 | -9.08E-04 | <0.05 | 0.03 |
| BA32 | 1.32E-01 | 2.09E-13 | -1.16E-03 | <0.0001 | 0.24 |
| BA33 | 6.03E-02 | 7.87E-03 | -4.65E-04 | <0.05 | 0.09 |
| BA44 | 1.29E-01 | 1.84E-13 | -1.04E-03 | <0.0001 | 0.32 |
| BA45 | 1.31E-01 | 2.22E-11 | -1.12E-03 | <0.0001 | 0.22 |
| BA46 | 1.25E-01 | 9.79E-12 | -1.05E-03 | <0.0001 | 0.23 |
| BA47 | 1.08E-01 | 4.92E-06 | -8.49E-04 | <0.001 | 0.16 |
| BA1 | 1.32E-01 | 1.35E-12 | -1.19E-03 | <0.0001 | 0.19 |
| BA2 | 1.36E-01 | 2.85E-14 | -1.21E-03 | <0.0001 | 0.23 |
| BA3 | 8.66E-02 | 1.63E-06 | -7.93E-04 | <0.0001 | 0.09 |
| BA5 | 1.09E-01 | 2.45E-11 | -9.49E-04 | <0.0001 | 0.20 |
| BA7 | 1.21E-01 | 5.65E-13 | -1.05E-03 | <0.0001 | 0.23 |
| BA23 | 1.26E-01 | 1.63E-06 | -1.05E-03 | <0.0001 | 0.13 |
| BA29 | 7.21E-02 | 3.77E-02 | -5.10E-04 | >0.05 | 0.05 |
| BA30 | 1.11E-01 | 1.52E-06 | -9.56E-04 | <0.0001 | 0.11 |
| BA31 | 1.29E-01 | 1.64E-08 | -1.10E-03 | <0.0001 | 0.16 |
| BA39 | 1.36E-01 | 1.16E-12 | -1.16E-03 | <0.0001 | 0.24 |
| BA40 | 1.27E-01 | 1.01E-13 | -1.08E-03 | <0.0001 | 0.27 |
| BA20 | 1.06E-01 | 7.45E-06 | -7.60E-04 | <0.01 | 0.22 |
| BA21 | 1.13E-01 | 1.25E-06 | -8.69E-04 | <0.001 | 0.20 |
| BA22 | 1.37E-01 | 7.71E-15 | -1.13E-03 | <0.0001 | 0.33 |
| BA37 | 1.43E-01 | 1.19E-12 | -1.20E-03 | <0.0001 | 0.26 |
| BA38 | 8.27E-02 | 3.17E-04 | -5.96E-04 | <0.01 | 0.16 |
| BA41 | 1.26E-01 | 5.15E-10 | -1.07E-03 | <0.0001 | 0.19 |
| BA42 | 1.56E-01 | 3.22E-13 | -1.33E-03 | <0.0001 | 0.26 |
| BA17 | 1.09E-01 | 9.17E-07 | -9.79E-04 | <0.0001 | 0.09 |
| BA18 | 1.37E-01 | 3.56E-12 | -1.20E-03 | <0.0001 | 0.21 |
| BA19 | 1.23E-01 | 6.05E-12 | -1.04E-03 | <0.0001 | 0.24 |
| BA26 | 5.38E-02 | 1.79E-01 | -2.24E-04 | >0.05 | 0.08 |
| BA27 | 1.20E-01 | 5.71E-05 | -1.01E-03 | <0.001 | 0.08 |
| BA28 | 1.11E-01 | 7.83E-05 | -9.39E-04 | <0.001 | 0.10 |
| BA35 | 1.01E-01 | 1.17E-03 | -7.53E-04 | <0.05 | 0.10 |
| BA36 | 1.23E-01 | 5.24E-06 | -1.02E-03 | <0.001 | 0.12 |
| BA43 | 1.36E-01 | 3.36E-14 | -1.15E-03 | <0.0001 | 0.28 |
| **R2\*-SWM** | | | | | |
| BA4 | 1.62E-01 | 9.63E-18 | -1.48E-03 | <0.0001 | 0.27 |
| BA6 | 1.45E-01 | 1.85E-17 | -1.38E-03 | <0.0001 | 0.24 |
| BA8 | 1.54E-01 | 2.85E-16 | -1.51E-03 | <0.0001 | 0.21 |
| BA9 | 1.49E-01 | 4.12E-16 | -1.53E-03 | <0.0001 | 0.22 |
| BA10 | 1.35E-01 | 4.86E-12 | -1.42E-03 | <0.0001 | 0.19 |
| BA11 | 1.33E-01 | 1.81E-09 | -1.35E-03 | <0.0001 | 0.14 |
| BA24 | 1.16E-01 | 2.71E-08 | -1.12E-03 | <0.0001 | 0.16 |
| BA25 | 1.09E-01 | 1.03E-03 | -1.23E-03 | <0.001 | 0.07 |
| BA32 | 1.55E-01 | 4.85E-15 | -1.62E-03 | <0.0001 | 0.24 |
| BA33 | 1.16E-01 | 1.56E-06 | -1.22E-03 | <0.0001 | 0.13 |
| BA44 | 1.17E-01 | 2.60E-09 | -1.13E-03 | <0.0001 | 0.13 |
| BA45 | 1.31E-01 | 1.40E-08 | -1.30E-03 | <0.0001 | 0.11 |
| BA46 | 1.14E-01 | 8.50E-09 | -1.17E-03 | <0.0001 | 0.11 |
| BA47 | 1.37E-01 | 2.70E-07 | -1.37E-03 | <0.0001 | 0.09 |
| BA1 | 1.47E-01 | 2.61E-15 | -1.32E-03 | <0.0001 | 0.23 |
| BA2 | 1.31E-01 | 1.61E-12 | -1.22E-03 | <0.0001 | 0.18 |
| BA3 | 1.24E-01 | 3.84E-13 | -1.10E-03 | <0.0001 | 0.22 |
| BA5 | 1.24E-01 | 2.32E-11 | -1.18E-03 | <0.0001 | 0.17 |
| BA7 | 1.35E-01 | 3.26E-13 | -1.28E-03 | <0.0001 | 0.20 |
| BA23 | 1.39E-01 | 1.65E-07 | -1.41E-03 | <0.0001 | 0.09 |
| BA29 | 9.89E-02 | 9.13E-04 | -9.23E-04 | <0.01 | 0.04 |
| BA30 | 1.12E-01 | 2.45E-06 | -1.14E-03 | <0.0001 | 0.08 |
| BA31 | 1.34E-01 | 4.54E-09 | -1.29E-03 | <0.0001 | 0.13 |
| BA39 | 1.39E-01 | 1.32E-12 | -1.34E-03 | <0.0001 | 0.18 |
| BA40 | 1.36E-01 | 7.60E-13 | -1.31E-03 | <0.0001 | 0.19 |
| BA20 | 1.39E-01 | 1.77E-09 | -1.27E-03 | <0.0001 | 0.14 |
| BA21 | 1.62E-01 | 3.73E-11 | -1.56E-03 | <0.0001 | 0.15 |
| BA22 | 1.55E-01 | 3.87E-15 | -1.48E-03 | <0.0001 | 0.20 |
| BA37 | 1.61E-01 | 6.19E-15 | -1.56E-03 | <0.0001 | 0.19 |
| BA38 | 1.38E-01 | 3.56E-08 | -1.29E-03 | <0.0001 | 0.11 |
| BA41 | 1.17E-01 | 2.78E-08 | -1.16E-03 | <0.0001 | 0.10 |
| BA42 | 1.45E-01 | 2.58E-12 | -1.33E-03 | <0.0001 | 0.19 |
| BA17 | 1.10E-01 | 2.25E-07 | -9.71E-04 | <0.0001 | 0.12 |
| BA18 | 1.43E-01 | 1.21E-13 | -1.27E-03 | <0.0001 | 0.24 |
| BA19 | 1.33E-01 | 3.75E-12 | -1.23E-03 | <0.0001 | 0.18 |
| BA26 | 8.61E-02 | 7.57E-03 | -7.83E-04 | <0.05 | 0.03 |
| BA27 | 1.19E-01 | 9.95E-05 | -9.75E-04 | <0.01 | 0.10 |
| BA28 | 1.49E-01 | 1.01E-07 | -1.34E-03 | <0.0001 | 0.12 |
| BA35 | 1.18E-01 | 7.27E-05 | -1.04E-03 | <0.001 | 0.07 |
| BA36 | 1.28E-01 | 2.36E-06 | -1.12E-03 | <0.0001 | 0.10 |
| BA43 | 1.48E-01 | 9.94E-13 | -1.43E-03 | <0.0001 | 0.16 |
| **R2\*-TWMB** | | | | | |
| ARCL | 1.12E-01 | 1.74E-06 | -1.21E-03 | <0.0001 | 0.11 |
| ARCR | 5.74E-02 | 1.15E-02 | -6.45E-04 | <0.01 | 0.07 |
| ATRL | 1.22E-01 | 5.14E-07 | -1.46E-03 | <0.0001 | 0.22 |
| ATRR | 1.09E-01 | 1.16E-05 | -1.32E-03 | <0.0001 | 0.20 |
| CGCL | 1.11E-01 | 2.75E-06 | -1.17E-03 | <0.0001 | 0.12 |
| CGCR | 1.09E-01 | 5.63E-06 | -1.16E-03 | <0.0001 | 0.10 |
| CSTL | 1.15E-01 | 3.79E-07 | -1.22E-03 | <0.0001 | 0.12 |
| CSTR | 6.71E-02 | 2.46E-03 | -6.92E-04 | <0.01 | 0.09 |
| FA | 6.25E-02 | 1.44E-02 | -8.44E-04 | <0.001 | 0.13 |
| FP | 3.61E-02 | 2.35E-01 | -6.01E-04 | <0.05 | 0.08 |
| pARCL | 1.23E-01 | 4.71E-08 | -1.23E-03 | <0.0001 | 0.12 |
| pARCR | 1.04E-01 | 1.23E-05 | -1.08E-03 | <0.0001 | 0.10 |
| IFOL | 9.67E-02 | 4.63E-05 | -1.15E-03 | <0.0001 | 0.15 |
| IFOR | 7.41E-02 | 2.01E-03 | -9.56E-04 | <0.0001 | 0.16 |
| ILFL | 9.37E-02 | 2.06E-04 | -1.06E-03 | <0.0001 | 0.09 |
| ILFR | 7.51E-02 | 2.03E-03 | -9.12E-04 | <0.001 | 0.11 |
| SLFL | 1.18E-01 | 9.09E-08 | -1.23E-03 | <0.0001 | 0.12 |
| SLFR | 7.16E-02 | 7.81E-04 | -7.75E-04 | <0.001 | 0.10 |
| **Susceptibility-CGM** | | | | | |
| BA4 | 3.35E-04 | 7.08E-12 | -3.02E-06 | <0.0001 | 0.18 |
| BA6 | 2.47E-04 | 4.91E-07 | -2.23E-06 | <0.0001 | 0.09 |
| BA8 | 1.64E-04 | 1.19E-04 | -1.38E-06 | <0.01 | 0.09 |
| BA9 | 1.35E-04 | 4.15E-04 | -1.24E-06 | <0.01 | 0.04 |
| BA10 | 5.44E-05 | 1.02E-01 | -5.46E-07 | >0.05 | 0.00 |
| BA11 | 1.78E-05 | 4.20E-01 | -2.27E-07 | >0.05 | 0.01 |
| BA24 | 1.65E-04 | 3.33E-04 | -1.83E-06 | <0.0001 | 0.07 |
| BA25 | -1.50E-05 | 8.00E-01 | 2.37E-08 | >0.05 | 0.01 |
| BA32 | 2.18E-04 | 2.43E-06 | -2.04E-06 | <0.0001 | 0.08 |
| BA33 | 2.34E-05 | 7.02E-01 | -5.69E-07 | >0.05 | 0.04 |
| BA44 | 2.01E-04 | 3.94E-04 | -1.83E-06 | <0.01 | 0.04 |
| BA45 | 1.50E-04 | 7.32E-04 | -1.43E-06 | <0.01 | 0.03 |
| BA46 | 1.13E-04 | 5.02E-03 | -1.15E-06 | <0.01 | 0.02 |
| BA47 | 8.25E-05 | 5.73E-02 | -6.89E-07 | >0.05 | 0.01 |
| BA1 | 1.50E-04 | 6.89E-07 | -1.31E-06 | <0.0001 | 0.11 |
| BA2 | 2.64E-04 | 1.93E-09 | -2.60E-06 | <0.0001 | 0.11 |
| BA3 | 5.88E-05 | 2.52E-01 | -8.20E-07 | >0.05 | 0.03 |
| BA5 | 2.66E-04 | 9.34E-08 | -2.44E-06 | <0.0001 | 0.10 |
| BA7 | 2.14E-04 | 9.92E-09 | -2.11E-06 | <0.0001 | 0.10 |
| BA23 | 3.61E-04 | 7.50E-09 | -3.25E-06 | <0.0001 | 0.13 |
| BA29 | 1.29E-04 | 2.34E-01 | -1.60E-06 | >0.05 | 0.01 |
| BA30 | 1.28E-04 | 3.13E-02 | -1.40E-06 | <0.05 | 0.05 |
| BA31 | 2.78E-04 | 2.33E-07 | -2.68E-06 | <0.0001 | 0.08 |
| BA39 | 1.72E-04 | 2.23E-05 | -1.78E-06 | <0.0001 | 0.06 |
| BA40 | 2.00E-04 | 7.40E-09 | -1.94E-06 | <0.0001 | 0.10 |
| BA20 | 2.50E-05 | 3.61E-01 | -2.63E-07 | >0.05 | 0.00 |
| BA21 | 3.20E-05 | 3.46E-01 | -3.14E-07 | >0.05 | -0.01 |
| BA22 | 1.61E-04 | 3.35E-05 | -1.70E-06 | <0.0001 | 0.07 |
| BA37 | 8.56E-05 | 1.16E-02 | -9.77E-07 | <0.01 | 0.06 |
| BA38 | 2.56E-05 | 4.99E-01 | -3.13E-07 | >0.05 | 0.04 |
| BA41 | 2.54E-04 | 2.34E-05 | -2.98E-06 | <0.0001 | 0.14 |
| BA42 | 2.64E-04 | 6.61E-07 | -2.62E-06 | <0.0001 | 0.11 |
| BA17 | 8.60E-05 | 2.21E-02 | -9.09E-07 | <0.05 | 0.01 |
| BA18 | 1.18E-04 | 4.11E-05 | -1.28E-06 | <0.0001 | 0.09 |
| BA19 | 1.56E-04 | 1.23E-05 | -1.69E-06 | <0.0001 | 0.10 |
| BA26 | 1.23E-04 | 3.65E-01 | -1.19E-06 | >0.05 | -0.01 |
| BA27 | -2.65E-06 | 9.66E-01 | 1.82E-07 | >0.05 | 0.01 |
| BA28 | 6.48E-05 | 2.93E-01 | -5.02E-07 | >0.05 | 0.00 |
| BA35 | -1.82E-05 | 7.79E-01 | 9.76E-08 | >0.05 | 0.03 |
| BA36 | 4.79E-05 | 2.67E-01 | -3.67E-07 | >0.05 | 0.01 |
| BA43 | 1.57E-04 | 9.17E-04 | -1.90E-06 | <0.0001 | 0.10 |
| **Susceptibility-SWM** | | | | | |
| BA4 | 1.08E-04 | 3.23E-02 | -5.79E-07 | >0.05 | 0.12 |
| BA6 | 1.66E-04 | 3.53E-05 | -1.51E-06 | <0.001 | 0.07 |
| BA8 | 1.39E-04 | 6.50E-04 | -1.33E-06 | <0.001 | 0.06 |
| BA9 | 1.48E-04 | 1.74E-04 | -1.62E-06 | <0.0001 | 0.09 |
| BA10 | 1.10E-04 | 1.25E-02 | -1.49E-06 | <0.001 | 0.15 |
| BA11 | -1.92E-05 | 5.36E-01 | -9.53E-09 | >0.05 | 0.06 |
| BA24 | 3.78E-04 | 1.09E-09 | -3.88E-06 | <0.0001 | 0.12 |
| BA25 | 7.18E-05 | 1.92E-01 | -1.20E-06 | <0.05 | 0.10 |
| BA32 | 2.97E-04 | 9.24E-08 | -3.11E-06 | <0.0001 | 0.10 |
| BA33 | 2.10E-04 | 1.90E-03 | -2.54E-06 | <0.001 | 0.09 |
| BA44 | 8.67E-05 | 7.68E-02 | -9.04E-07 | >0.05 | 0.01 |
| BA45 | 8.95E-05 | 7.55E-02 | -1.02E-06 | <0.05 | 0.02 |
| BA46 | 1.01E-04 | 2.30E-02 | -1.25E-06 | <0.01 | 0.08 |
| BA47 | -6.05E-05 | 2.97E-01 | 2.71E-07 | >0.05 | 0.04 |
| BA1 | 5.67E-05 | 1.62E-01 | -1.79E-07 | >0.05 | 0.11 |
| BA2 | 1.50E-04 | 4.39E-04 | -1.24E-06 | <0.01 | 0.07 |
| BA3 | 7.55E-05 | 5.22E-02 | -5.29E-07 | >0.05 | 0.06 |
| BA5 | 1.99E-04 | 2.15E-04 | -1.67E-06 | <0.01 | 0.07 |
| BA7 | 9.05E-05 | 3.97E-02 | -8.07E-07 | >0.05 | 0.01 |
| BA23 | 3.94E-04 | 6.34E-07 | -3.75E-06 | <0.0001 | 0.08 |
| BA29 | 1.78E-04 | 1.27E-01 | -1.77E-06 | >0.05 | 0.00 |
| BA30 | -3.80E-05 | 6.43E-01 | 3.13E-07 | >0.05 | -0.01 |
| BA31 | 2.44E-04 | 3.40E-04 | -2.38E-06 | <0.001 | 0.04 |
| BA39 | -8.15E-07 | 9.87E-01 | 4.06E-08 | >0.05 | -0.01 |
| BA40 | 5.28E-05 | 2.42E-01 | -5.12E-07 | >0.05 | 0.00 |
| BA20 | 2.71E-05 | 4.39E-01 | -1.79E-07 | >0.05 | 0.00 |
| BA21 | -2.45E-05 | 5.25E-01 | 1.84E-07 | >0.05 | 0.02 |
| BA22 | 2.02E-05 | 6.79E-01 | -4.33E-07 | >0.05 | 0.04 |
| BA37 | -8.56E-06 | 8.33E-01 | 6.13E-08 | >0.05 | 0.02 |
| BA38 | 3.63E-05 | 3.98E-01 | -4.43E-07 | >0.05 | 0.04 |
| BA41 | -2.79E-05 | 5.67E-01 | 4.33E-07 | >0.05 | 0.01 |
| BA42 | 1.79E-04 | 2.90E-04 | -1.46E-06 | <0.01 | 0.11 |
| BA17 | 1.50E-04 | 4.36E-04 | -1.44E-06 | <0.001 | 0.05 |
| BA18 | 1.07E-04 | 2.53E-03 | -1.02E-06 | <0.01 | 0.03 |
| BA19 | 3.88E-05 | 3.53E-01 | -3.30E-07 | >0.05 | 0.00 |
| BA26 | 1.43E-04 | 3.00E-01 | -1.24E-06 | >0.05 | 0.00 |
| BA27 | 2.62E-04 | 5.93E-03 | -2.73E-06 | <0.01 | 0.02 |
| BA28 | 1.53E-04 | 4.08E-02 | -1.80E-06 | <0.05 | 0.03 |
| BA35 | 2.68E-04 | 5.82E-03 | -2.86E-06 | <0.01 | 0.03 |
| BA36 | 2.36E-04 | 1.07E-04 | -2.29E-06 | <0.001 | 0.08 |
| BA43 | 3.55E-05 | 5.20E-01 | -5.95E-07 | >0.05 | 0.03 |
| **Susceptibility-TWMB** | | | | | |
| ARCL | -2.36E-04 | 1.01E-07 | 2.48E-06 | <0.0001 | 0.10 |
| ARCR | -2.95E-04 | 4.23E-09 | 3.13E-06 | <0.0001 | 0.14 |
| ATRL | -1.87E-04 | 3.24E-04 | 2.14E-06 | <0.0001 | 0.09 |
| ATRR | -2.21E-04 | 1.47E-05 | 2.46E-06 | <0.0001 | 0.12 |
| CGCL | 6.07E-05 | 3.25E-01 | -8.13E-07 | >0.05 | 0.01 |
| CGCR | -1.46E-05 | 8.22E-01 | 2.09E-07 | >0.05 | -0.01 |
| CSTL | -1.92E-04 | 3.06E-06 | 1.80E-06 | <0.0001 | 0.08 |
| CSTR | -2.02E-04 | 1.19E-05 | 2.04E-06 | <0.0001 | 0.08 |
| FA | -2.16E-04 | 1.70E-03 | 2.77E-06 | <0.0001 | 0.16 |
| FP | -5.75E-05 | 4.17E-01 | 1.22E-06 | >0.05 | 0.11 |
| pARCL | -3.47E-04 | 1.16E-07 | 3.72E-06 | <0.0001 | 0.11 |
| pARCR | -2.94E-04 | 3.26E-05 | 3.22E-06 | <0.0001 | 0.08 |
| IFOL | -4.87E-05 | 3.54E-01 | 4.94E-07 | >0.05 | -0.01 |
| IFOR | -1.97E-04 | 2.49E-03 | 1.94E-06 | <0.01 | 0.05 |
| ILFL | -2.63E-05 | 6.49E-01 | 6.99E-07 | >0.05 | 0.07 |
| ILFR | -2.01E-04 | 4.57E-03 | 2.36E-06 | <0.001 | 0.07 |
| SLFL | -2.67E-04 | 1.46E-09 | 2.67E-06 | <0.0001 | 0.12 |
| SLFR | -3.25E-04 | 3.19E-12 | 3.31E-06 | <0.0001 | 0.16 |

CGM: Cortical Grey Matter, SWM: Superficial White Matter, TWMB: Tractography White Matter Bundle, BA: Brodmann Area, FA: Callosum Forceps Minor, FP: Callosum Forceps Major, pARC: Posterior Arcuate Fasciculus, ILF: Inferior Longitudinal Fasciculus, SLF: Superior Longitudinal Fasciculus, IFO: Inferior Fronto-Occipital Fasciculus, ARC: Arcuate, ATR: Thalamic Radiation, CGC: Cingulum Cingulate, CST: Corticospinal, L: Left, R: Right.

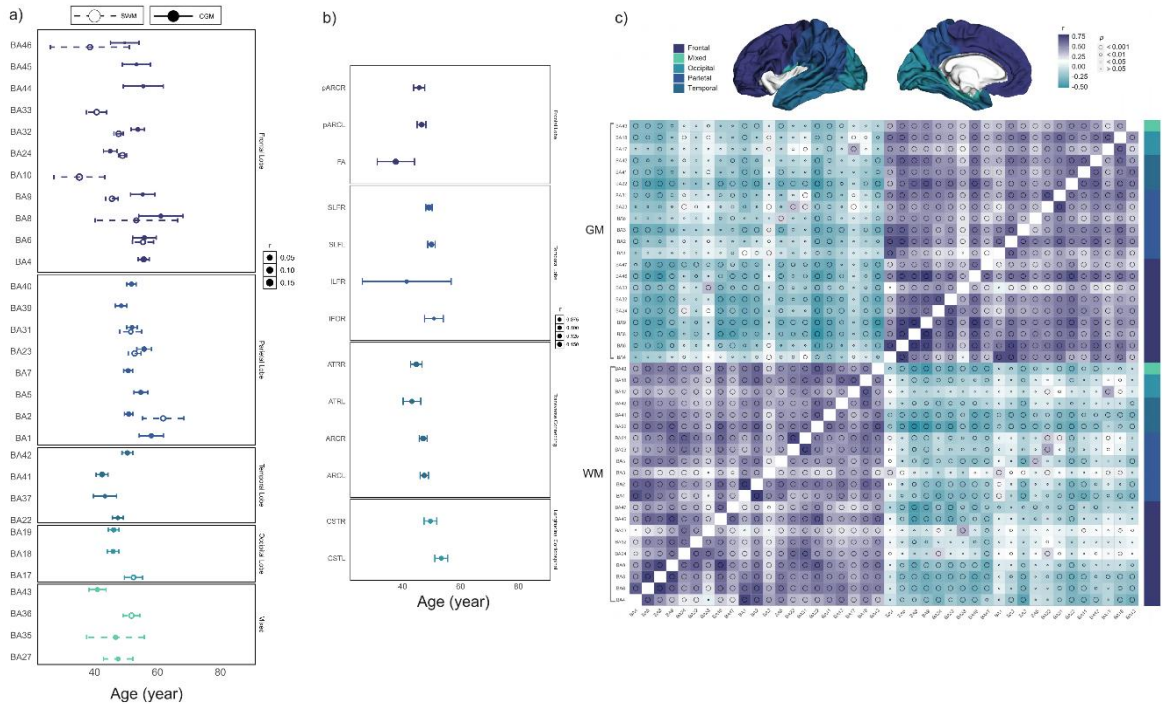

**S. 4 Peak Ages and Quantitative Connectivity Analysis in QSM**

Panel a shows the QSM peak ages determined by significant LTRs from quadratic models (p < 0.05) in CGM and SWM across BAs. Each data point represents the mean value of a specific BA, with an error bar showing the standard error derived from 1000 bootstrap iterations. Panel B shows peak ages extracted among TWMBs. The data are organized by different brain lobes or the location of white matter tracts. Panel C shows the Pearson correlation matrix for QSM in both intra- and inter-regional connectivity patterns within and between CGM and SWM. In each matrix, the upper triangles depict Z-scores derived from the residuals of significant quadratic models adjusted for aging effects, illustrating connections that are independent of age. The lower triangles display the raw average values for each region, highlighting the overall connectivity. Only regions with significant quadratic regression fits were included in the matrix calculations. Each matrix organizes data into five lobes based on the anatomical locations of the selected. CGM: Cortical Grey Matter, SWM: Superficial White Matter, BA: Brodmann Area, FA: Callosum Forceps Minor, FP: Callosum Forceps Major, pARC: Posterior Arcuate Fasciculus, ILF: Inferior Longitudinal Fasciculus, SLF: Superior Longitudinal Fasciculus, IFO: Inferior Fronto-Occipital Fasciculus, ARC: Arcuate, ATR: Thalamic Radiation, CGC: Cingulum Cingulate, CST:
Corticospinal, L: Left, R: Right. QSM: quantitative susceptibility mapping.

**S. 5 Data Harmonization Comparison**

| qMRI | Structure | Method | Mean (SD) | CV |
|---|---|---|---|---|
| R1 | | EBS | 0.69 (0.016) | 0.02 |
| | CGM | HBRbsbs | 0.69 (0.017) | 0.03 |
| | | Raw | 0.69 (0.024) | 0.03 |
| | | EBS | 1.05 (0.039) | 0.04 |
| | SMB | HBRbs | 1.05 (0.039) | 0.04 |
| | | Raw | 1.05 (0.041) | 0.04 |
| | | EBS | 1.16 (0.042) | 0.04 |
| | TWMB | HBRbs | 1.16 (0.041) | 0.04 |
| | | Raw | 1.16 (0.041) | 0.04 |
| R2* | | EBS | 17.98 (1.10) | 0.06 |
| | CGM | HBRbs | 17.53 (1.06) | 0.06 |
| | | Raw | 17.97 (1.36) | 0.08 |
| | | EBS | 20.05 (1.08) | 0.05 |
| | SMB | HBRbs | 19.97 (1.04) | 0.05 |
| | | Raw | 20.05 (1.21) | 0.06 |
| | | EBS | 20.67 (1.06) | 0.05 |
| | TWMB | HBRbs | 20.82 (1.09) | 0.05 |
| | | Raw | 20.67 (1.08) | 0.05 |
| Susceptibility | | EBS | 0.002 (0.002) | 19.00 |
| | CGM | HBRbs | 0.002 (0.002) | 1.61 |
| | | Raw | 0.002 (0.002) | -37.48 |
| | | EBS | 0.002 (0.003) | 0.91 |
| | SMB | HBRbs | 0.002 (0.003) | 0.42 |
| | | Raw | 0.002 (0.003) | 0.91 |
| | | EBS | -0.011 (0.003) | -0.27 |
| | TWMB | HBRbs | -0.012 (0.003) | -0.26 |
| | | Raw | -0.011 (0.003) | -0.30 |

CGM: Cortical Grey Matter, SWM: Superficial White Matte, TWMB: Tractography White Matter Bundle, SD: Standard Deviation, CV: coefficient of variation, EBS: Empirical Bayes Statistics Harmonization, HBRbs: Hierarchical Bayesian-based B-spline Harmonization.

## S. 6 Polynomial Regression Analysis and Peak Age for Multi-Center Cohorts

| Region | α | β(age) | $p$ (age) | β(age$^2$) | $p$ (age$^2$) | $p$ (Sex) | $R^2$ | Peak Age | $p$ LTR |
|---|---|---|---|---|---|---|---|---|---|
| **R1-CGM** | | | | | | | | | |
| **Frontal Lobe** | | | | | | | | | |
| BA4 | 0.68 | 2.83E-03 | -2.43E-05 | 1.36E-52 | 3.40E-33 | 0.95 | 0.56 | 58.37 | <0.0001 |
| BA6 | 0.64 | 2.93E-03 | -2.50E-05 | 4.22E-85 | 1.02E-55 | 0.35 | 0.72 | 58.5 | <0.0001 |
| BA8 | 0.62 | 3.04E-03 | -2.61E-05 | 3.27E-99 | 1.30E-66 | 0.38 | 0.76 | 58.27 | <0.0001 |
| BA9 | 0.62 | 2.79E-03 | -2.47E-05 | 8.58E-99 | 5.22E-69 | 0.19 | 0.74 | 56.59 | <0.0001 |
| BA10 | 0.64 | 2.21E-03 | -1.96E-05 | 4.70E-57 | 3.51E-38 | 0.12 | 0.56 | 56.46 | <0.0001 |
| BA11 | 0.63 | 2.40E-03 | -2.14E-05 | 1.42E-73 | 1.03E-50 | 0.01 | 0.64 | 56.15 | <0.0001 |
| BA24 | 0.62 | 1.57E-03 | -1.30E-05 | 9.74E-50 | 1.00E-29 | 0.39 | 0.58 | 60.27 | <0.0001 |
| BA32 | 0.63 | 1.86E-03 | -1.55E-05 | 1.28E-57 | 7.62E-35 | 0.52 | 0.62 | 60.19 | <0.0001 |
| BA33 | 0.61 | 1.47E-03 | -1.24E-05 | 3.50E-41 | 6.00E-25 | 0.40 | 0.51 | 59.51 | <0.0001 |
| BA44 | 0.64 | 2.23E-03 | -1.83E-05 | 2.52E-72 | 1.70E-43 | 0.41 | 0.70 | 61.13 | <0.0001 |
| BA45 | 0.64 | 2.42E-03 | -2.01E-05 | 2.39E-74 | 8.72E-46 | 0.52 | 0.70 | 60.28 | <0.0001 |
| BA46 | 0.64 | 2.47E-03 | -2.11E-05 | 1.72E-77 | 3.54E-50 | 0.14 | 0.69 | 58.54 | <0.0001 |
| BA47 | 0.64 | 2.26E-03 | -1.85E-05 | 1.66E-61 | 2.11E-36 | 0.10 | 0.66 | 61.08 | <0.0001 |
| **Parietal Lobe** | | | | | | | | | |
| BA1 | 0.66 | 3.40E-03 | -3.09E-05 | 1.79E-75 | 1.34E-53 | 0.44 | 0.63 | 55.04 | <0.0001 |
| BA2 | 0.63 | 3.60E-03 | -3.35E-05 | 6.54E-11 | 5.23E-87 | 0.10 | 0.74 | 53.72 | <0.0001 |
| BA3 | 0.67 | 2.82E-03 | -2.56E-05 | 7.01E-62 | 2.88E-43 | 0.51 | 0.56 | 55.23 | <0.0001 |
| BA5 | 0.65 | 2.48E-03 | -2.31E-05 | 6.43E-64 | 1.99E-46 | 0.01 | 0.56 | 53.93 | <0.0001 |
| BA7 | 0.64 | 3.04E-03 | -2.82E-05 | 8.26E-99 | 2.90E-74 | 0.15 | 0.70 | 53.85 | <0.0001 |
| BA23 | 0.63 | 2.01E-03 | -1.73E-05 | 5.06E-66 | 1.41E-42 | 0.46 | 0.63 | 58.24 | <0.0001 |
| BA29 | 0.64 | 1.41E-03 | -1.13E-05 | 5.93E-22 | 3.69E-12 | 0.87 | 0.36 | 62.49 | <0.0001 |
| BA30 | 0.63 | 1.96E-03 | -1.68E-05 | 1.30E-43 | 3.60E-27 | 0.32 | 0.51 | 58.5 | <0.0001 |
| BA31 | 0.63 | 2.11E-03 | -1.85E-05 | 1.09E-74 | 2.42E-50 | 0.30 | 0.66 | 56.98 | <0.0001 |
| BA39 | 0.64 | 2.23E-03 | -1.95E-05 | 1.14E-75 | 1.78E-50 | 0.27 | 0.67 | 57.43 | <0.0001 |
| BA40 | 0.64 | 2.22E-03 | -1.92E-05 | 3.69E-76 | 4.26E-50 | 0.06 | 0.68 | 57.82 | <0.0001 |
| **Temporal Lobe** | | | | | | | | | |
| BA20 | 0.62 | 1.48E-03 | -1.20E-05 | 2.68E-49 | 3.52E-28 | 0.16 | 0.60 | 61.9 | <0.0001 |
| BA21 | 0.62 | 1.73E-03 | -1.44E-05 | 2.03E-57 | 1.48E-34 | 0.46 | 0.62 | 60.25 | <0.0001 |
| BA22 | 0.63 | 2.30E-03 | -1.90E-05 | 5.55E-82 | 6.96E-51 | 0.54 | 0.73 | 60.35 | <0.0001 |
| BA37 | 0.64 | 1.75E-03 | -1.49E-05 | 1.68E-57 | 5.94E-36 | 0.65 | 0.6 | 58.97 | <0.0001 |
| BA38 | 0.62 | 1.17E-03 | -8.73E-06 | 9.89E-32 | 2.06E-15 | 0.50 | 0.53 | 67.31 | <0.0001 |
| BA41 | 0.65 | 1.76E-03 | -1.43E-05 | 5.57E-41 | 2.50E-23 | 0.79 | 0.53 | 61.76 | <0.0001 |
| BA42 | 0.66 | 2.36E-03 | -2.02E-05 | 7.81E-46 | 1.80E-28 | 0.26 | 0.53 | 58.51 | <0.0001 |
| **Occipital Lobe** | | | | | | | | | |
| BA17 | 0.69 | 1.57E-03 | -1.44E-05 | 4.96E-25 | 3.50E-17 | 0.12 | 0.29 | 54.95 | <0.0001 |
| BA18 | 0.68 | 2.06E-03 | -1.91E-05 | 1.13E-49 | 2.11E-35 | 0.64 | 0.47 | 54.17 | <0.0001 |
| BA19 | 0.66 | 2.19E-03 | -2.00E-05 | 7.35E-69 | 4.97E-49 | 0.94 | 0.59 | 54.76 | <0.0001 |
| **Mixed** | | | | | | | | | |
| BA26 | 0.64 | 1.78E-03 | -1.54E-05 | 1.45E-15 | 8.53E-10 | 0.87 | 0.23 | 58.25 | <0.0001 |
| BA27 | 0.65 | 1.51E-03 | -1.18E-05 | 9.36E-27 | 5.09E-14 | 0.03 | 0.45 | 64.44 | <0.0001 |
| BA28 | 0.62 | 2.61E-03 | -2.45E-05 | 4.47E-48 | 4.19E-35 | 0.00 | 0.46 | 53.31 | <0.0001 |
| BA35 | 0.64 | 1.07E-03 | -7.72E-06 | 3.35E-16 | 1.60E-07 | 0.59 | 0.36 | 70.87 | <0.0001 |
| BA36 | 0.64 | 1.69E-03 | -1.58E-05 | 3.10E-45 | 9.35E-33 | 0.00 | 0.44 | 53.46 | <0.0001 |
| BA43 | 0.64 | 2.56E-03 | -2.22E-05 | 1.45E-70 | 3.81E-46 | 0.95 | 0.65 | 57.8 | <0.0001 |
| **R1-SWM** | | | | | | | | | |
| **Frontal Lobe** | | | | | | | | | |
| BA4 | 1.08 | 1.10E-03 | -2.27E-05 | 8.50E-04 | 2.23E-09 | 0.67 | 0.28 | 23.7 | <0.0001 |
| BA6 | 1.06 | 2.00E-03 | -3.22E-05 | 1.68E-08 | 4.66E-15 | 0.51 | 0.26 | 30.94 | <0.0001 |
| BA8 | 1.04 | 2.17E-03 | -3.26E-05 | 1.33E-07 | 5.70E-12 | 0.53 | 0.17 | 33.06 | <0.0001 |
| BA9 | 1.06 | 1.49E-03 | -2.73E-05 | 7.22E-05 | 3.59E-10 | 0.14 | 0.25 | 26.87 | <0.0001 |
| BA10 | 1.04 | 2.38E-03 | -3.54E-05 | 4.34E-10 | 1.22E-15 | 0.17 | 0.22 | 33.56 | <0.0001 |
| BA11 | 0.99 | 3.75E-03 | -4.69E-05 | 1.82E-20 | 6.44E-24 | 0.03 | 0.2 | 39.87 | <0.0001 |
| BA24 | 1.03 | 3.04E-03 | -4.07E-05 | 1.78E-14 | 4.66E-19 | 0.00 | 0.19 | 37.23 | <0.0001 |
| BA32 | 1.05 | 3.02E-03 | -4.19E-05 | 1.66E-15 | 1.18E-21 | 0.45 | 0.23 | 35.99 | <0.0001 |
| BA33 | 0.9 | 6.41E-03 | -6.98E-05 | 2.23E-39 | 1.71E-36 | 0.15 | 0.29 | 45.92 | <0.0001 |
| BA44 | 1.08 | 2.05E-03 | -3.27E-05 | 1.03E-08 | 2.75E-15 | 0.86 | 0.26 | 31.14 | <0.0001 |
| BA45 | 1.02 | 2.44E-03 | -3.35E-05 | 8.21E-11 | 9.61E-15 | 0.49 | 0.16 | 36.45 | <0.0001 |
| BA46 | 1.07 | 1.47E-03 | -2.55E-05 | 7.14E-05 | 2.00E-09 | 0.87 | 0.2 | 28.49 | <0.0001 |
| BA47 | 0.98 | 4.45E-03 | -5.29E-05 | 5.16E-30 | 4.32E-32 | 0.34 | 0.24 | 42.05 | <0.0001 |
| **Parietal Lobe** | | | | | | | | | |
| BA1 | 1.02 | 1.53E-03 | -2.02E-05 | 1.90E-05 | 7.54E-07 | 0.77 | 0.06 | 37.3 | <0.0001 |
| BA2 | 1.06 | 2.03E-03 | -3.06E-05 | 7.80E-08 | 2.39E-12 | 0.55 | 0.18 | 33.09 | <0.001 |
| BA3 | 1.03 | 1.20E-03 | -1.58E-05 | 1.79E-04 | 1.57E-05 | 0.86 | 0.04 | 37.28 | <0.0001 |
| BA5 | 0.99 | 3.15E-03 | -4.18E-05 | 1.75E-15 | 4.67E-20 | 0.20 | 0.19 | 37.66 | <0.001 |
| BA7 | 1.04 | 1.59E-03 | -2.51E-05 | 4.01E-05 | 1.78E-08 | 0.46 | 0.14 | 31.2 | <0.0001 |
| BA23 | 1.03 | 2.95E-03 | -4.13E-05 | 6.06E-13 | 2.69E-18 | 0.56 | 0.21 | 35.6 | <0.0001 |
| BA29 | 0.89 | 5.69E-03 | -6.24E-05 | 7.59E-26 | 4.12E-24 | 0.81 | 0.19 | 45.54 | <0.0001 |
| BA30 | 1.03 | 2.90E-03 | -4.05E-05 | 7.45E-11 | 3.37E-15 | 0.77 | 0.17 | 35.65 | <0.0001 |
| BA31 | 1.04 | 2.86E-03 | -4.06E-05 | 2.16E-13 | 2.64E-19 | 0.64 | 0.23 | 35.27 | <0.0001 |
| BA39 | 1.06 | 2.02E-03 | -3.11E-05 | 9.97E-08 | 1.07E-12 | 0.03 | 0.2 | 32.28 | <0.0001 |
| BA40 | 1.01 | 2.77E-03 | -3.66E-05 | 2.39E-13 | 3.55E-17 | 0.01 | 0.17 | 37.67 | <0.0001 |
| **Temporal Lobe** | | | | | | | | | |
| BA20 | 0.91 | 5.77E-03 | -6.33E-05 | 7.96E-40 | 2.78E-37 | 0.38 | 0.28 | 45.7 | <0.0001 |
| BA21 | 0.97 | 4.95E-03 | -5.73E-05 | 9.99E-32 | 1.69E-32 | 0.05 | 0.24 | 43.2 | <0.0001 |
| BA22 | 1.04 | 3.00E-03 | -3.69E-05 | 5.76E-14 | 6.71E-16 | 0.91 | 0.12 | 40.58 | <0.0001 |
| BA37 | 1.01 | 3.92E-03 | -4.87E-05 | 4.62E-23 | 9.56E-27 | 0.05 | 0.21 | 40.18 | <0.0001 |
| BA38 | 0.88 | 4.90E-03 | -5.07E-05 | 3.65E-29 | 8.48E-25 | 0.11 | 0.24 | 48.35 | <0.0001 |
| BA41 | 1.07 | 2.28E-03 | -3.46E-05 | 1.59E-09 | 2.71E-15 | 0.10 | 0.23 | 32.72 | <0.0001 |
| BA42 | 1.1 | 8.69E-04 | -2.08E-05 | 1.81E-02 | 8.53E-07 | 0.32 | 0.27 | 19.86 | <0.0001 |
| **Occipital Lobe** | | | | | | | | | |
| BA17 | 1 | 9.13E-04 | -1.38E-05 | 8.76E-03 | 5.25E-04 | 0.54 | 0.05 | 31.43 | <0.001 |
| BA18 | 1.02 | 1.76E-03 | -2.39E-05 | 2.11E-08 | 3.06E-11 | 0.37 | 0.11 | 36.56 | <0.0001 |
| BA19 | 1.06 | 2.23E-03 | -3.29E-05 | 3.60E-11 | 3.64E-17 | 0.39 | 0.23 | 33.89 | <0.0001 |
| **Mixed** | | | | | | | | | |
| BA26 | 0.93 | 4.37E-03 | -4.69E-05 | 5.45E-15 | 1.52E-13 | 0.38 | 0.11 | 46.65 | <0.0001 |
| BA27 | 0.87 | 3.76E-03 | -3.21E-05 | 1.10E-12 | 7.00E-08 | 0.23 | 0.19 | 58.89 | <0.0001 |
| BA28 | 0.93 | 4.39E-03 | -4.46E-05 | 1.03E-30 | 2.19E-25 | 0.19 | 0.25 | 49.2 | <0.0001 |
| BA35 | 0.93 | 4.95E-03 | -5.29E-05 | 5.49E-40 | 6.09E-36 | 0.10 | 0.30 | 46.79 | <0.0001 |
| BA36 | 0.86 | 4.38E-03 | -4.30E-05 | 2.25E-32 | 3.84E-25 | 0.12 | 0.29 | 51.07 | <0.0001 |
| BA43 | 1.04 | 2.05E-03 | -3.05E-05 | 1.40E-06 | 4.15E-10 | 0.48 | 0.14 | 33.33 | <0.0001 |
| **R1-TWMB** | | | | | | | | | |
| ARCL | 1.13 | 2.29E-03 | -3.33E-05 | 8.62E-09 | 5.19E-13 | 0.14 | 0.17 | 34.17 | <0.0001 |
| ARCR | 1.13 | 2.81E-03 | -3.92E-05 | 8.60E-13 | 8.97E-18 | 0.38 | 0.2 | 35.7 | <0.0001 |
| ATRL | 1.13 | 3.21E-03 | -4.68E-05 | 5.74E-14 | 5.56E-21 | 0.34 | 0.27 | 34.23 | <0.0001 |
| ATRR | 1.13 | 3.24E-03 | -4.59E-05 | 3.79E-14 | 2.84E-20 | 0.28 | 0.24 | 35.22 | <0.0001 |
| CGCL | 1.1 | 3.03E-03 | -4.18E-05 | 1.38E-13 | 1.23E-18 | 0.05 | 0.2 | 36.13 | <0.0001 |
| CGCR | 1.08 | 4.05E-03 | -5.25E-05 | 2.30E-23 | 1.17E-28 | 0.01 | 0.26 | 38.58 | <0.0001 |
| CSTL | 1.11 | 2.26E-03 | -3.42E-05 | 1.11E-10 | 4.62E-17 | 0.40 | 0.25 | 32.93 | <0.0001 |
| CSTR | 1.15 | 1.11E-03 | -2.29E-05 | 2.48E-03 | 6.61E-08 | 0.47 | 0.25 | 23.05 | <0.0001 |
| FA | 1.19 | 2.04E-03 | -3.57E-05 | 2.72E-06 | 1.75E-12 | 0.58 | 0.28 | 28.36 | <0.0001 |
| FP | 1.2 | -1.78E-04 | -1.53E-05 | 6.56E-01 | 9.54E-04 | 0.10 | 0.42 | * | <0.001 |
| IFOL | 1.16 | 1.55E-03 | -3.27E-05 | 2.32E-04 | 2.48E-11 | 0.93 | 0.35 | 23.2 | <0.0001 |
| IFOR | 1.14 | 2.41E-03 | -3.97E-05 | 4.16E-09 | 1.25E-16 | 0.61 | 0.3 | 30.19 | <0.0001 |
| ILFL | 1.13 | 2.49E-03 | -3.94E-05 | 2.02E-10 | 6.24E-18 | 0.55 | 0.29 | 31.31 | <0.0001 |
| ILFR | 1.11 | 3.74E-03 | -5.12E-05 | 1.15E-20 | 9.04E-28 | 0.50 | 0.28 | 36.52 | <0.0001 |
| pARCL | 1.09 | 3.49E-03 | -4.41E-05 | 5.17E-18 | 2.55E-21 | 0.03 | 0.18 | 39.48 | <0.0001 |
| pARCR | 1.09 | 4.20E-03 | -5.26E-05 | 7.42E-26 | 6.85E-30 | 0.16 | 0.24 | 39.98 | <0.0001 |
| SLFL | 1.12 | 2.70E-03 | -3.77E-05 | 2.12E-11 | 7.92E-16 | 0.17 | 0.18 | 35.7 | <0.0001 |
| SLFR | 1.13 | 3.01E-03 | -4.22E-05 | 9.84E-14 | 3.15E-19 | 0.30 | 0.22 | 35.61 | <0.0001 |
| **R2\*-CGM** | | | | | | | | | |
| **Frontal Lobe** | | | | | | | | | |
| BA4 | 14.60 | 0.17 | -1.42E-03 | 2.75E-24 | 3.64E-17 | 0.64 | 0.42 | 60.79 | <0.0001 |
| BA6 | 13.30 | 0.15 | -1.26E-03 | 2.60E-27 | 3.13E-19 | 0.17 | 0.46 | 60.88 | <0.0001 |
| BA8 | 12.90 | 0.14 | -1.21E-03 | 4.68E-26 | 4.07E-19 | 0.15 | 0.42 | 59.31 | <0.0001 |
| BA9 | 13.00 | 0.14 | -1.24E-03 | 4.67E-27 | 1.14E-21 | 0.12 | 0.36 | 56.23 | <0.0001 |
| BA10 | 13.90 | 0.13 | -1.12E-03 | 1.24E-17 | 2.92E-14 | 0.00 | 0.24 | 56.07 | <0.0001 |
| BA11 | 14.50 | 0.12 | -9.77E-04 | 8.23E-09 | 1.61E-06 | 0.08 | 0.16 | 59.83 | <0.0001 |
| BA24 | 13.00 | 0.11 | -8.81E-04 | 3.08E-09 | 1.54E-06 | 0.04 | 0.20 | 61.86 | <0.0001 |
| BA25 | 13.80 | 0.10 | -8.88E-04 | 2.96E-03 | 7.20E-03 | 0.20 | 0.03 | 54.93 | <0.01 |
| BA32 | 12.80 | 0.14 | -1.22E-03 | 7.89E-19 | 6.88E-15 | 0.03 | 0.28 | 56.71 | <0.0001 |
| BA33 | 13.00 | 0.07 | -5.16E-04 | 7.25E-04 | 9.50E-03 | 0.00 | 0.10 | * | <0.01 |
| BA44 | 14.00 | 0.14 | -1.11E-03 | 2.16E-20 | 3.32E-14 | 0.16 | 0.38 | 61.20 | <0.0001 |
| BA45 | 14.00 | 0.14 | -1.19E-03 | 6.61E-17 | 8.82E-13 | 0.65 | 0.27 | 58.34 | <0.0001 |
| BA46 | 14.10 | 0.13 | -1.09E-03 | 3.86E-17 | 8.08E-13 | 0.16 | 0.28 | 58.85 | <0.0001 |
| BA47 | 14.70 | 0.11 | -8.41E-04 | 2.05E-07 | 4.09E-05 | 0.45 | 0.16 | 63.44 | <0.0001 |
| **Parietal Lobe** | | | | | | | | | |
| BA1 | 15.10 | 0.13 | -1.19E-03 | 4.53E-17 | 4.70E-14 | 0.22 | 0.22 | 55.34 | <0.0001 |
| BA2 | 14.50 | 0.14 | -1.24E-03 | 2.72E-20 | 3.97E-16 | 0.68 | 0.28 | 56.52 | <0.0001 |
| BA3 | 15.80 | 0.09 | -8.57E-04 | 3.06E-09 | 8.00E-08 | 0.41 | 0.11 | 54.61 | <0.0001 |
| BA5 | 14.30 | 0.11 | -9.70E-04 | 1.10E-15 | 2.99E-12 | 0.02 | 0.24 | 57.23 | <0.0001 |
| BA7 | 14.80 | 0.13 | -1.11E-03 | 1.32E-19 | 2.57E-15 | 0.22 | 0.28 | 57.31 | <0.0001 |
| BA23 | 14.20 | 0.13 | -1.13E-03 | 4.65E-09 | 7.04E-07 | 0.98 | 0.15 | 58.80 | <0.0001 |
| BA29 | 15.40 | 0.08 | -5.65E-04 | 9.06E-03 | 6.06E-02 | 0.24 | 0.07 | - | 0.061 |
| BA30 | 14.30 | 0.13 | -1.09E-03 | 1.12E-09 | 1.31E-07 | 0.47 | 0.14 | 57.42 | <0.0001 |
| BA31 | 14.20 | 0.13 | -1.13E-03 | 1.61E-11 | 1.11E-08 | 0.22 | 0.19 | 58.77 | <0.0001 |
| BA39 | 14.40 | 0.14 | -1.19E-03 | 5.08E-18 | 1.88E-13 | 0.33 | 0.29 | 58.90 | <0.0001 |
| BA40 | 14.10 | 0.13 | -1.11E-03 | 2.76E-18 | 6.94E-14 | 0.15 | 0.29 | 58.24 | <0.0001 |
| **Temporal Lobe** | | | | | | | | | |
| BA20 | 13.20 | 0.12 | -8.87E-04 | 2.82E-09 | 1.03E-05 | 0.31 | 0.26 | 67.78 | <0.0001 |
| BA21 | 13.40 | 0.12 | -9.41E-04 | 6.44E-10 | 1.99E-06 | 0.04 | 0.25 | 64.95 | <0.0001 |
| BA22 | 13.80 | 0.14 | -1.17E-03 | 8.66E-22 | 2.38E-15 | 0.23 | 0.38 | 60.91 | <0.0001 |
| BA37 | 14.40 | 0.15 | -1.22E-03 | 5.82E-17 | 2.36E-12 | 0.20 | 0.30 | 59.63 | <0.0001 |
| BA38 | 12.70 | 0.09 | -6.95E-04 | 4.13E-06 | 5.62E-04 | 0.09 | 0.17 | 68.02 | <0.001 |
| BA41 | 14.80 | 0.12 | -1.06E-03 | 1.25E-12 | 1.56E-09 | 0.39 | 0.21 | 58.54 | <0.0001 |
| BA42 | 14.30 | 0.16 | -1.35E-03 | 7.85E-18 | 2.40E-13 | 0.04 | 0.30 | 58.65 | <0.0001 |
| **Occipital Lobe** | | | | | | | | | |
| BA17 | 17.00 | 0.11 | -9.99E-04 | 5.51E-09 | 2.03E-07 | 0.63 | 0.11 | 55.82 | <0.0001 |
| BA18 | 16.20 | 0.14 | -1.25E-03 | 3.47E-17 | 2.14E-13 | 0.21 | 0.26 | 57.28 | <0.0001 |
| BA19 | 15.60 | 0.13 | -1.13E-03 | 7.80E-18 | 2.14E-13 | 0.65 | 0.29 | 58.62 | <0.0001 |
| Mixed | | | | | | | | | |
| BA26 | 16.80 | 0.05 | -2.29E-04 | 1.17E-01 | 5.08E-01 | 0.20 | 0.09 | - | 0.508 |
| BA27 | 14.20 | 0.14 | -1.17E-03 | 1.12E-07 | 5.89E-06 | 0.84 | 0.12 | 58.40 | <0.0001 |
| BA28 | 13.50 | 0.12 | -9.79E-04 | 2.65E-06 | 7.50E-05 | 0.04 | 0.11 | 59.47 | <0.0001 |
| BA35 | 14.70 | 0.12 | -9.40E-04 | 9.81E-06 | 6.05E-04 | 0.83 | 0.13 | 65.34 | <0.001 |
| BA36 | 13.40 | 0.13 | -1.10E-03 | 2.15E-08 | 2.89E-06 | 0.78 | 0.14 | 59.72 | <0.0001 |
| BA43 | 13.90 | 0.14 | -1.17E-03 | 7.31E-20 | 1.08E-14 | 0.18 | 0.33 | 59.03 | <0.0001 |
| **R2*-SWM** | | | | | | | | | |
| **Frontal Lobe** | | | | | | | | | |
| BA4 | 16.10 | 0.17 | -1.53E-03 | 1.97E-25 | 1.82E-21 | 0.01 | 0.30 | 54.48 | <0.0001 |
| BA6 | 16.10 | 0.15 | -1.42E-03 | 2.43E-25 | 1.13E-22 | 0.08 | 0.27 | 52.67 | <0.0001 |
| BA8 | 15.80 | 0.16 | -1.53E-03 | 1.28E-22 | 1.95E-21 | 0.29 | 0.22 | 50.80 | <0.0001 |
| BA9 | 16.10 | 0.15 | -1.57E-03 | 5.99E-23 | 3.38E-23 | 0.18 | 0.22 | 48.80 | <0.0001 |
| BA10 | 17.20 | 0.14 | -1.43E-03 | 1.47E-16 | 1.16E-17 | 0.01 | 0.18 | 47.23 | <0.0001 |
| BA11 | 17.10 | 0.14 | -1.37E-03 | 1.94E-12 | 2.50E-12 | 0.09 | 0.12 | 49.53 | <0.0001 |
| BA24 | 16.80 | 0.12 | -1.12E-03 | 2.40E-10 | 1.45E-09 | 0.00 | 0.14 | 51.75 | <0.0001 |
| BA25 | 16.70 | 0.11 | -1.23E-03 | 1.83E-04 | 2.80E-05 | 0.04 | 0.06 | 43.48 | <0.0001 |
| BA32 | 16.40 | 0.16 | -1.66E-03 | 1.45E-20 | 2.01E-21 | 0.03 | 0.21 | 47.92 | <0.0001 |
| BA33 | 16.00 | 0.12 | -1.27E-03 | 5.97E-09 | 1.94E-09 | 0.00 | 0.11 | 47.45 | <0.0001 |
| BA44 | 17.70 | 0.13 | -1.23E-03 | 4.19E-14 | 3.62E-13 | 0.08 | 0.14 | 51.24 | <0.0001 |
| BA45 | 17.40 | 0.14 | -1.37E-03 | 3.32E-12 | 8.19E-12 | 0.28 | 0.12 | 50.15 | <0.0001 |
| BA46 | 18.00 | 0.12 | -1.19E-03 | 6.60E-12 | 4.72E-12 | 0.15 | 0.12 | 48.90 | <0.0001 |
| BA47 | 17.50 | 0.13 | -1.36E-03 | 6.64E-09 | 7.56E-09 | 0.33 | 0.08 | 49.39 | <0.0001 |
| **Parietal Lobe** | | | | | | | | | |
| BA1 | 16.60 | 0.15 | -1.36E-03 | 1.40E-20 | 6.26E-17 | 0.70 | 0.26 | 55.40 | <0.0001 |
| BA2 | 16.90 | 0.14 | -1.28E-03 | 2.46E-18 | 6.10E-16 | 0.08 | 0.21 | 53.51 | <0.0001 |
| BA3 | 17.10 | 0.13 | -1.16E-03 | 7.09E-19 | 2.46E-15 | 0.07 | 0.25 | 55.79 | <0.0001 |
| BA5 | 16.40 | 0.13 | -1.22E-03 | 1.79E-15 | 4.00E-14 | 0.00 | 0.17 | 52.09 | <0.0001 |
| BA7 | 16.80 | 0.14 | -1.35E-03 | 1.03E-18 | 1.40E-16 | 0.00 | 0.21 | 52.97 | <0.0001 |
| BA23 | 17.00 | 0.15 | -1.47E-03 | 1.61E-10 | 2.10E-10 | 0.59 | 0.09 | 49.46 | <0.0001 |
| BA29 | 16.50 | 0.10 | -9.72E-04 | 3.95E-05 | 1.52E-04 | 0.12 | 0.05 | 53.75 | <0.001 |

| Region | | | | | | | | | |
|---|---|---|---|---|---|---|---|---|---|
| BA30 | 16.80 | 0.13 | -1.29E-03 | 2.51E-10 | 3.09E-10 | 0.11 | 0.10 | 49.49 | <0.0001 |
| BA31 | 17.10 | 0.14 | -1.35E-03 | 1.05E-12 | 1.18E-11 | 0.01 | 0.14 | 51.94 | <0.0001 |
| BA39 | 17.10 | 0.15 | -1.43E-03 | 5.58E-19 | 3.13E-17 | 0.00 | 0.21 | 52.28 | <0.0001 |
| BA40 | 16.80 | 0.14 | -1.37E-03 | 1.72E-18 | 6.20E-17 | 0.00 | 0.21 | 51.90 | <0.0001 |
| **Temporal Lobe** | | | | | | | | | |
| BA20 | 15.80 | 0.15 | -1.38E-03 | 2.86E-14 | 4.67E-12 | 0.59 | 0.17 | 54.55 | <0.0001 |
| BA21 | 16.50 | 0.16 | -1.59E-03 | 4.67E-15 | 5.40E-14 | 0.06 | 0.15 | 51.45 | <0.0001 |
| BA22 | 17.10 | 0.16 | -1.56E-03 | 4.22E-22 | 5.63E-20 | 0.13 | 0.23 | 52.34 | <0.0001 |
| BA37 | 17.00 | 0.16 | -1.59E-03 | 2.85E-19 | 7.53E-18 | 0.05 | 0.20 | 51.47 | <0.0001 |
| BA38 | 15.10 | 0.15 | -1.43E-03 | 1.35E-12 | 6.75E-11 | 0.27 | 0.14 | 53.79 | <0.0001 |
| BA41 | 17.80 | 0.12 | -1.16E-03 | 1.80E-10 | 2.85E-10 | 0.75 | 0.10 | 49.70 | <0.0001 |
| BA42 | 16.90 | 0.15 | -1.39E-03 | 2.44E-17 | 1.65E-14 | 0.02 | 0.23 | 54.80 | <0.0001 |
| **Occipital Lobe** | | | | | | | | | |
| BA17 | 17.60 | 0.12 | -1.03E-03 | 2.20E-10 | 2.42E-08 | 0.24 | 0.15 | 56.67 | <0.0001 |
| BA18 | 17.30 | 0.15 | -1.33E-03 | 2.35E-19 | 1.51E-15 | 0.01 | 0.27 | 56.03 | <0.0001 |
| BA19 | 17.40 | 0.14 | -1.31E-03 | 2.43E-17 | 5.44E-15 | 0.01 | 0.21 | 53.66 | <0.0001 |
| **Mixed** | | | | | | | | | |
| BA26 | 16.80 | 0.10 | -9.17E-04 | 2.77E-04 | 1.03E-03 | 0.12 | 0.04 | 56.03 | <0.01 |
| BA27 | 15.50 | 0.14 | -1.21E-03 | 5.60E-08 | 5.24E-06 | 0.34 | 0.14 | 59.62 | <0.0001 |
| BA28 | 15.20 | 0.16 | -1.47E-03 | 9.00E-11 | 5.90E-09 | 0.26 | 0.13 | 55.17 | <0.0001 |
| BA35 | 16.40 | 0.13 | -1.16E-03 | 1.38E-07 | 3.92E-06 | 0.49 | 0.10 | 56.49 | <0.0001 |
| BA36 | 15.10 | 0.14 | -1.25E-03 | 1.42E-09 | 1.26E-07 | 0.72 | 0.13 | 57.06 | <0.0001 |
| BA43 | 16.70 | 0.15 | -1.47E-03 | 2.58E-17 | 3.07E-16 | 0.13 | 0.17 | 51.17 | <0.0001 |
| **R2\*-TWMB** | | | | | | | | | |
| CGCL | 18.00 | 0.12 | -1.21E-03 | 1.78E-08 | 7.24E-09 | 0.00 | 0.10 | 47.71 | <0.0001 |
| FP | 21.40 | 0.04 | -6.55E-04 | 1.15E-01 | 1.44E-02 | 0.74 | 0.08 | 23.36 | <0.05 |
| SLFR | 19.20 | 0.09 | -9.10E-04 | 3.24E-06 | 1.20E-06 | 0.00 | 0.09 | 46.80 | <0.0001 |
| CSTL | 17.10 | 0.12 | -1.27E-03 | 9.13E-10 | 2.54E-10 | 0.03 | 0.11 | 47.18 | <0.0001 |
| IFOR | 19.90 | 0.08 | -1.04E-03 | 6.27E-05 | 1.05E-06 | 0.02 | 0.13 | 39.59 | <0.0001 |
| ARCR | 19.20 | 0.07 | -7.46E-04 | 4.99E-04 | 1.78E-04 | 0.01 | 0.05 | 45.05 | <0.001 |
| ILFR | 19.40 | 0.09 | -1.02E-03 | 4.50E-05 | 3.01E-06 | 0.03 | 0.09 | 42.29 | <0.0001 |
| ATRR | 19.30 | 0.10 | -1.26E-03 | 2.56E-06 | 1.37E-08 | 0.00 | 0.17 | 40.05 | <0.0001 |
| ATRL | 19.00 | 0.13 | -1.51E-03 | 1.32E-09 | 2.00E-12 | 0.08 | 0.20 | 41.98 | <0.0001 |
| SLFL | 18.40 | 0.12 | -1.23E-03 | 5.40E-10 | 1.75E-10 | 0.03 | 0.11 | 47.66 | <0.0001 |
| pARCL | 17.50 | 0.13 | -1.29E-03 | 3.74E-11 | 7.39E-11 | 0.02 | 0.12 | 49.87 | <0.0001 |
| CGCR | 18.20 | 0.10 | -1.08E-03 | 1.26E-06 | 4.00E-07 | 0.00 | 0.08 | 46.70 | <0.0001 |
| ARCL | 18.80 | 0.11 | -1.17E-03 | 5.81E-08 | 9.08E-09 | 0.02 | 0.10 | 46.15 | <0.0001 |
| FA | 20.30 | 0.07 | -8.92E-04 | 2.03E-03 | 6.83E-05 | 0.10 | 0.11 | 36.25 | <0.0001 |
| ILFL | 19.10 | 0.09 | -1.07E-03 | 9.43E-06 | 8.23E-07 | 0.15 | 0.08 | 43.79 | <0.0001 |
| CSTR | 18.00 | 0.07 | -7.62E-04 | 1.18E-04 | 9.85E-05 | 0.00 | 0.07 | 48.56 | <0.0001 |
| pARCR | 17.70 | 0.12 | -1.21E-03 | 6.61E-09 | 5.06E-09 | 0.00 | 0.10 | 48.62 | <0.0001 |
| IFOL | 19.50 | 0.10 | -1.18E-03 | 8.82E-07 | 1.26E-08 | 0.03 | 0.14 | 41.81 | <0.0001 |
| **Susceptibility-CGM** | | | | | | | | | |
| **Frontal Lobe** | | | | | | | | | |
| BA4 | -3.39E-04 | 3.56E-04 | -3.21E-06 | 1.54E-17 | 2.81E-14 | 0.01 | 0.24 | 55.73 | <0.0001 |
| BA6 | -2.49E-03 | 2.53E-04 | -2.32E-06 | 3.31E-09 | 9.44E-08 | 0.85 | 0.11 | 54.88 | <0.0001 |
| BA8 | -4.02E-03 | 1.81E-04 | -1.53E-06 | 1.20E-06 | 5.27E-05 | 0.13 | 0.12 | 59.94 | <0.0001 |
| BA9 | -3.01E-03 | 1.24E-04 | -1.13E-06 | 2.27E-04 | 9.24E-04 | 0.34 | 0.04 | 55.44 | <0.001 |
| BA10 | -1.31E-03 | 7.36E-05 | -7.20E-07 | 1.55E-02 | 1.99E-02 | 0.09 | 0.02 | 51.54 | <0.05 |
| BA11 | -1.27E-03 | 5.38E-06 | -1.15E-07 | 7.74E-01 | 5.45E-01 | 0.93 | 0.00 | - | 0.545 |
| BA24 | -2.38E-03 | 1.74E-04 | -1.91E-06 | 9.22E-06 | 1.87E-06 | 0.26 | 0.06 | 45.4 | <0.0001 |
| BA25 | -2.91E-03 | 1.44E-05 | -2.21E-07 | 7.68E-01 | 6.57E-01 | 0.03 | 0.01 | - | 0.657 |
| BA32 | -3.08E-03 | 2.38E-04 | -2.20E-06 | 3.99E-09 | 9.10E-08 | 0.48 | 0.11 | 54.4 | <0.0001 |
| BA33 | 1.36E-03 | 3.17E-05 | -6.32E-07 | 5.46E-01 | 2.36E-01 | 0.37 | 0.03 | - | 0.236 |
| BA44 | -7.35E-04 | 2.19E-04 | -2.00E-06 | 9.48E-06 | 6.77E-05 | 0.67 | 0.06 | 55.67 | <0.0001 |
| BA45 | -3.60E-04 | 1.60E-04 | -1.51E-06 | 1.87E-05 | 6.72E-05 | 0.06 | 0.05 | 53.41 | <0.0001 |
| BA46 | -1.60E-03 | 1.24E-04 | -1.25E-06 | 5.87E-04 | 6.55E-04 | 0.09 | 0.03 | 49.75 | <0.001 |
| BA47 | -1.59E-03 | 8.15E-05 | -6.62E-07 | 2.79E-02 | 7.83E-02 | 0.96 | 0.02 | - | 0.078 |
| **Parietal Lobe** | | | | | | | | | |
| BA1 | -1.52E-03 | 1.56E-04 | -1.38E-06 | 2.74E-09 | 1.83E-07 | 0.08 | 0.13 | 56.97 | <0.0001 |
| BA2 | -1.32E-03 | 2.75E-04 | -2.71E-06 | 4.57E-13 | 2.20E-12 | 0.03 | 0.13 | 50.84 | <0.0001 |
| BA3 | 4.25E-03 | 6.31E-05 | -8.64E-07 | 1.54E-01 | 5.49E-02 | 0.07 | 0.03 | - | 0.055 |
| BA5 | -4.88E-03 | 2.90E-04 | -2.63E-06 | 1.50E-11 | 1.09E-09 | 0.99 | 0.14 | 55.2 | <0.0001 |
| BA7 | -2.37E-03 | 2.15E-04 | -2.12E-06 | 9.07E-12 | 3.60E-11 | 0.73 | 0.11 | 50.77 | <0.0001 |
| BA23 | -2.06E-03 | 3.87E-04 | -3.47E-06 | 4.13E-13 | 1.14E-10 | 0.46 | 0.17 | 55.83 | <0.0001 |
| BA29 | -1.79E-03 | 1.53E-04 | -1.82E-06 | 9.79E-02 | 5.32E-02 | 0.24 | 0.01 | - | 0.053 |
| BA30 | 1.07E-03 | 1.63E-04 | -1.73E-06 | 1.98E-03 | 1.28E-03 | 0.01 | 0.04 | 47.01 | <0.01 |
| BA31 | -3.67E-04 | 3.00E-04 | -2.88E-06 | 3.75E-10 | 3.15E-09 | 0.29 | 0.10 | 52.18 | <0.0001 |
| BA39 | 1.56E-04 | 1.82E-04 | -1.86E-06 | 1.25E-07 | 1.04E-07 | 0.01 | 0.08 | 48.97 | <0.0001 |
| BA40 | -2.11E-03 | 2.15E-04 | -2.07E-06 | 2.33E-13 | 3.27E-12 | 0.08 | 0.14 | 51.87 | <0.0001 |
| **Temporal Lobe** | | | | | | | | | |
| BA20 | -2.85E-03 | 3.63E-05 | -3.48E-07 | 1.40E-01 | 1.64E-01 | 0.80 | 0.00 | - | 0.164 |
| BA21 | -1.58E-03 | 1.97E-05 | -2.06E-07 | 4.99E-01 | 4.88E-01 | 0.50 | -0.01 | - | 0.488 |
| BA22 | -2.15E-03 | 1.85E-04 | -1.91E-06 | 1.51E-08 | 9.54E-09 | 0.01 | 0.09 | 48.56 | <0.0001 |
| BA37 | -1.09E-03 | 9.87E-05 | -1.10E-06 | 7.04E-04 | 2.18E-04 | 0.00 | 0.06 | 44.86 | <0.001 |
| BA38 | -4.62E-03 | 4.55E-05 | -5.07E-07 | 1.63E-01 | 1.26E-01 | 0.00 | 0.02 | - | 0.126 |
| BA41 | -7.53E-04 | 2.70E-04 | -3.10E-06 | 2.24E-07 | 6.56E-09 | 0.21 | 0.11 | 43.45 | <0.0001 |
| BA42 | 2.54E-03 | 2.69E-04 | -2.64E-06 | 5.03E-09 | 1.68E-08 | 0.00 | 0.10 | 51.08 | <0.0001 |
| **Occipital Lobe** | | | | | | | | | |
| BA17 | 4.04E-03 | 9.93E-05 | -1.05E-06 | 2.24E-03 | 1.49E-03 | 0.77 | 0.02 | 47.05 | <0.01 |
| BA18 | 2.63E-03 | 1.44E-04 | -1.54E-06 | 8.76E-09 | 1.35E-09 | 0.00 | 0.11 | 46.57 | <0.0001 |
| BA19 | 1.81E-03 | 1.63E-04 | -1.78E-06 | 1.12E-07 | 1.50E-08 | 0.01 | 0.10 | 45.8 | <0.0001 |
| **Mixed** | | | | | | | | | |
| BA26 | -2.74E-03 | 8.39E-05 | -8.24E-07 | 4.86E-01 | 5.01E-01 | 0.25 | 0.00 | - | 0.501 |
| BA27 | 1.61E-03 | 3.69E-05 | -2.14E-07 | 4.91E-01 | 6.96E-01 | 0.1 | 0.01 | - | 0.696 |
| BA28 | 5.87E-04 | 6.42E-05 | -4.67E-07 | 2.21E-01 | 3.81E-01 | 0.27 | 0.01 | - | 0.381 |
| BA35 | 1.60E-04 | 4.82E-05 | -4.65E-07 | 4.01E-01 | 4.26E-01 | 0.00 | 0.02 | - | 0.426 |
| BA36 | -4.52E-04 | 3.72E-05 | -2.52E-07 | 3.08E-01 | 4.96E-01 | 0.29 | 0.01 | - | 0.496 |
| BA43 | -6.43E-04 | 1.88E-04 | -2.18E-06 | 2.41E-06 | 9.57E-08 | 0.45 | 0.10 | 43.03 | <0.0001 |
| **Susceptibility-SWM** | | | | | | | | | |
| **Frontal Lobe** | | | | | | | | | |
| BA4 | -9.81E-03 | 1.21E-04 | -6.88E-07 | 5.05E-03 | 1.17E-01 | 0.95 | 0.15 | - | 0.117 |
| BA6 | -6.66E-03 | 1.67E-04 | -1.50E-06 | 1.54E-06 | 1.87E-05 | 0.00 | 0.09 | 55.85 | <0.0001 |
| BA8 | -2.97E-03 | 1.33E-04 | -1.27E-06 | 1.00E-04 | 2.85E-04 | 0.00 | 0.07 | 53.26 | <0.001 |
| BA9 | -1.29E-03 | 1.71E-04 | -1.78E-06 | 4.09E-07 | 2.33E-07 | 0.00 | 0.1 | 48.07 | <0.0001 |
| BA10 | 4.06E-03 | 1.03E-04 | -1.42E-06 | 6.54E-03 | 2.32E-04 | 0.00 | 0.14 | 34.81 | <0.001 |
| BA11 | 1.57E-03 | -2.71E-05 | 7.82E-08 | 3.14E-01 | 7.74E-01 | 0.22 | 0.05 | - | 0.774 |
| BA24 | -1.65E-03 | 3.65E-04 | -3.76E-06 | 1.87E-12 | 1.05E-12 | 0.4 | 0.12 | 48.64 | <0.0001 |
| BA25 | -6.50E-04 | 7.66E-05 | -1.24E-06 | 9.65E-02 | 8.25E-03 | 0.79 | 0.1 | 24.05 | <0.01 |
| BA32 | -1.06E-03 | 2.81E-04 | -2.95E-06 | 5.34E-09 | 1.87E-09 | 0.94 | 0.08 | 47.72 | <0.0001 |
| BA33 | 4.50E-03 | 2.09E-04 | -2.52E-06 | 2.52E-04 | 1.57E-05 | 0.32 | 0.08 | 41.2 | <0.0001 |
| BA44 | 4.46E-04 | 8.73E-05 | -9.20E-07 | 3.80E-02 | 3.17E-02 | 0.05 | 0.01 | 48.64 | <0.05 |
| BA45 | 2.19E-03 | 9.04E-05 | -1.08E-06 | 3.59E-02 | 1.40E-02 | 0.27 | 0.02 | 38.75 | <0.05 |
| BA46 | 3.07E-03 | 8.80E-05 | -1.14E-06 | 2.05E-02 | 3.22E-03 | 0.00 | 0.09 | 37.28 | <0.01 |
| BA47 | 5.57E-03 | -4.85E-05 | 1.65E-07 | 3.34E-01 | 7.47E-01 | 0.74 | 0.04 | - | 0.747 |
| **Parietal Lobe** | | | | | | | | | |
| BA1 | -5.10E-03 | 5.91E-05 | -2.15E-07 | 9.64E-02 | 5.51E-01 | 0.92 | 0.11 | - | 0.551 |
| BA2 | -4.75E-03 | 1.37E-04 | -1.13E-06 | 2.11E-04 | 2.63E-03 | 0.15 | 0.08 | 62.4 | <0.01 |
| BA3 | -3.82E-03 | 2.95E-05 | -1.12E-07 | 4.17E-01 | 7.61E-01 | 0.04 | 0.03 | - | 0.761 |
| BA5 | -5.29E-03 | 1.93E-04 | -1.65E-06 | 5.87E-05 | 6.86E-04 | 0.59 | 0.07 | 60.23 | <0.001 |
| BA7 | -2.54E-03 | 1.06E-04 | -9.66E-07 | 6.46E-03 | 1.37E-02 | 0.68 | 0.02 | 55.64 | <0.05 |
| BA23 | -2.68E-03 | 4.04E-04 | -3.79E-06 | 3.46E-09 | 4.66E-08 | 0.35 | 0.1 | 53.41 | <0.0001 |
| BA29 | 1.24E-03 | 1.26E-04 | -1.26E-06 | 2.06E-01 | 2.11E-01 | 0.24 | 0 | - | 0.211 |
| BA30 | 4.65E-03 | -6.88E-05 | 5.97E-07 | 3.39E-01 | 4.13E-01 | 0.88 | 0 | - | 0.413 |
| BA31 | -6.86E-05 | 2.40E-04 | -2.35E-06 | 4.03E-05 | 7.18E-05 | 0.15 | 0.04 | 51.1 | <0.0001 |
| BA39 | 3.37E-03 | -8.88E-06 | 9.83E-08 | 8.34E-01 | 8.19E-01 | 0.03 | 0.01 | - | 0.819 |
| BA40 | 1.05E-03 | 4.39E-05 | -4.43E-07 | 2.64E-01 | 2.68E-01 | 0.04 | 0.01 | - | 0.268 |
| **Temporal Lobe** | | | | | | | | | |
| BA20 | -7.02E-04 | 8.89E-06 | -5.89E-09 | 7.72E-01 | 9.85E-01 | 0.99 | 0 | - | 0.985 |
| BA21 | 3.29E-03 | -4.51E-05 | 3.69E-07 | 1.76E-01 | 2.76E-01 | 0.02 | 0.02 | - | 0.276 |
| BA22 | 3.80E-03 | 3.77E-05 | -5.87E-07 | 3.72E-01 | 1.72E-01 | 0.00 | 0.06 | - | 0.172 |
| BA37 | 1.27E-04 | 6.38E-05 | -5.74E-07 | 7.37E-02 | 1.12E-01 | 0.00 | 0.04 | - | 0.112 |
| BA38 | 5.65E-04 | 5.49E-05 | -6.01E-07 | 1.45E-01 | 1.17E-01 | 0.01 | 0.02 | - | 0.117 |
| BA41 | -6.24E-04 | -2.94E-05 | 4.50E-07 | 4.96E-01 | 3.03E-01 | 0.44 | 0.01 | - | 0.303 |
| BA42 | -3.88E-03 | 1.70E-04 | -1.35E-06 | 6.59E-05 | 1.73E-03 | 0.00 | 0.13 | 64.87 | <0.01 |
| **Occipital Lobe** | | | | | | | | | |
| BA17 | -1.22E-03 | 1.47E-04 | -1.41E-06 | 5.37E-05 | 1.29E-04 | 0.03 | 0.05 | 52.2 | <0.001 |
| BA18 | -4.99E-04 | 1.11E-04 | -1.06E-06 | 2.94E-04 | 6.75E-04 | 0.02 | 0.04 | 52.93 | <0.001 |
| BA19 | -7.27E-04 | 6.70E-05 | -5.90E-07 | 6.24E-02 | 1.06E-01 | 0.94 | 0.01 | - | 0.106 |
| **Mixed** | | | | | | | | | |
| BA26 | 1.81E-03 | 7.91E-05 | -6.96E-07 | 5.07E-01 | 5.66E-01 | 0.72 | -0.01 | - | 0.566 |
| BA27 | 8.64E-04 | 2.97E-04 | -3.06E-06 | 2.81E-04 | 2.27E-04 | 0.10 | 0.03 | 48.31 | <0.001 |
| BA28 | 2.15E-03 | 1.62E-04 | -1.87E-06 | 1.11E-02 | 4.02E-03 | 0.17 | 0.03 | 42.47 | <0.01 |
| BA35 | -9.22E-04 | 3.03E-04 | -3.11E-06 | 3.12E-04 | 2.77E-04 | 0.17 | 0.03 | 48.8 | <0.001 |
| BA36 | -1.94E-03 | 2.44E-04 | -2.36E-06 | 7.62E-06 | 2.01E-05 | 0.01 | 0.06 | 51.99 | <0.0001 |
| BA43 | 1.80E-03 | 5.46E-05 | -7.66E-07 | 2.66E-01 | 1.25E-01 | 0.18 | 0.02 | - | 0.125 |
| **Susceptibility-TWMB** | | | | | | | | | |
| CGCL | -5.53E-03 | 5.00E-05 | -7.40E-07 | 3.52E-01 | 1.77E-01 | 0.42 | 0.02 | - | 0.177 |
| FP | -2.11E-02 | -8.66E-05 | 1.50E-06 | 1.62E-01 | 1.81E-02 | 0.71 | 0.1 | 29.89 | <0.05 |
| SLFR | -2.60E-03 | -3.33E-04 | 3.38E-06 | 6.57E-17 | 6.82E-17 | 0.6 | 0.17 | 49.31 | <0.0001 |
| CSTL | -1.20E-02 | -1.94E-04 | 1.84E-06 | 1.67E-08 | 1.46E-07 | 0.08 | 0.09 | 52.95 | <0.0001 |
| IFOR | -1.21E-02 | -1.92E-04 | 1.90E-06 | 5.19E-04 | 7.52E-04 | 0.02 | 0.04 | 49.22 | <0.001 |
| ARCR | -3.89E-03 | -3.21E-04 | 3.36E-06 | 9.13E-14 | 2.22E-14 | 0.27 | 0.14 | 47.76 | <0.0001 |
| ILFR | -1.04E-02 | -2.08E-04 | 2.44E-06 | 7.29E-04 | 1.11E-04 | 0.34 | 0.05 | 42.29 | <0.001 |
| ATRR | -7.78E-03 | -2.23E-04 | 2.48E-06 | 2.77E-07 | 2.38E-08 | 0.00 | 0.11 | 44.86 | <0.0001 |
| ATRL | -8.01E-03 | -1.88E-04 | 2.15E-06 | 2.47E-05 | 2.46E-06 | 0.07 | 0.08 | 43.56 | <0.0001 |
| SLFL | -3.78E-03 | -2.65E-04 | 2.67E-06 | 5.95E-13 | 1.11E-12 | 0.37 | 0.12 | 49.64 | <0.0001 |
| pARCL | 3.06E-04 | -3.62E-04 | 3.82E-06 | 1.84E-10 | 5.04E-11 | 0.25 | 0.11 | 47.35 | <0.0001 |
| CGCR | -4.09E-03 | -3.78E-05 | 4.31E-07 | 4.97E-01 | 4.47E-01 | 0.12 | 0.00 | - | 0.447 |
| ARCL | -3.47E-03 | -2.72E-04 | 2.83E-06 | 2.77E-12 | 1.17E-12 | 0.91 | 0.12 | 48.13 | <0.0001 |
| FA | -9.45E-03 | -2.21E-04 | 2.85E-06 | 1.66E-04 | 2.16E-06 | 0.00 | 0.17 | 38.19 | <0.0001 |
| ILFL | -1.41E-02 | -5.91E-05 | 1.01E-06 | 2.27E-01 | 4.44E-02 | 0.62 | 0.07 | 29.39 | <0.05 |
| CSTR | -1.29E-02 | -2.15E-04 | 2.17E-06 | 4.84E-08 | 6.30E-08 | 0.02 | 0.08 | 49.48 | <0.0001 |
| pARCR | -1.01E-03 | -3.15E-04 | 3.39E-06 | 1.45E-07 | 3.14E-08 | 0.16 | 0.08 | 46.34 | <0.0001 |
| IFOL | -1.46E-02 | -1.00E-04 | 9.68E-07 | 3.16E-02 | 4.16E-02 | 0.17 | 0.01 | 53.83 | <0.05 |

CGM: Cortical Grey Matter, SWM: Superficial White Matter, TWMB: Tractography White Matter Bundle, BA: Brodmann Area, FA: Callosum Forceps Minor, FP: Callosum Forceps Major, pARC: Posterior Arcuate Fasciculus, ILF: Inferior Longitudinal Fasciculus, SLF: Superior Longitudinal Fasciculus, IFO: Inferior Fronto-Occipital Fasciculus, ARC: Arcuate, ATR: Thalamic Radiation, CGC: Cingulum Cingulate, CST: Corticospinal, L: Left, R: Right, *: Age peak was outside the studied age range with continuous increases or decreases.

## S. 7 Correlation Analysis for Peak Ages in Quantitative Trajectories

| qMRI | Group Comparison | Structure | Spearman's ρ | p |
|---|---|---|---|---|
| R1 | | | | |
| | Raw - HBRbs | CGM | 0.70 | < 0.0001 |
| | | SWM | 0.87 | < 0.0001 |
| | | TWMB | 0.77 | < 0.001 |
| | Raw - EBS | CGM | 0.85 | < 0.0001 |
| | | SWM | 0.89 | < 0.0001 |
| | | TWMB | 0.52 | < 0.05 |
| | HBRbs - EBS | CGM | 0.89 | < 0.0001 |
| | | SWM | 0.85 | < 0.0001 |
| | | TWMB | 0.74 | < 0.01 |
| R2* | | | | |
| | Raw - HBRbs | CGM | 0.96 | < 0.0001 |
| | | SWM | 0.99 | < 0.0001 |
| | | TWMB | 0.97 | < 0.0001 |
| | Raw - EBS | CGM | 0.91 | < 0.0001 |
| | | SWM | 0.98 | < 0.0001 |
| | | TWMB | -0.28 | > 0.05 |
| | HBRbs - EBS | CGM | 0.90 | < 0.0001 |
| | | SWM | 0.98 | < 0.0001 |
| | | TWMB | -0.27 | > 0.05 |
| Susceptibility | | | | |
| | Raw - HBRbs | CGM | 0.99 | < 0.0001 |
| | | SWM | 0.98 | < 0.0001 |
| | | TWMB | 0.99 | < 0.0001 |
| | Raw - EBS | CGM | 0.98 | < 0.0001 |
| | | SWM | 0.97 | < 0.0001 |
| | | TWMB | -0.23 | > 0.05 |
| | HBRbs - EBS | CGM | 0.96 | < 0.0001 |
| | | SWM | 0.96 | < 0.0001 |
| | | TWMB | -0.02 | > 0.05 |

CGM: Cortical Grey Matter, SWM: Superficial White Matte, TWMB: Tractography White Matter Bundle, EBS: Empirical Bayes Statistics Harmonization, HBRbs: Hierarchical Bayesian-based B-spline Harmonization.